\documentclass{ptephy_v1}

\usepackage{mathrsfs}

\begin{document}

\title{Evaporation of a nonsingular Reissner-Nordstr\"{o}m black hole\\
  and information loss problem}

\author{Kensuke Sueto${}^{1}$}
\author{Hirotaka Yoshino${}^{1,2}$}

\affil{${}^1$Department of Physics, Osaka Metropolitan University, Osaka 558-8585, Japan}

\affil{${}^2$Nambu Yoichiro Institute of Theoretical and Experimental Physics (NITEP),
Osaka Metropolitan University, Osaka 558-8585, Japan}



%
%
\begin{abstract}
  One of the attractive solutions to the information loss problem
  is that the event horizon does not appear
  in the process of gravitational collapse and
  subsequent evaporation once the spacetime singularity
  is regularized by some mechanism, as pointed out by Hayward
  and Frolov. 
  In this paper, we examine whether this Hayward-Frolov scenario
  holds for the evaporation of a charged black hole.
  The process of collapse and evaporation is modeled with the
  charged Vaidya spacetime and 
  two kinds of regularization of the central singularity
  are considered.
  Analyzing the spacetime structure of the evaporating black hole,
  we find that the appropriately regularized
  evaporating Reissner-Nordstr\"{o}m
  ``black hole'' has no event and Cauchy horizons, indicating the
  possibility that the Hayward-Frolov scenario may have sufficient
  generality as the solution to the information loss problem. In addition, the properties of the non-singular  evaporating Reissner-Nordstr\"{o}m black hole are examined in detail.
\end{abstract}

\subjectindex{E31, E03, E05}

\maketitle

%
%

\section{Introduction}
\label{Sec:Introduction}

The theory of general relativity 
has broad applicability to gravitational phenomena such as 
cosmology, black holes and gravitational waves,
but it is not expected to be
a complete theory of gravity because the singularity theorems
predict the formation of spacetime singularities
in fairly generic situations such as inside of black holes.
In a Schwarzschild spacetime,
a spacelike singularity is located at $r=0$,
and the spacetime is not extendible beyond that singularity.

%
\begin{figure}
  \centering
  \includegraphics[width=0.5\textwidth,bb=0 0 409 453]{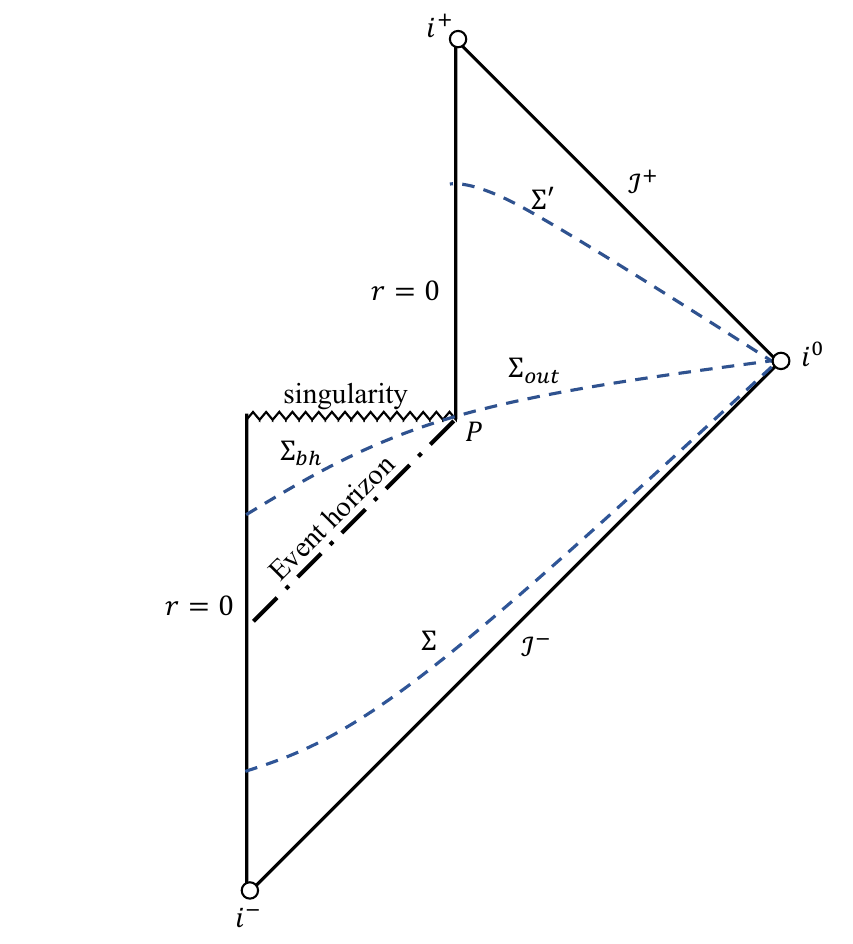}
  \caption{Penrose diagram for the formation and the evaporation
  of a black hole.}
  \label{Penrose-diagram-standard}
\end{figure}
%

Another problem of general relativity arises
if we consider the effects
of quantum fields in a curved spacetime.
Hawking has shown that the black hole emit quantum particles
with a thermal spectrum,
which is called the Hawking radiation \cite{Hawking:1974,Hawking:1975}.
Due to this process, a
black hole 
is expected to evaporate if the backreaction
of the radiation is taken into account. This process appears to
break the unitarity of quantum theories.
Here, we briefly describe the reason following the argument of
Ref.~\cite{Chakraborty:2017pmn}.
Figure \ref{Penrose-diagram-standard} shows the typical
Penrose diagram of the spacetime
for the formation and the subsequent evaporation
of a Schwarzschild black hole \cite{Hawking:1975,Kodama:1979,Kuroda:1984}.
Let $\Sigma$, $\Sigma_P$, $\Sigma'$ be spacelike hypersurfaces
before the event horizon is formed,
at the moment the black hole evaporates completely,
and after the evaporation, respectively,
as shown in the figure.
Since there is no causal connection between the inside and the outside
of the event horizon, the hypersurface $\Sigma_P$ can be divided
as $\Sigma_P=\Sigma_{\rm bh}+\Sigma_{\rm out}$, where $\Sigma_{\rm bh}$
and $\Sigma_{\rm out}$
denote parts of the hypersurface in
the inside and outside regions of the horizon, respectively.
Since $\Sigma_{\rm bh}$ disappears as the black hole evaporates,
the evolution from $\Sigma$ to $\Sigma'$ implies
a transition from the pure state to the mixed state.
This breaks the unitarity, which is the
basic principle of quantum theories.  
This problem is called the black hole information loss problem
pointed out by Hawking in 1976 \cite{Hawking:1976}, and 
since then, the information loss problem has been actively
discussed.

Many solutions to the information loss problem have been  
proposed so far (see, e.g., \cite{Chakraborty:2017pmn,Raju:2020,Calmet:2022} for reviews).
In this paper, we would like to focus attention
to the scenario proposed by Hayward \cite{Hayward:2005}.
The basic idea is as follows. 
In the example of the evaporation of a Schwarzschild black hole,
the violation of unitarity is caused by the disappearance
of the spacelike hypersurface inside of the event horizon
due to the formation of a spacelike singularity.
The singularity formation would occur
because Einstein's equations, i.e. the equations
of the classical theory of gravity,
are applied to the regime where 
quantum gravity effects would become important.
Many researchers would believe that
the paradox should be resolved if a complete theory of quantum gravity
appears.
In such a complete theory, the spacetime singularity is expected to
be resolved, and Hayward assumed that the spacetime structure
is described with an effective metric without divergence
of the curvature tensor. Hayward called such spacetime structure
``the nonsingular black hole'' and in this paper, we
also call it ``the regularized black hole'' .
Modeling the gravitational collapse and the
subsequent evaporation of a black hole
by combining Vaidya spacetimes with a 
naturally regularized metric at $r=0$, 
the spacetime structure was shown to 
drastically change from that of Fig.~\ref{Penrose-diagram-standard}.
The resultant spacetime has no event horizon
although an apparent horizon is present
(see Fig.~5 of \cite{Hayward:2005}).
This model was independently reconstructed by Frolov
(see arXiv 1st version of \cite{Frolov:2014jva}), and hence, 
we call this scenario
the {\it Hayward-Frolov scenario} in this paper. 
This scenario has an attractive feature in that the
information loss problem can be solved with a minimal
change in the metric at the central singularity.
See also Refs.~\cite{Frolov:1981,Roman:1983,Bambi:2013,Rovelli:2014,Bambi:2016,Frolov:2016,Frolov:2017,Kawai:2017,Binetruy:2018,Ho:2019,Kawai:2020} for models that share the similar features.\footnote{See also Ref.~\cite{Brahma:2019}
for the discussion that pointed out the difficulty in realizing
the Hayward-Frolov scenario within the framework of general relativity.}

Here, we point out that
whether this scenario holds for generic spacetimes
must be questioned. 
For example, the Hayward-Frolov scenario has been studied
just for spherically symmetric systems without electric charge. 
In realistic situations, 
a star under the gravitational collapse would be rotating, and hence, 
would have angular momentum, and 
the region outside of a collapsing star
is expected to settle to a Kerr spacetime.
The Kerr spacetime has two horizons at
$r=r_{\pm} := M\pm \sqrt{M^2-a^2}$, where $M$ is the
mass and $Ma$ is the angular momentum.
Here, $r=r_+$ is the event horizon, and $r=r_-$
corresponds to the Cauchy horizon. At the Cauchy horizon, the spacetime
singularity is expected to develop
if a perturbation is added due to the 
mass inflation \cite{Hiscock:1981,Poisson:1990,Ori:1991-1}, which corresponds to the
fact that an observer crossing the Cauchy horizon
would see  phenomena of infinite duration of time
in the outside region
at an instant.
Then, do we have to regularize the spacetime
at a larger scale (i.e., near $r=r_-$) in the Hayward-Frolov scenario
when the spacetime has the angular momentum?
If so, does the resultant spacetime conserve information?
Or, the resolution of the spacetime singularity
at $r=0$ is sufficient?
Such a question is the basic motivation of this paper.


To study this problem, it is convenient to
begin with a simpler model.
The Reissner-Nordstr\"om spacetime
with the mass $M$ and the charge $Q$
also possesses both the event horizon and the
Cauchy horizon at $r=r_+$ and $r=r_-$, respectively,
where $r_{\pm} = M\pm \sqrt{M^2-Q^2}$.
Since the Reissner-Nordstr\"om spacetime
is spherically symmetric, it can be
handled more easily compared to the Kerr spacetime,
while it possesses the similar feature to the Kerr spacetime.
For this reason,
we consider the gravitational collapse
and the subsequent evaporation of charged matter.
The spacetime is modeled by combining charged
Vaidya solutions 
with a regularized metric at the center.


Here, we have to comment on the existing works by Kaminaga \cite{Kaminaga:1988pg}
and Levin and Ori \cite{Levin:1996qt} which investigated
the spacetime structure of a collapsing charged matter 
and its subsequent evaporation, without
regularizing the metric (see also \cite{Parikh:1998ux,Hong:2008}).
They considered the cases where the spacetime
has self-similarity. Kaminaga studied the 
simple case (called the model (A) in  Ref.~\cite{Levin:1996qt}) 
where the collapsing charged matter and the charged negative-energy
Hawking particles continue to fall into the center, $r=0$.
Levin and Ori took account of the possibility that  
infalling charged matter and charged 
negative-energy Hawking particles might experience
bounce towards the outward direction 
due to electromagnetic interactions
that had been pointed out in Ref.~\cite{Ori:1991-2}
(see also Refs.~\cite{Booth:2015,Chatterjee:2015,Creelman:2016}), and  
studied the Penrose diagram for such a case
(called the model (B) in \cite{Levin:1996qt}).
In both cases, a naked singularity appears
in a spacetime.

Motivated by such existing studies and discussions, the purposes of this
paper are the following three. First, 
we discuss the simplest regularization
of the spacetime singularity at $r=0$
of an ingoing charged Vaidya spacetime.
Since the same discussion can be applied
for the static spacetime,
we first discuss the regularization of the
Reissner-Nordstr\"om spacetime.
We consider fairly general form of the metric functions,
and derive the simplest one.
The derived simplest metric turns out to
correspond to the metric
that is presented in Ref.~\cite{Frolov:2016pav}
as an example of the regularized metric
of the Reissner-Nordstr\"om spacetime.

Next, we study the spacetime structure for 
the collapse and the subsequent evaporation
of a charged star with the regularized metric at the center.
Due to a technical reason, we 
do not consider the bounce of the null charged matter,
and hence, focus only on Kaminaga's model, or the model (A). 
We consider two kinds of regularizations at the center.
One is the regularization such that
the self-similarity of Kaminaga's model is maintained. 
In this case, the typical scale $\ell$ of the regularization
is time dependent. It turns out that this method is not
sufficient to resolve the information loss problem
since the spacetime singularity still appears.
The other is that the typical scale of the regularization $\ell$
is constant and does not depend on time.
Since the constancy of $\ell$
violates the self-similarity of the spacetime,
this case cannot be handled analytically and 
numerical calculations are required.
In this case, the spacetime does not possess both
the event horizon and the Cauchy horizon,
and hence, has the ideal feature for resolving the
information loss problem. Our result indicates that
the Hayward-Frolov scenario would hold if an appropriate
regularization is assumed.

Finally, we study the properties of evaporating non-singular Reissner-Nordstr\"om spacetime.  Frolov pointed out interesting properties in an evaporating Hayward ``black hole'' spacetime without an event horizon, i.e. the existence of the repeller (or equivalently, the quasi-horizon) and the attractor of outgoing null geodesics \cite{Frolov:2014jva}. Due to the existence of the attractor, extremely strong blueshift of outcoming photons occurs at the last stage of evaporation. We study how such properties depend on the charge in our system. 

This paper is organized as follows.
In the next section, we review the spacetime structures
of the collapse of the charged matter and its evaporation
studied by Kaminaga~\cite{Kaminaga:1988pg},
since it is closely related to our study in this paper.
In Sect.~\ref{Sec:regularization},
we discuss how to regularize the metric of
a Reissner-Nordstr\"om spacetime, and explain 
two methods of the regularization of charged Vaidya spacetimes
that will be used in this paper.
In Sect.~\ref{Sec:Self-similar-model}, we study
the self-similar model of the collapse and the evaporation
of charged matter where the scale $\ell$ of the regularization
of the central singularity is time dependent. In
Sect.~\ref{Sec:Non-self-similar}, 
the case where $\ell$ is time independent is analyzed.
Section~\ref{Sec:summary} is devoted to a summary
and discussions.
In Appendix~\ref{Appendix:Self-similar}, 
the detailed calculation of the homothetic Killing vector field
in the self-similar model is presented, and 
in Appendix~\ref{Appendix:AH}, how to
solve for the location of the apparent horizons in Vaidya spacetimes
is explained.
We present the detailed calculations for the structures of
the collapse domain in Appendix~\ref{Appendix:Collapse-domain}.
In Appendix~\ref{Appendix:Geometrical-quantities}, 
calculations for the Riemann invariants 
of regularized charged Vaidya spacetimes are presented.
Furthermore, we examine the properties
of the Einstein tensor of regularized charged Vaidya spacetimes
and in particular, discuss
in which regions the null and dominant energy conditions are
satisfied if these spacetimes are
realized in the framework of general relativity.
The characteristics of the apparent horizons in these spacetimes
are also discussed.
In Appendix~\ref{Appendix:Outer-domain}, we discuss
how to connect the inner and outer domains, in which ingoing
and outgoing Hawking fluxes are present, respectively,
along the pair-creation surface.
In Appendix~\ref{Appendix:Penrose_different},
we present the Penrose diagram that is drawn in a different
method from that of the main text. 
Throughout the paper, we use the unit in which
the speed of light and the gravitational constant
are unity, $c=G=1$.

%
%
\section{Review of evaporation of a Reissner-Nordstr\"om black hole}
\label{Sec:review-part}

A toy model for the collapse and evaporation
of a charged star were presented by
Kaminaga~\cite{Kaminaga:1988pg} 
(see also Levin and Ori~\cite{Levin:1996qt})
by cutting and gluing the ingoing and outgoing charged Vaidya spacetimes
and the Reissner-Nordstr\"om spacetime. 
Here, we briefly review this model  
since it is closely related to
our analysis in Sect.~\ref{Sec:Self-similar-model}.

\subsection{Charged Vaidya spacetimes}

We begin with reviewing the charged Vaidya spacetimes \cite{Bonnor:1970}. 
The ingoing charged Vaidya spacetime is
described by the metric
\begin{equation}
  ds^2 = -F_-(v,r)dv^2+2dr dv+r^2d\Omega^2
  \label{Metric:IngoingVaidya}
\end{equation}
with
\begin{equation}
  \label{eq-Kaminaga-f}
F_-(v,r)=1-\frac{2M(v)}{r}+\frac{Q^2(v)}{r^2},
\end{equation}
where $d\Omega^2$ is the standard metric of a unit sphere,
$d\Omega^2=d\theta^2+\sin^2\theta d\phi^2$, 
in the spherical-polar coordinates $(\theta,\phi)$.
Here, the mass and charge functions, $M(v)$ and $Q(v)$,
can be freely chosen. 
This metric describes the structure of
a spherically symmetric spacetime 
with infalling charged null-matter fluid.
The electromagnetic field is generated by the charged matter,
and the only nonzero component of the electromagnetic field strength
$\mathcal{F}_{\mu\nu}$ is
\begin{equation}
  \mathcal{F}_{rv}\,=\, -\mathcal{F}_{vr} \, = \,
  \frac{Q(v)}{r^2}.
  \label{electromagnetic-field-strength-tensor}
\end{equation}
The energy-momentum tensor $T_{\mu\nu}$ is decomposed into two parts,
\begin{equation}
  T_{\mu\nu}=T^{\rm (e)}_{\mu\nu}+T^{\rm (m)}_{\mu\nu},
  \label{charged-ingoing-Vaidya-Tmunu-spilit}
\end{equation}
where $T^{\rm (e)}_{\mu\nu}$ is the standard
energy-momentum tensor for the electromagnetic field,
\begin{equation}
  T^{\rm (e)}_{\mu\nu}\ = \
  \frac{1}{4\pi}\left(\mathcal{F}_{\mu\rho}{\mathcal{F}_\nu}^\rho-\frac14g_{\mu\nu}\mathcal{F}_{\rho\sigma}\mathcal{F}^{\rho\sigma}\right),
\end{equation}
and
$T^{\rm (m)}_{\mu\nu}$ is the matter part. For the field
strength given by Eq.~\eqref{electromagnetic-field-strength-tensor},
$T^{\rm (e)}_{\mu\nu}$ takes the form
\begin{equation}
  {T^{\rm (e)\,\mu}}_{\nu}
  \,=\,  \mathrm{diag}(-\tilde{\rho},\,\tilde{p}_r,\,\tilde{p}_\theta,\,\tilde{p}_\phi)
  \label{energy-momentum-tensor-EM}
\end{equation}
with
\begin{equation}
  \tilde{\rho} \,=\, -\tilde{p}_r \,=\, \tilde{p}_\theta \,=\, \tilde{p}_\phi
  \,=\, \frac{Q^2}{8\pi r^4},
\end{equation}
where $\tilde{\rho}$, $\tilde{p}_r$, $\tilde{p}_\theta$, and $\tilde{p}_\phi$
denote the energy density and the pressures in the $r$, $\theta$, and $\phi$
directions due to the electromagnetic field, respectively. 
In order to express the matter part, it is convenient to introduce
the ingoing radial null vector
\begin{equation}
  k^\mu=-(\partial_r)^\mu,
  \label{ingoing-radial-null-vector}
\end{equation}
and its dual vector $k_{\mu} = -(dv)_\mu$. In terms of $k^\mu$,
$T^{\rm (m)}_{\mu\nu}$ is expressed as 
\begin{equation}
  T^{\rm (m)}_{\mu\nu} = \tilde{\nu}\, k_{\mu}k_{\nu},
  \label{energy-momentum-tensor-matter}
\end{equation}
with
\begin{equation}
  \tilde{\nu} =
  \frac{1}{4\pi r^2}\left({M}^\prime-\frac{Q{Q}^\prime}{r}\right).
\end{equation}
Here, the prime ($\prime$) denotes the
ordinary derivative, i.e. $M^\prime=dM/dv$ in this case.
This matter is the charged null fluid having the energy density
$\tilde{\nu}$
(for the observer whose four-velocity $v^\mu$ satisfies $k^\mu v_\mu = -1$)
which propagates from $r=\infty$ to $r=0$ along $v=\mathrm{constant}$
surfaces. 
From the Maxwell equation, the four-electric current is given as
\begin{align}
  j^\mu\, =\, -\frac{1}{4\pi}\nabla_\nu \mathcal{F}^{\nu\mu}\,=\,\frac{Q^\prime}{4\pi r^2}\, k^\mu,
\end{align}
which is a null vector.
This reflects the fact that 
the charged null matter carries the electric charge at the speed of light.

Similarly, the metric of the outgoing Vaidya spacetime is given
by
\begin{equation}
  ds^2 = -F_+(u,r)du^2-2dr du+r^2d\Omega^2,
  \label{Metric:OutgoingVaidya}
\end{equation}
where
\begin{equation}
F_+(u,r)=1-\frac{2M(u)}{r}+\frac{Q^2(u)}{r^2}.
\end{equation}
This metric describes the structure of
a spacetime 
with outgoing charged null-matter fluid.

\subsection{Kaminaga's self-similar model}
\label{Subsec:Evaporation_model_A}

A model of an evaporating charged black hole
due to Hawking radiation was proposed by Kaminaga~\cite{Kaminaga:1988pg}.
This model was called the model (A) by Levin and Ori \cite{Levin:1996qt},
and here, we give a detailed review on this model
because the extension of this model will be discussed
in Sect.~\ref{Sec:Self-similar-model}.
The setup of the system in this paper is slightly different from those
of Refs.~\cite{Kaminaga:1988pg} and \cite{Levin:1996qt}.

\subsubsection{Setup}

In this model, a charged black hole is formed by
the gravitational collapse of charged null-matter fluid
that is described by the charged Vaidya solution.
After the collapse, Hawking particles
are assumed to be created at a timelike hypersurface
that is slightly outside of the apparent horizon.
Outside of that timelike hypersurface,
the Hawking particles constitute an outgoing null matter fluid
with positive outgoing energy flux,
while inside of that timelike hypersurface,
they constitute ingoing null-matter fluid
with (initially) ingoing negative energy flux.\footnote{
  We commented ``initially'' because 
  due to the matter part of the energy-momentum tensor
  of Eq.~\eqref{energy-momentum-tensor-matter},
  the energy density changes its sign at some certain radius.
This property plays an important role in the discussion of Levin and Ori \cite{Levin:1996qt}.}
The outside and inside regions are modeled by the
outgoing and ingoing charged Vaidya spacetimes, respectively.

It is convenient to study the ingoing charged Vaidya spacetime
with the mass function\footnote{Precisely speaking,
  Kaminaga adopted the case of $v_{\rm i}=0$ (i.e., instantaneous collapse)
  in Ref.~\cite{Kaminaga:1988pg}. Also, the period where the mass 
  function linearly decreases is assumed to end before $v=v_{\rm f}$, and $M(v)$ suddenly jumps from a positive value to zero.
  For this reason, the presented Penrose diagram is different from
  Fig.~\ref{Fig:Penrose-Vaidya-w-HKH} of this paper.
  The mass function of Levin and Ori of Ref.~\cite{Levin:1996qt}
  is the same as Eq.~\eqref{Linear-mass-function} for $v>0$,
  but they assumed 
  the spacetime to be initially the Reissner-Nordstr\"om spacetime
  (without the gravitational collapse).}
\begin{equation}
  M(v)=
  \begin{cases}
    0\quad &(v\le-v_{\rm i}),\\
    M_0(1+v/v_{\rm i})\quad &\left(-v_{\rm i}\le v\le 0 \right),\\
    M_0(1-v/v_{\rm f})\quad &\left(0\le v\le v_{\rm f} \right),\\
    0\quad  &(v\ge v_{\rm f}),\\
  \end{cases}
  \label{Linear-mass-function}
\end{equation}
and the charge function, 
\begin{equation}
  Q(v)\,=\,qM(v).
  \label{Linear-charge-function}
\end{equation}
Since we consider the formation of a sub-extremal black hole,
$|q|<1$ is assumed. 
Once the structure of this ingoing charged Vaidya spacetime
is clarified, the spacetime of
an evaporating charged black hole can be easily
constructed with a minor modification
by cutting and gluing this spacetime,
the outgoing charged Vaidya spacetime, and the Reissner-Nordstr\"om
spacetime (interested readers
are referred to Fig.~\ref{Fig:Penrose-Vaidya-wo-HKH} in advance).
Note that the linear dependence of $M(v)$ and $Q(v)$ on
the advanced time $v$ is assumed by a technical reason.
With this choice of functions, the
spacetime becomes self-similar and the analysis becomes
easy (see below).
In the mass function of Eq.~\eqref{Linear-mass-function}, on the one hand,
the gravitational collapse occurs in the domain
$-v_{\rm i}\leq v\leq 0$,
and we call this domain the collapsing phase. On the other hand,
the negative energy flux of the ingoing Hawking
particles is present in the domain $0\le v\le v_{\rm f}$,
and we call this domain the evaporating phase.
The other domains are the Minkowski spacetimes.
We focus our attention particularly to the domain
$0\le v\le v_{\rm f}$ where 
the ingoing energy flux is negative,
because that domain is expected to describe
the spacetime structure of the evaporating
black hole with backreaction.

\subsubsection{Double null coordinates}

In the domain $0\le v\le v_{\rm f}$,
we introduce two new coordinates $\tilde{v}$ and $R$ by
\begin{equation}
  \tilde{v}=\frac{M(v)}{M_0}, \quad R=\frac{r}{M(v)}.
  \label{Eq:tildev-R}
\end{equation}
Note that $0\le v\le v_{\rm f}$ corresponds to $1\ge \tilde{v}\ge 0$. 
Then, the function $F_-(v,r)$ in the metric of Eq.~\eqref{Metric:IngoingVaidya} is rewritten as
 \begin{align}
    F_-(v,r)
    &=1-\frac{2}{R}+\frac{q^2}{R^2}=:f(R).
    \label{Eq:definition-fR}
 \end{align}
 In these coordinates, the metric becomes
\begin{align}
  ds^2
  &=-M_0^2\alpha\tilde{v}(f\alpha+2R)d\tilde{v}\left(\frac{d\tilde{v}}{\tilde{v}}+\frac{2dR}{f\alpha+2R}\right) + r^2d\Omega^2,
  \label{Eq:metric-in-tildev-R}
\end{align}
where $\alpha$ denotes the duration of evaporation normalized by 
the initial mass,
\begin{equation}
  \alpha=v_{\rm f}/M_0.
\end{equation}
From this formula, the retarded time $u$ can be introduced by
\begin{align}
  u=\ln{\tilde{v}}+\int{\frac{2dR}{f\alpha+2R}}.
  \label{Eq:Introduction-u}
\end{align}
Then, in the double null coordinates $(u,\tilde{v})$, 
the metric is expressed as
\begin{equation}
  ds^2=-M_0^2\alpha\tilde{v}(f\alpha+2R)dud\tilde{v}+ r^2d\Omega^2.
  \label{Eq:metric-in-double-null}
\end{equation}

%
\begin{figure}[tb]
  \centering
  \includegraphics[width=0.4\textwidth,bb=0 0 260 171]{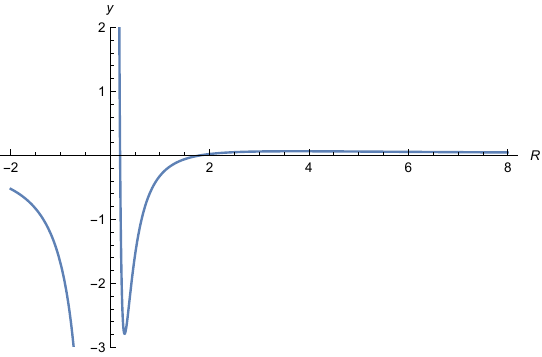}
  \caption{The graph of $y(R)$ for the case $q=0.6$.}
  \label{Fig:y}
\end{figure}
%

The spacetime structure
changes drastically depending on the number of zeros
of the 
function $f\alpha + 2R$.
The equation $f\alpha + 2R=0$
is rewritten as 
\begin{align}
  \label{eq-formation-evaporation}
-\frac{1}{\alpha}\,=\,\frac{R^2-2R+q^2}{2R^3}=:y(R).
\end{align}
The graph of the function $y(R)$ is shown in Fig.~\ref{Fig:y}
for the case $q = 0.6$. 
$y(R)$ takes extremal values at $R=R_\pm:=2\pm\sqrt{4-3q^2}$,
and $y(R_-)<0$ and $y(R_+)>0$
are satisfied.
Since $\alpha$ is a positive constant, the left-hand side
of Eq.~\eqref{eq-formation-evaporation} is negative, and 
Eq.~\eqref{eq-formation-evaporation}
has one, two, and three solutions
in the cases that $1/\alpha>|y(R_-)|$,
$1/\alpha=|y(R_-)|$,
and $1/\alpha<|y(R_-)|$, respectively, where
\begin{equation}
  |y(R_-)|
  \,=\,
  -\frac{2-q^2-\sqrt{4-3q^2}}{(2-\sqrt{4-3q^2})^3}.
\end{equation}
When $f\alpha + 2R=0$ has one root,
on the one hand, the root $R_0$ is negative, $R_0<0$, and 
the function $f\alpha + 2R$ is always positive 
in the range $0< R<\infty$. 
In this case, the null coordinate $u$ spans the
whole domain of $1\ge \tilde{v}\ge 0$.
On the other hand, if $f\alpha + 2R=0$ has three roots, which are
denoted by $R_0$,
$R_{\rm C}^-$ and $R_{\rm C}^+$ with
$R_0<0<R_{\rm C}^{-}<R_{\rm C}^{+}$, 
the coordinate $u$ diverges
to $\mp\infty$ in the limit $R\to R_{\rm C}^{\pm}$
with a fixed $\tilde{v}$. 
Explicitly, the relation between $u$, $\tilde{v}$, and
$R$ is written as
\begin{equation}
  u\,=\,\ln\tilde{v}
  +\frac{1}{\kappa_0}\ln|R-R_0|
  -\frac{1}{\kappa_-}\ln|R-R_{\rm C}^-|
  +\frac{1}{\kappa_+}\ln|R-R_{\rm C}^+|,
  \label{Eq:formula-for-u}
\end{equation}
where
\begin{subequations}
\begin{eqnarray}
\kappa_0 &=& \frac{(R_{\rm C}^+-R_{0})(R_{\rm C}^--R_{0})}{R_0^2},\\
\kappa_\pm &=& \frac{(R_{\rm C}^+-R_{\rm C}^-)(R_{\rm C}^\pm-R_{0})}{\left(R_{\rm C}^\pm\right)^2}.
\end{eqnarray}
\end{subequations}
The divergence of $u$ in the limit $R\to R_{\rm C}^\pm$ 
is just the coordinate effect. In fact, 
the continuous null coordinates $U_+$ and $U_-$
across $R=R_{\rm C}^+$ and $R=R_{\rm C}^-$,
respectively, 
can be locally introduced by
\begin{equation}
  U_\pm=\left\{\begin{array}{cc}
  \exp(\pm\kappa_\pm u) & (R>R_{\rm C}^\pm),\\
  -\exp(\pm\kappa_\pm u) & (R<R_{\rm C}^\pm),
\end{array}
  \right.
  \label{Continuous-coordinate-across-Rpm}
\end{equation}
in which the metric takes the regular form
\begin{equation}
ds^2 \,=\, - \frac{2M_0^2\alpha\tilde{v}^{1\mp\kappa_\pm}(R-R_0)^{1\mp\kappa_\pm/\kappa_0}\left|R-R_{\rm C}^\mp\right|^{1+\kappa_\pm/\kappa_\mp}}{\kappa_{\pm}R^2}dU_{\pm}d\tilde{v}.
\end{equation}
The important feature 
is that if we decrease $\tilde{v}$ to zero 
for a fixed $u$ in the domain $0<R<R_{\rm C}^+$,
the value of $R$ approaches $R_{\rm C}^-$
because $\kappa_0$, 
$\kappa_+$ and $\kappa_-$ are all positive in Eq.~\eqref{Eq:formula-for-u}.
Since this limit corresponds to $r=0$ because
of the definition of $R$ of Eq.~\eqref{Eq:tildev-R},
there is a null singularity.  
This feature is used in drawing the Penrose diagram later.
In Section~\ref{Appendix:diagram of linear mass},
we present the behavior of outgoing null geodesics,
each of which gives a $u$-constant surface,
in the $(v,r)$-plane for the cases that
the homothetic Killing horizons are present and absent.
This would help us to understand the properties of this spacetime.

\subsubsection{Homothetic Killing vector field}

In the domain where $M(v)$ changes linearly,
a homothetic Killing vector field $\xi^\mu$ is
present. The homothetic Killing vector field is the
vector field that satisfies the equation
\begin{equation}
  \nabla_{\mu}\xi_\nu+\nabla_\nu\xi_\mu=K g_{\mu\nu}
  \label{Eq:homothetic_Killing_equation}
\end{equation}
for some constant $K$ (if $K$ is not a constant, $\xi^\mu$ is
called the conformal Killing vector field).
The presence of the homothetic Killing field 
is proven in Appendix~\ref{Appendix:Self-similar}.
The contravariant components of the homothetic Killing vector is given by
\begin{equation}
  \xi^\mu=-(1,\tilde{v}, 0, 0),
  \label{homothetic_Killing_contravariant}
\end{equation}
in the $(u,\tilde{v},\theta,\phi)$ coordinates 
and satisfies
Eq.~\eqref{Eq:homothetic_Killing_equation} with $K=-2$. 
The norm of this vector is
\begin{align}
  g_{\mu\nu}\xi^{\mu}\xi^{\nu}&=-M_0^2\alpha\tilde{v}^2(f\alpha+2R).
\end{align}
A homothetic Killing horizon is defined
as a surface on which $\xi^\mu$ becomes null,
and its location satisfies $f\alpha+2R=0$
or $\tilde{v}=0$ in the present system.
Whether there exist homothetic Killing horizons
except for $\tilde{v}=0$ 
depends on the parameter $\alpha$ and $q$.
If $\alpha$ is sufficiently large and $q^2$ is
sufficiently small, there exist two solutions
for $f\alpha+2R=0$ in the domain $R>0$
which are $R_{\rm C}^{\pm}$ introduced above.

Here, we point out that the vector field $\xi^\mu$ is tangent to the
$R$-constant hypersurfaces as can be checked from
Eq.~\eqref{homothetic_Killing_contravariant} and 
the definition of $u$ of Eq.~\eqref{Eq:Introduction-u}.
%
Due to the spherical symmetry, the characteristic of $\xi^\mu$
(i.e., timelike, null, or spacelike) and that of an $R$-constant hypersurface
are identical to each other. Therefore, we find that 
$\xi^\mu$ and $R$-constant hypersurfaces are 
(i) timelike if $f\alpha+2R>0$;
(ii) null if $f\alpha+2R=0$;
and (iii) spacelike if  $f\alpha+2R<0$.
In the case that a homothetic Killing horizon does not exist
except for $\tilde{v}=0$,
all $R$-constant hypersurfaces are timelike. 
In the case that the two homothetic Killing horizons are present
except for $\tilde{v}=0$,
$R$-constant hypersurfaces are
timelike for $0<R\le R_{\rm C}^-$ and $R_{\rm C}^+<R$,
null for $R=R_{\rm C}^{\pm}$, and spacelike for $R_{\rm C}^-<R<R_{\rm C}^+$.

\subsubsection{Trapped region}

Suppose photons are emitted from $r$-constant surface
at some moment in both inward and outward radial directions. 
The region where both ingoing 
and outgoing null geodesic congruences
have negative expansion 
is called the trapped region,
and the outermost boundary of that region is called
the apparent horizon.
In the present spacetime, the expansion of ingoing null geodesic congruence
is always negative. 
As shown in Appendix~\ref{Appendix:AH},
the expansion of outgoing null geodesic congruence
is proportional to $f(R)$, which is defined in Eq.~\eqref{Eq:definition-fR}. 
The boundaries of the trapped region are
obtained by solving the equation $f(R)=0$. 
If $q^2<1$, there are two solutions
for this equation, which are
\begin{equation}
  R_{\rm A}^{\pm} \,=\, 1\pm\sqrt{1-q^2}.
  \label{Eq:Location-AH-charged-Vaidya}
\end{equation}
The surfaces $R=R_{\rm A}^{-}$
and $R=R_{\rm A}^{+}$ correspond to
the inner boundary 
of the trapped region and the apparent horizon, respectively.
Following Ref.~\cite{Frolov:2014jva}, we call
them the inner and outer apparent horizons in this paper.

\subsubsection{$(v,r)$-diagram}
\label{subsubsec:Numerical-method-linear-mass}

In order to understand the spacetime structure,
it is very helpful 
to present the outgoing null geodesics, homothetic Killing horizons,
and the apparent horizons in the $(v,r)$-plane.
We call such a diagram the ``$(v,r)$-diagram''.
Although the outgoing null geodesics
are given by $u$-constant surfaces determined by Eq.~\eqref{Eq:Introduction-u},
it is useful to develop a code to generate outgoing
null geodesics numerically, since non-self-similar spacetimes,
for which no analytic solution to the geodesic equation is present, 
will be studied later. 
From the null condition, radial outgoing null geodesics satisfy
\begin{align}
  \label{outgoing}
  \frac{d{r}}{d{v}}=\frac{F_-({v},{r})}{2}.
\end{align}
This differential equation can be solved numerically
in the domain ${r}\ge 0$ using the Runge-Kutta method.
Note that there is a difficulty in solving for the outgoing null geodesics 
that is emitted from the curvature singularity at $r=0$.
In fact, the equation is singular at the center $r=0$ as
\begin{equation}
\frac{dr}{dv} \ = \ \frac12\left(1-\frac{2M(v)}{r}+\frac{q^2M(v)^2}{r^2}\right),
\end{equation}
in the range $-v_{\rm i}<v<v_{\rm f}$.
We avoid this singularity by adopting $x=r^3$. The equation
for $x$ is 
\begin{equation}
\frac{dx}{dv} \ = \ \frac32\left[x^{2/3}-2M(v)x^{1/3}+q^2M(v)^2\right],
\end{equation}
and after solving this equation, we obtain $r(v)=x(v)^{1/3}$.

%
\begin{figure}[tb]
  \centering
  \includegraphics[width=0.49\textwidth,bb= 0 0 360 252]{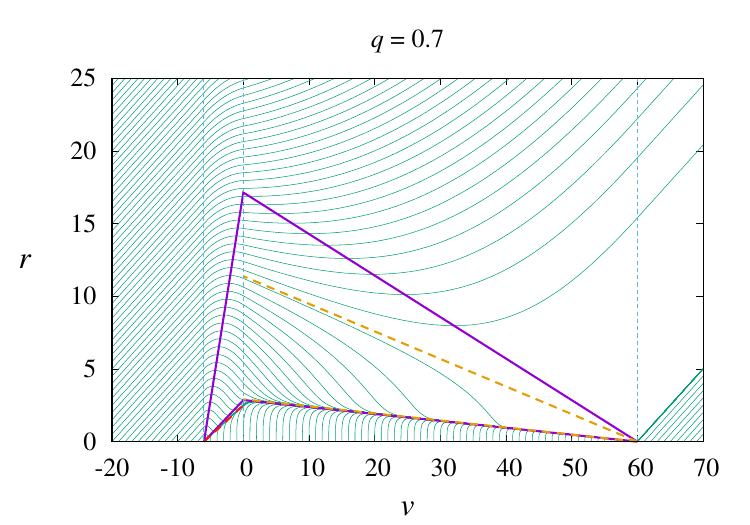}
  \includegraphics[width=0.49\textwidth,bb= 0 0 407 286]{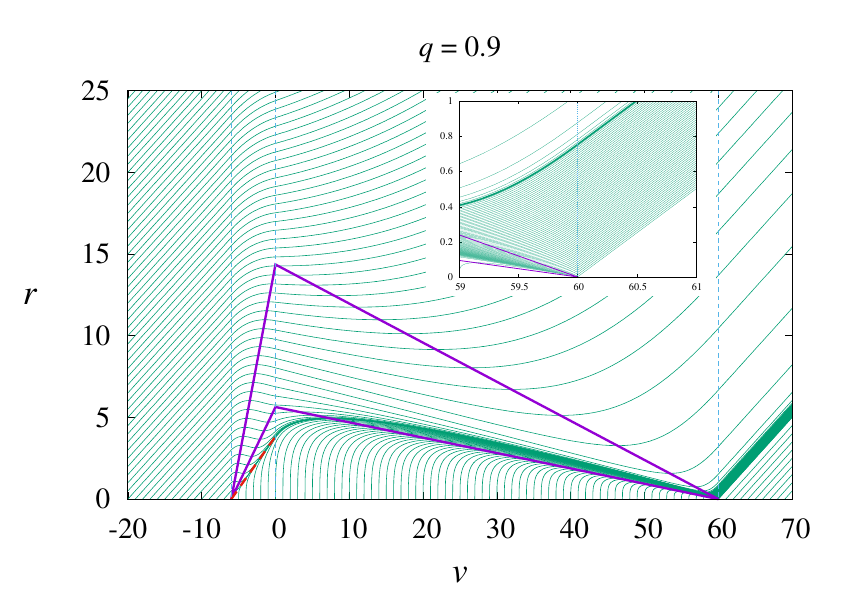}
  \caption{Outgoing null geodesics (thin green curves) and the boundary of the trapped region (closed purple lines) 
    for the ingoing charged Vaidya spacetime (Kaminaga's model) 
    in the cases $q=0.7$ (left panel) and  $q=0.9$ (right panel).
    The parameters $M_0=10$, $v_{\rm i}=6$, and $v_{\rm f}=60$ are adopted. 
    The red dashed line in each panel indicate the homothetic
    Killing horizon in the collapsing phase.
    The orange dashed lines in the left panel shows the homothetic
    Killing horizons in the evaporating phase (no homothetic Killing horizon
    exists in the evaporating phase in the right panel). 
    The vertical dotted lines in each panel indicate
    $v=-v_{\rm i}$, $0$, and $v_{\rm f}$ from left to right.
  }
  \label{geodesic_AH_ell0}
\end{figure}
%

Figure~\ref{geodesic_AH_ell0} shows the $(v,r)$-diagrams
for $q=0.7$ (left panel) and $0.9$ (right panel).
In the diagrams, 
behavior of outgoing null geodesics (thin green curves)
and the boundary of the trapped region (thick purple lines or curve)
are shown. In both panels, 
the parameters $M_0=10$, $v_{\rm i}=6$, and $v_{\rm f}=60$ are adopted.
In the left panel for $q=0.7$, there are two homothetic Killing horizons
in the evaporating phase,
and they are indicated by orange dashed lines. 
These two homothetic Killing horizons coincide with
two of the outgoing null geodesics. 
The outer homothetic Killing horizon clearly separates the outgoing
null geodesics that escape to infinity and those trapped inside
of it, and hence, it is the event horizon as well.
Since the null geodesics recede from the
outer homothetic Killing horizon, it 
plays a role of the repeller of outgoing null geodesics.
Frolov called such a repeller ``quasi-horizon''
(see Fig.~2 of \cite{Frolov:2014jva}) 
and showed that it is well approximated by
the positions where $d^2r/dv^2$ becomes zero
in general ingoing Vaidya spacetimes,
where the outgoing null geodesics are given by $r=r(v)$
(see also Ref.~\cite{Binetruy:2018} for more detailed discussions
on such a separatrix). 
The trapped outgoing null geodesics are strongly attracted
by the inner homothetic Killing horizon, and therefore,
the inner homothetic Killing horizon plays a role of the attractor.
All outgoing null geodesics within the outer homothetic
Killing horizon 
plunge into the singularity at $(v,r)=(v_{\rm f},0)$. This means that
the singularity at $(v,r)=(v_{\rm f},0)$ has an extended structure
in the $u$ direction, and thus, that singularity is a null singularity.

In the right panel for the case of $q=0.9$,
the homothetic Killing horizons do not exist. 
As a result, there is no event horizon in this spacetime, and
all outgoing null geodesics escape to infinity.
The absence of the inner homothetic Killing horizon
implies the non-existence of the attractor.
Although outgoing null geodesics tend to gather
around the inner boundary of the trapped region,
the gathering is not sufficiently strong to form an attractor.
Since no outgoing geodesic plunges into the singularity
at $(v,r)=(v_{\rm f},0)$ from $v<v_{\rm f}$, it does not have an extended structure
in the $u$ direction.

In each of the left and right panels, the red dashed line
represents the homothetic Killing horizon
in the collapsing phase (see Appendix~\ref{Appendix:Collapse-domain}
for an analytic study for the spacetime structure of the collapsing phase).
This homothetic Killing horizon is an attractor
of outgoing null geodesics in the future direction.
As a result, only one outgoing null geodesic
is emitted from the point $(v,r)=(-v_{\rm i},0)$.
This means that the singularity at this point
does not have an extended structure
in the $u$-direction in both cases of $q=0.7$ and $0.9$.

\subsubsection{Penrose diagram}

%
\begin{figure}[tb]
  \centering
  \includegraphics[scale=0.45]{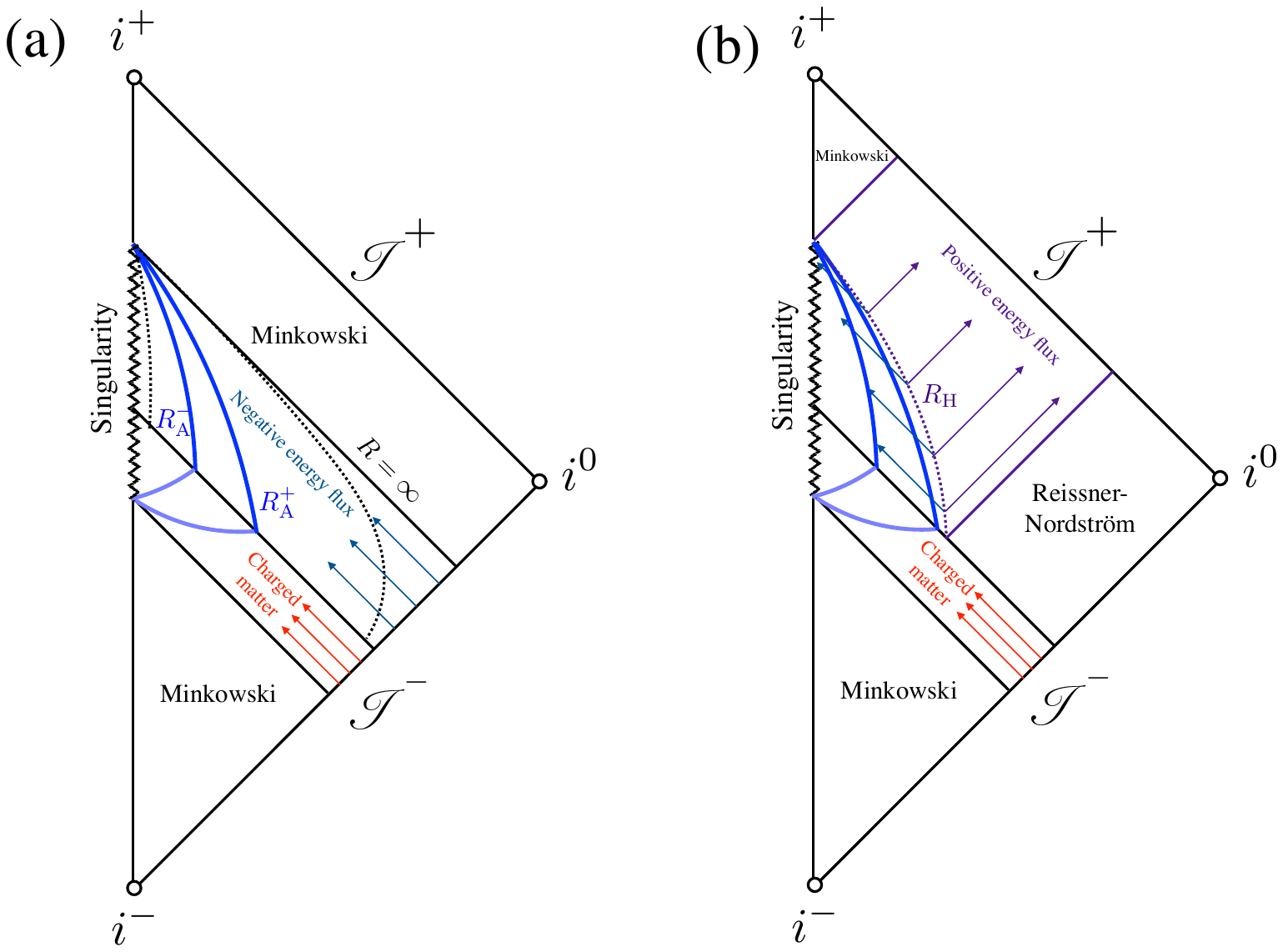}
  \caption{Penrose diagrams for the charged Vaidya spacetime
    (left panel)
    and the model of the Hawking radiation (right panel)
    in the case without 
    a homothetic Killing horizon.}
  \label{Fig:Penrose-Vaidya-wo-HKH}
\end{figure}
%

Summarizing the above informations,
we can now draw the Penrose diagram. We suppose that 
$q^2$ 
is smaller than one, and hence, 
the apparent horizon forms.
The existence of the homothetic Killing horizon 
depends on the value of $\alpha$.
Below, we first show the Penrose diagram 
given by the ingoing charged Vaidya spacetime 
with the mass function of Eq.~\eqref{Linear-mass-function},
and then, discuss how to modify that diagram 
in order to obtain the diagram for  
an evaporating charged black hole.

%
\begin{figure}[tb]
  \centering
  \includegraphics[scale=0.5]{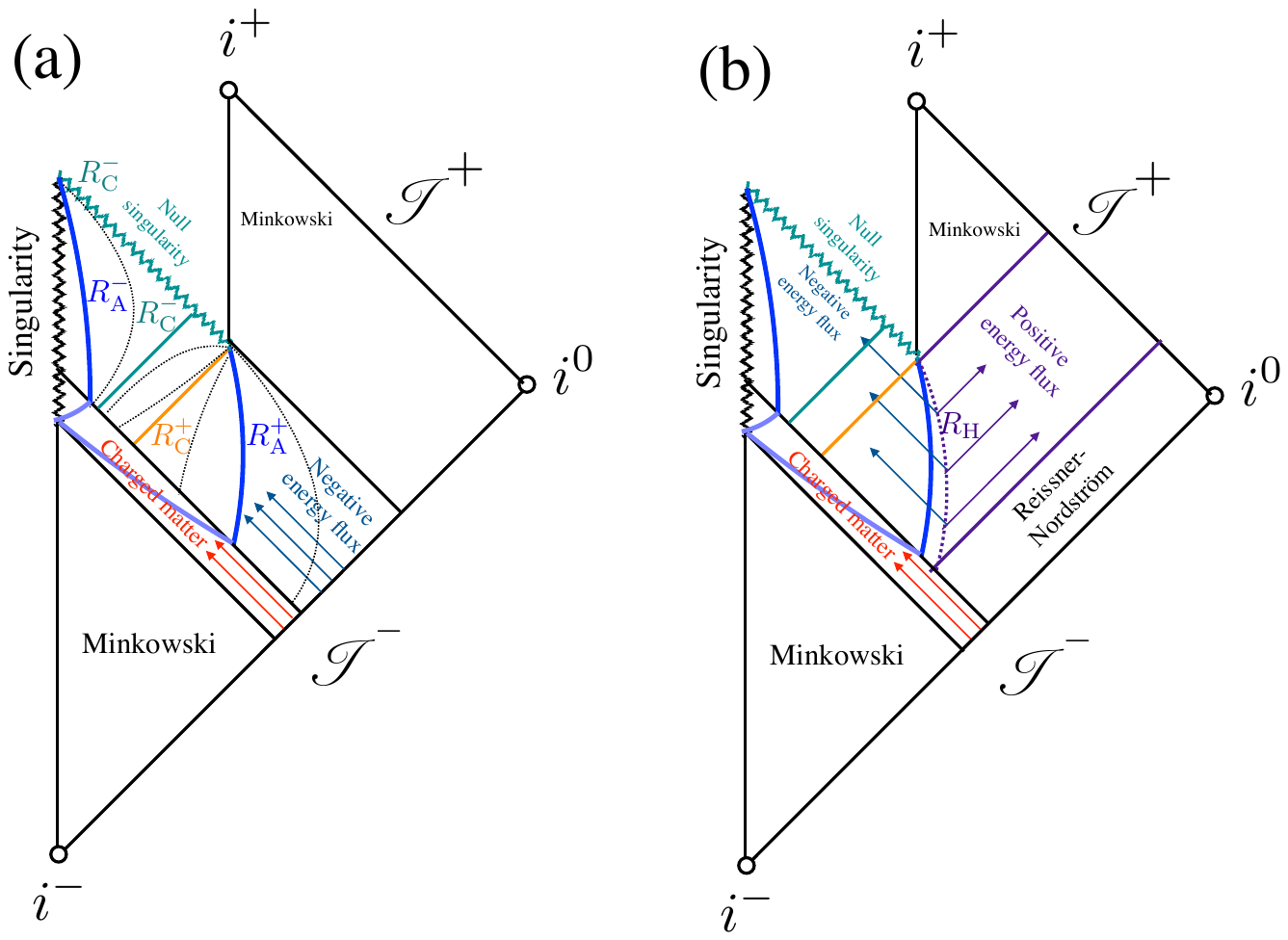}
  \caption{Penrose diagrams for the charged Vaidya spacetime (left panel)
    and the model of the Hawking radiation (right panel) in the case that
    homothetic Killing horizons form.}
  \label{Fig:Penrose-Vaidya-w-HKH}
\end{figure}
%

The left panel of Fig.~\ref{Fig:Penrose-Vaidya-wo-HKH} shows the Penrose diagram for the ingoing charged Vaidya spacetime
in the case that
the homothetic Killing horizon does not exist except for $\tilde{v}=0$,
which corresponds to the right panel of Fig.~\ref{geodesic_AH_ell0}. 
The $R$-constant surfaces are shown by dotted curves,
and they are all timelike in the evaporating phase.
All of the $R$-constant surfaces
shrink to one point, $r=\tilde{v}=0$.
The boundaries of the trapped region, $R=R_{\rm A}^{\pm}$, 
are shown by solid curves.\footnote{In the collapsing phase, the both inner and outer boundaries of the trapped region are spacelike, and in the evaporating phase they are both timelike. One might wonder whether this result is consistent with Hayward's theorem \cite{Hayward:1994} which states that if the null energy condition holds in general relativity, an outer trapping horizon is spacelike and an inner trapping horizon is timelike.  Our result is consistent with Hayward's result, as presented in Appendix~\ref{Appendix:Geometrical-quantities}. }
The trapped region shrinks as $\tilde{v}$ is decreased
and becomes pointlike at $\tilde{v}=r=0$. 
There appears 
a timelike singularity at the center, $r=0$. 
The event horizon is absent, and the 
spacetime possesses the naked singularity that is 
visible to distant observers.

We now turn our attention to the evaporating 
charged black hole. We 
adopt the idealized model such that Hawking particles 
are created at $R=R_{\rm H}$ which is located slightly 
outside of the apparent horizon. 
The created particles propagate outward and inward radial directions,
and the outwardly propagating particles carry positive
energy flux while the inward propagating particles carry negative
energy flux. To realize this situation, we remove the
region $R_{\rm H}< R<\infty$ and the upper Minkowski region
except for the inside of the future light cone of 
$r=\tilde{v}=0$ from the figure of the left panel. 
Then, we glue an appropriate outgoing charged Vaidya solution 
at $R=R_{\rm H}$, and this method is explained 
in Appendix~\ref{Appendix:Outer-domain}. 
For the region outside of the domain swept by the outgoing Hawking particles,
we place the Reissner-Nordstr\"om solution  
with the mass $M_0$ and the total electric charge $Q_0=q M_0$.
The resultant diagram is shown in 
the right panel of Fig.~\ref{Fig:Penrose-Vaidya-wo-HKH}.

The left panel of Fig.~\ref{Fig:Penrose-Vaidya-w-HKH} is the same as Fig.~\ref{Fig:Penrose-Vaidya-wo-HKH}(a) but for the case that two 
homothetic Killing horizons exist in
the domain $1>\tilde{v}>0$.
This case corresponds to the left panel of Fig.~\ref{geodesic_AH_ell0}. 
In this case,
$R$-constant surfaces 
with $R^{-}_{\rm C}< R<\infty$ 
shrink to one point,
while those with $0<R<R^{-}_{\rm C}$
shrink to a different point.
The homothetic Killing horizons, $R=R_{\rm C}^{\pm}$, are null surfaces,
and the surface $R=R_{\rm C}^{-}$ is located
to future of the surface $R=R_{\rm C}^{+}$.
There is a timelike singularity at 
$R=0$, which is located to future of the surface $R=R_{\rm C}^{-}$.
As discussed above, if we decrease $\tilde{v}$ to zero
fixing the retarded time $u$, the value of $R$
approaches $R_{\rm C}^{-}$.
Therefore, the surface $\tilde{v}=0$ 
is composed of two parts,
$R=\infty$ and $R=R_{\rm C}^-$,
and the part $R=R_{\rm C}^-$ is the null singularity.
The region $0<R<R_{\rm C}^+$
is invisible to outside observers.
With the same procedure as Fig.~\ref{Fig:Penrose-Vaidya-wo-HKH},
it is possible to draw the diagram for an 
evaporating charged black hole as shown in the right panel
of Fig.~\ref{Fig:Penrose-Vaidya-w-HKH}. 
A similar diagram is presented in Ref.~\cite{Levin:1996qt}.

In Sects.~\ref{Sec:Self-similar-model}
and \ref{Sec:Non-self-similar},
we consider how these diagrams are changed
if we regularize the singularity at the center
keeping and breaking the self-similar property of the spacetime,
respectively.

%
%
\section{Regularization of Reissner-Nordstr\"{o}m metric}
\label{Sec:regularization}

In this section, we discuss how to regularize the
singularity at the center of electrically charged spacetimes.
First we study the regularization of a 
static Reissner-Nordstr\"{o}m spacetime,
since the same regularization method can be applied
to the ingoing charged Vaidya spacetime as well.
Then, we present the regularization of the
ingoing charged Vaidya spacetime that will be used in
subsequent sections.

\subsection{A method of constructing nonsingular static Reissner-Nordstr\"{o}m spacetime}

Let us consider the metric of a static spacetime in the ingoing Eddington-Finkelstein coordinates,
 \begin{align}
   ds^2=-F(r)dv^2+2dv dr+r^2d\Omega^2.
   \label{Static-spacetime-metric-ingoing-EF}
 \end{align}
 For a Schwarzschild spacetime, the function $F(r)$ takes the well-known
 form, $F(r)=1-2M/r$. This function diverges at $r=0$, and it
 is the origin of the curvature singularity at $r=0$.
 Hayward~\cite{Hayward:2005} considered the regularized form as
 \begin{equation}
   F(r) = 1-\frac{2Mr^2}{r^3+2M\ell^2}.
   \label{Hayward-form}
 \end{equation}
 Then, the central singularity disappears, and there is a regular center
 at $r=0$. Let us consider the charged version of the regularized
 function of $F(r)$.

In Ref.~\cite{Frolov:2016pav}, Frolov presented an example of a nonsingular Reissner-Nordstr\"{o}m metric. Here, considering a rather general form of the metric function, we shall show that Frolov's metric is the simplest one
(see Ref.~\cite{Binetruy:2018} for a similar discussion
in the uncharged case). 
The Reissner-Nordstr\"{o}m metric is given by Eq.~\eqref{Static-spacetime-metric-ingoing-EF} with
 \begin{align}
   F(r)=1-\frac{2M}{r}+\frac{Q^2}{r^2},
 \end{align}
 where $M$ is the Arnowitt-Deser-Misner mass and $Q$ is the total charge.
 Here, we would like to modify the function $F(r)$ in order to make
 the spacetime nonsingular at $r=0$.
 Following Ref.~\cite{Frolov:2016pav}, the function $F(r)$ is assumed to behave  near $r\sim 0$ as
\begin{align}
  \label{restrict F}
  F(r)\sim 1+\varepsilon \frac{r^2}{\ell^2},
\end{align}
where $\varepsilon$ is $+1$ or $-1$ and $\ell$ is the cut-off parameter
that gives the typical scale of regularization.
We assume $F(r)$ to be a ratio of polynomials with respect to $r$ of the order $n$.
\begin{align}
  F(r)=1-\frac{\sum\limits_{k=0}^{n-1}b_k r^{k}}{\sum\limits_{j=0}^{n}a_jr^j}.
\end{align}
From the condition of Eq.~\eqref{restrict F}, we have $F(0)=1$, $F'(0)=0$, $F''(0)=2\varepsilon/{\ell^2}$, and this implies $b_0=0,b_1=0,a_0=-\varepsilon b_2 \ell^2$. For large $r$, $F(r)$ can be expanded as
\begin{align}
  F(r)\approx 1-\frac{b_{n-1}}{r}+\frac{a_{n-1}b_{n-1}-b_{n-2}}{r^2}-\frac{b_{n-3}-a_{n-1}b_{n-2}-b_{n-1}(a_{n-2}-a_{n-1}^2)}{r^3}+O\left(\frac{1}{r^4}\right).
\end{align}
Requiring this formula to coincide with the Reissner-Nordstr\"{o}m metric up to the order of $1/r^3$ for large $r$, we obtain
\begin{subequations}
\begin{align}
    &b_{n-1}=2M,
  \label{condition1}\\
    a_{n-1}b_{n-1}&-b_{n-2}=Q^2,
  \label{condition2}\\
    b_{n-3}-a_{n-1}b_{n-2}&-b_{n-1}(a_{n-2}-a_{n-1}^2)=0.
  \label{condition3}
\end{align}
\end{subequations}
These conditions constitute a nonsingular Reissner-Nordstr\"{o}m metric.

First, we consider the case of $n=3$. Since $b_1=b_0=0$ holds in general,
Eqs.~\eqref{condition1}--\eqref{condition3} fixes the coefficients
as
\begin{align}
  a_1=\left(\frac{Q^2}{2M}\right)^2, \quad
    a_2=\frac{Q^2}{2M}, \quad
  b_2=2M, \quad
\end{align}
and thus, the function $F(r)$ is determined as
\begin{align}
  F(r)=1-\frac{2Mr^2}{r^3+\frac{Q^2}{2M}r^2+\left(\frac{Q^2}{2M}\right)^2r-2\varepsilon M\ell^2}.
\end{align}
However, this function is not reduced to the Reissner-Nordstr\"{o}m metric in the limit of $\ell\to 0$. Therefore, we cannot construct an appropriate nonsingular Reissner-Nordstr\"{o}m metric in the case of $n=3$.

Next, we consider the case of $n=4$. The conditions of Eqs.~\eqref{condition1}--\eqref{condition3} are written as
\begin{align}
    a_2=\frac{Q^2}{2M}a_3, \quad
    b_2=2Ma_3-Q^2,\quad
    b_3=2M, 
\end{align}
and this leads to
\begin{align}
  F(r)\,=\,1-\frac{2Mr^3-(Q^2-2Ma_3)r^2}{r^4+a_3r^3+\frac{Q^2}{2M}a_3r^2+a_1r+\varepsilon(Q^2-2Ma_3)\ell^2}.
  \label{Fr-trial-n4}
\end{align}
There are two unfixed parameters $a_1$ and $a_3$, and we require that $F(r)$ coincide with Eq.~\eqref{Hayward-form} of the Hayward model 
in the uncharged case. Since $F(r)$ is 
\begin{align}
  F(r)=1-\frac{2Mr^3+2Ma_3r^2}{r^4+a_3r^3+a_1r-2\varepsilon Ma_3\ell^2},
\end{align}
for $Q=0$, we have $a_3=0$, and thus
\begin{align}
  F(r)=1-\frac{2Mr^2}{r^3+a_1}.
\end{align}
Comparing this equation with the Hayward model, the value of $a_1$ is determined as $a_1=2M\ell^2$. As a result, we have to choose $\varepsilon=+1$
in Eq.~\eqref{Fr-trial-n4} to avoid divergence of $F(r)$, and
\begin{align}
  \label{F of modified RN}
  F(r)=1-\frac{(2Mr-Q^2)r^2}{r^4+(2Mr+Q^2)\ell^2}
\end{align}
is the simplest model of the nonsingular Reissner-Nordstr\"{o}m spacetime. The function $F(r)$ behaves as
\begin{align}
  F(r) \ \approx \
  \begin{cases}
\displaystyle  1-\frac{r^2}{\ell^2}+\frac{r^5}{2\ell^4M}+O(r^8) & (Q=0);\\
\displaystyle  1+\frac{r^2}{\ell^2}-\frac{4Mr^3}{\ell^2Q^2}+O(r^4) & (Q\neq 0),
  \end{cases}
  \label{Static-F-expansion-center}
\end{align}
in the neighborhood of $r=0$.
In Hayward's spacetime with $Q=0$, the difference of the metric
from the exact dS metric is $O(r^5)$, and hence,
the geometry is well approximated by the dS spacetime.
In the case of $Q\neq 0$, 
although the metric around $r=0$ resembles that of the
anti-de Sitter (AdS) spacetime, 
its difference from the exact AdS metric
is $O(r^3)$, and hence, the appoximation
is not as good as the Hayward case.
We will come back to this point later.

\subsection{Global structure of nonsingular static Reissner-Nordstr\"{o}m spacetime }

%
\begin{figure}[tb]
  \centering
  \includegraphics[scale=0.2]{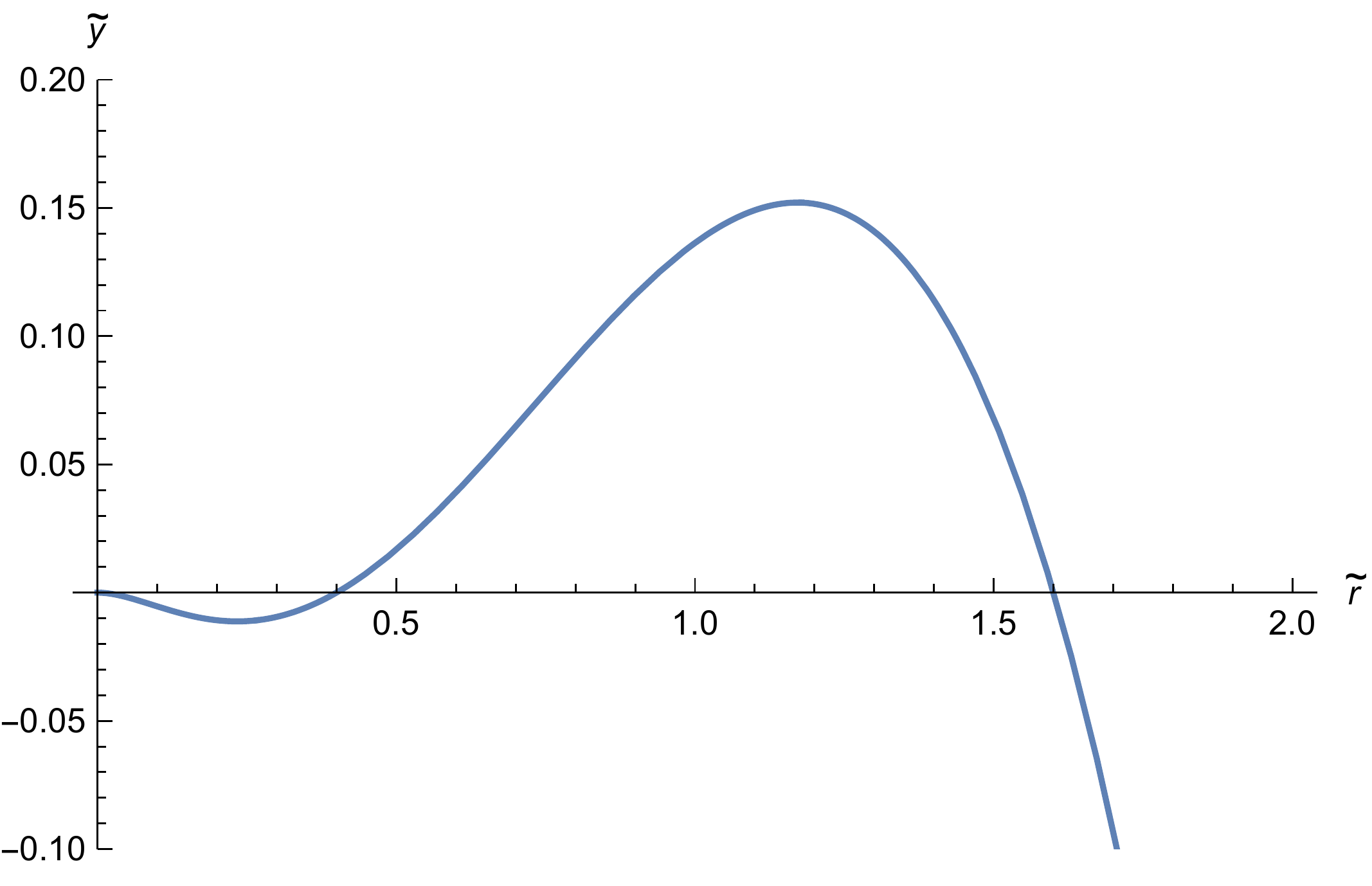}
  \caption{The behavior of the function $\tilde{y}(\tilde{r})$ for $q=0.8$.}
  \label{pic-of y}
\end{figure}
%

Here we analyze the locations of the horizons of the nonsingular Reissner-Nordstr\"{o}m spacetime. The event and Cauchy horizons exist at positions satisfying $F(r)=0$, and this equation is expressed as
\begin{align}
  \label{AH}
  r^4-2Mr^3+Q^2r^2+2M\ell^2r+Q^2\ell^2=0.
\end{align}
Introducing the normalized variables  $\tilde{r}={r}/{M}$, $\tilde{\ell}={\ell}/{M}$, $q={Q}/{M}$, we can easily rewrite Eq.~\eqref{AH} as
\begin{align}
  \tilde{\ell}^2=-\frac{\tilde{r}^2(\tilde{r}-\tilde{r}_0^+)(\tilde{r}-\tilde{r}_0^-)}{q^2+2\tilde{r}},
  \label{AH-eq-factorized}
\end{align}
where $\tilde{r}_0^{\pm}=1\pm\sqrt{1-q^2}$.
It can be seen that $\tilde{r}_0^{\pm}$ are the event and Cauchy horizons of the original Reissner-Nordstr\"{o}m spacetime (in the case of $\tilde{\ell}=0$). Denoting  the right-hand side
of Eq.~\eqref{AH-eq-factorized} as $\tilde{y}(\tilde{r})$, the graph of $\tilde{y}(\tilde{r})$ can be drawn as shown in Fig.~\ref{pic-of y} for the case of $q<1$.
With this figure, the radial positions of the event and Cauchy horizons,
$\tilde{r}_H^+$ and $\tilde{r}_H^-$, can be obtained as the values of $\tilde{r}$
of the intersection points
of the curve $\tilde{y}=\tilde{y}(\tilde{r})$ and the line $\tilde{y}=\tilde{\ell}^2$.
As a result, we have the relation
$\tilde{r}_0^{-}<\tilde{r}_H^{-}<\tilde{r}_H^{+}<\tilde{r}_0^{+}$.

Here we compare the Hayward model and the nonsingular
Reissner-Nordstr\"{o}m spacetime obtained here. 
The Hayward spacetime possesses two horizons, and the inner horizon is formed
around $r=\ell$ due to the
fact that the central region behaves like the de Sitter (dS) spacetime. By contrast, in the nonsingular Reissner-Nordstr\"{o}m  spacetime, the central region behaves like an AdS spacetime because $F(0)\to +\infty$
in the limit $r\to 0$ in the original Reissner-Nordstr\"{o}m spacetime, and $F(0)=1$ is required for nonsingular black holes.
As a result, $F(r)=0$ has two solutions,
$r=\tilde{r}_H^+M$ and $\tilde{r}_H^-M$,
in the vicinity of the two horizons of
the original Reissner-Nordstr\"{o}m  spacetime.
Therefore, the global structure of a nonsingular Reissner-Nordstr\"{o}m  spacetime is similar to that of the original Reissner-Nordstr\"{o}m  spacetime except that the point $r=0$ is no longer a curvature singularity.

If we choose a fine-tuned value of $q$,
we can also consider a situation with only one degenerate horizon, but we will not consider such a situation in this paper.

\subsection{Geometrical properties of the regularized spacetime}

In order to realize the regularized spacetimes, there are two possibilities.
One is that the theory of gravity is modified from the theory of
general relativity. 
The other is that the theory of gravity is given
by general relativity and the properties of matter (and therefore,
the energy-momentum tensor) are changed.
In this paper, we adopt the former possibility
and do not care about whether such a spacetime is realized
within the framework of general relativity.
Nevertheless, it would be interesting
to examine the properties of the energy-momentum tensor
of this spacetime in the case that this spacetime
is realized in general relativity, because 
such an analysis would deepen our understanding on
the geometrical properties of this spacetime.
For this reason, we examine the properties of the Einstein tensor here.

We adopt
the tetrad basis $(e^a)_\mu $ ($a=0,1,2,3$) as
\begin{subequations}
\begin{eqnarray}
(e^0)_\mu &=&\left(-\frac{1+F}{2},\ 1,\ 0, \ 0\right),\label{tetrad0}\\
(e^1)_\mu &=&\left(\frac{1-F}{2},\ 1,\ 0, \ 0\right),\\
(e^2)_\mu &=&\left(0,\ 0,\ r,\ 0\right),\\
(e^3)_\mu &=&\left(0,\ 0,\ 0,\ r\sin\theta\right),\label{tetrad3}
\end{eqnarray}
\end{subequations}
in the $(v,r,\theta,\phi)$ coordinates,
and consider the tetrad components
of the Einstein tensor $G^{ab}=G^{\mu\nu}(e^a)_\mu (e^b)_\nu$.
It is written in the form
\begin{equation}
G^{ab}=\mathrm{diag}(\rho,\, p_r,\, p_\theta,\, p_\phi), 
\end{equation}
(which is called the canonical form of type I \cite{Hawking:1973}, see also \cite{Maeda:2022}) where
\begin{subequations}
\begin{eqnarray}
\rho\, = \, -p_r &=& \frac{1-F-rF_{,r}}{r^2},\\
p_{\theta}\, = \, p_{\phi} &=& \frac{F_{,r}}{r}+\frac{F_{,rr}}{2}.
\end{eqnarray}
\end{subequations}
With this quantity, we can examine the property of the energy-momentum
tensor in the case that the regularized spacetime
is realized in the framework of general relativity with
unordinary matter. 
In particular, we focus on the null energy condition (NEC) and the dominant
energy condition (DEC). Assuming the Einstein equation, the NEC is 
$G_{\mu\nu}k^\mu k^\nu\ge 0$ for an arbitrary null vector $k^\mu$,
and the DEC is that $-{G^\mu}_\nu v^\mu$ is
a future-directed timelike or null vector for an arbitrary future-directed
timelike vector $v^\mu$ (e.g., \cite{Hawking:1973}). The NEC is equivalent to
\begin{equation}
\rho + p_i\ge 0 
\end{equation}
for $i=r, \theta, \phi$, and the DEC
is equivalent to
\begin{equation}
\rho \ge |p_i|.
\end{equation}

%
\begin{figure}[tb]
  \centering
  \includegraphics[width=0.4\textwidth,bb= 0 0 360 263]{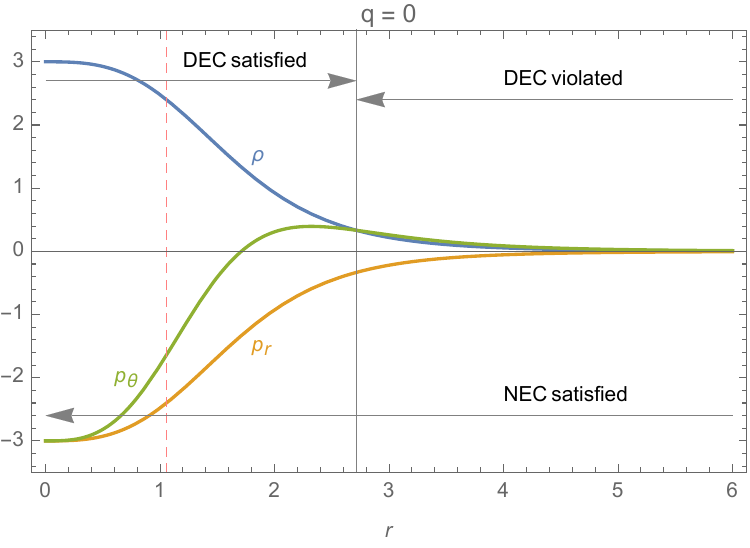}
  \caption{Behavior of $\rho$, $p_r$, and $p_{\theta}$ for
    the Hayward spacetime with $M=5$ and $\ell = 1$
    as functions of $r$. The regions where the NEC and DEC are
    satisfied/violated are also shown. The red vertical dashed line shows
    the location of the inner horizon, while the outer horizon is outside
  of this figure.}
  \label{EnergyCondition-Static-q000}
\end{figure}
%

Figure~\ref{EnergyCondition-Static-q000}
shows the behavior of $\rho$, $p_r$, and $p_{\theta}$
for the Hayward spacetime with $M=5$ and $\ell = 1$
as functions of $r$.
These quantities decay as $r$ is increased.
$\rho$ approaches $3/\ell^2$
and $p_\theta$ and $p_r$ approach $-3/\ell^2$ in the limit $r\to 0$,
and they are approximately constant in the neighborhood of $r=0$.
This is because the difference of Hayward's metric
from the dS metric is $O(r^5)$ from Eq.~\eqref{Static-F-expansion-center},
and hence, the deviation of the Einstein tensor from the
cosmological term is $O(r^3)$. 
This confirms the fact that the Hayward spacetime
is well approximated by a dS spacetime around the center.
Although the NEC is satisfied in all regions,
the DEC is violated at distant region
because the value of $p_{\theta}$
exceeds that of $\rho$ there.
See also Ref.~\cite{Maeda:2021} for more study on the energy
conditions of the Hayward spacetime and other regular black hole
spacetimes.

%
\begin{figure}[tb]
  \centering
  \includegraphics[width=0.4\textwidth,bb= 0 0 360 263]{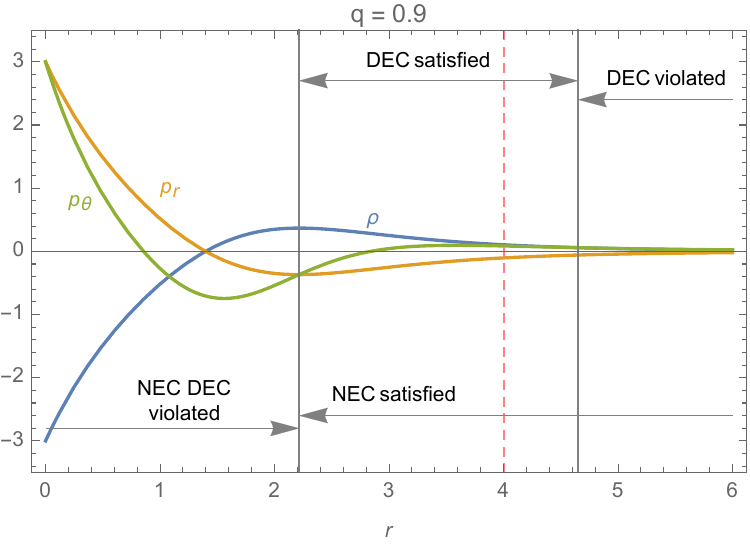}
  \caption{The same as Fig.~\ref{EnergyCondition-Static-q000}
    but for charged regularized spacetime with $q=0.90$.}
  \label{EnergyCondition-Static-q090}
\end{figure}
%

Figure~\ref{EnergyCondition-Static-q090}
shows the case of $q=0.9$
for the case $M=5$ and $\ell=1$. 
In contrast to the case $q=0.0$, 
$\rho$ approaches $-3/\ell^2$,
and $p_\theta$ and $p_r$ approach $3/\ell^2$ at $r\to 0$.
In this sense, the central point satisfies the
Einstein equation with a negative cosmological constant.
However, these three quantities
change their values rapidly as $r$ is increased.
This is because the difference of the present metric
from the AdS metric is $O(r^3)$ from Eq.~\eqref{Static-F-expansion-center},
and thus, the deviation of the Einstein tensor from the cosmological term
is $O(r)$. In this sense,   
the AdS spacetime does not approximate the
spacetime structure around the center of the
nonsingular Reissner-Nordstr\"om spacetime very well.
Around the central region, there appears
the region where the NEC and DEC are violated.
At the distant place, only the DEC is violated.
This can be confirmed to hold for arbitrary $M$, $Q$, and $\ell$,
by the asymptotic expansion,
\begin{subequations}
  \begin{eqnarray}
    \rho &=& \frac{Q^2}{r^4}+ \frac{12M^2\ell^2}{r^6} + \cdots,\\
    p_{\theta} &=& \frac{Q^2}{r^4}+ \frac{24M^2\ell^2}{r^6} + \cdots.
  \end{eqnarray}
\end{subequations}

\subsection{Regularization of ingoing charged Vaidya spacetime}

We now describe the methods of regularizing ingoing charged Vaidya
spacetime that will be used in the subsequent sections.

\subsubsection{Constant regularization}
\label{Constant-regularization}

The natural choice of the metric
for the regularized ingoing
charged Vaidya solution
would be to adopt the metric of Eq.~\eqref{Metric:IngoingVaidya}
with 
the regularized function
$F_-(v,r)$, which has the same formula as Eq.~\eqref{F of modified RN},
\begin{equation}
  \label{F-ingoingVaidya}
  F_-(v,r)=1-\frac{(2M(v)r-Q(v)^2)r^2}{r^4+(2M(v)r+Q(v)^2)\ell^2},
\end{equation}
except that the mass $M(v)$ and the charge $Q(v)$
are now dependent on the advanced time. 
Here, we introduced a constant $\ell$ with the dimension of length
that determines the scale of the regularization. 
This is a natural extension of the model by Hayward and Frolov.
The property of the spacetime with this choice of $F_-(v,r)$
will be analyzed in Sect.~\ref{Sec:Non-self-similar}.
The geometrical properties, such as Riemann invariants and
the Einstein tensor, are examined in Appendix~\ref{Appendix:Geometrical-quantities}.

\subsubsection{Time-dependent regularization}
\label{Time-dependent-regularization}

We would like to consider
a different choice where
the scale of the regularization $\ell$
depends on the advanced time as
\begin{equation}
  \ell(v)\,=\,M(v)b,
\end{equation}
  where
$b$ is a nondimensional constant. Then, the function
$F_-(v,r)$ is 
\begin{equation}
  \label{F-ingoingVaidya2}
  F_-(v,r)=1-\frac{(2M(v)r-Q(v)^2)r^2}{r^4+(2M(v)r+Q(v)^2)M(v)^2b^2}.
\end{equation}
The reason for considering the time-dependent regularization
is that the self-similar property
of the evaporation model (A), reviewed in
Sect.~\ref{Subsec:Evaporation_model_A},
is kept for the value of $b\neq 0$ as well,
and thus, the analysis can be carried out  
with a straightforward modification to Sect.~\ref{Subsec:Evaporation_model_A}. 
We present this analysis in the next section.
Examinations of geometrical properties in the constant regularization are done 
in Appendix~\ref{Appendix:Geometrical-quantities}.

%
%
\section{Self-similar model of evaporation}

\label{Sec:Self-similar-model}


In this section, we study the self-similar model
of an evaporating nonsingular charged black hole given by the time-dependent regularization
of Eq.~\eqref{F-ingoingVaidya2}.
The setup is the same as that of
the model (A) of an evaporating charged black hole 
reviewed in Sect.~\ref{Subsec:Evaporation_model_A}
except that the time-dependent regularized
function $F_-(v,r)$, Eq.~\eqref{F-ingoingVaidya2},
is used here. 
In the domain $0\le v\le v_{\rm f}$, 
the functions $M(v)$ and $Q(v)$ are assumed to be the same as
Eqs.~\eqref{Linear-mass-function} and \eqref{Linear-charge-function}.
In what follows, we briefly describe the results 
focusing attention to the difference
from the model (A) of Sect.~\ref{Subsec:Evaporation_model_A}
in the evaporation phase.
See also Appendix~\ref{Appendix:Collapse-domain}
for the study on the collapsing phase.


We introduce the two new coordinates $\tilde{v}$ and $R$ by
Eq.~\eqref{Eq:tildev-R}, and then, 
the function $F_-(v,r)$ of Eq.~\eqref{F-ingoingVaidya2} is rewritten as
 \begin{align}
    F_-(v,r)
    &=1-\frac{2R^3-q^2R^2}{R^4+2Rb^2+q^2b^2}=:f(R).
  \label{Definition-f(R)}
 \end{align}
 In these coordinates, the metric is expressed by the same formula
 as Eq.~\eqref{Eq:metric-in-tildev-R}. The
 retarded time $u$ can be introduced by Eq.~\eqref{Eq:Introduction-u},
 and 
 the metric is expressed as Eq.~\eqref{Eq:metric-in-double-null}
 in the double null coordinates $(u,\tilde{v})$.

 Similarly to the model (A) of Sect.~\ref{Subsec:Evaporation_model_A},
 zeros of the function $f\alpha+2R$ play important roles
 to determine the spacetime structure.
 Equation~\eqref{eq-formation-evaporation} is
 modified as
\begin{align}
  \label{eq-formation-evaporation-regularized}
-\frac{1}{\alpha}\,=\,\frac{R^4-2R^3+q^2R^2+2Rb^2+q^2b^2}{2R(R^4+2Rb^2+q^2b^2)}=:y(R).
\end{align}
Compared to the $b=0$ case, the divergence
of $y(R)$ in the neighborhood of $R=0$ 
becomes weaker. Except that point, 
the gross behavior 
of the function $y(R)$ does not change as long as $b$ is small:
The function $y(R)$ has the local minimum and local maximum
at $R=R_\pm$ in the domain $R>0$, respectively, and there is
zero, one, and two positive solutions
in the cases that $1/\alpha>|y(R_-)|$,
$1/\alpha=|y(R_-)|$,
and $1/\alpha<|y(R_-)|$, respectively. 
 If the equation $f\alpha+2R=0$ has only zero positive solution,
 the spacetime possesses 
 no homothetic Killing horizon except for $\tilde{v}=0$.
 On the other hand, if the equation $f\alpha+2R=0$ has two positive roots,
 $R_{\rm C}^-$ and $R_{\rm C}^+$, 
 they correspond to the homothetic Killing horizons.

 The outgoing null coordinate $u$ can be
 introduced by the same formula as Eq.~\eqref{Eq:Introduction-u}.
 When the equation $f\alpha+2R=0$ has only one (negative) solution,
 the coordinate $u$ spans the whole domain of $0\le R< \infty$,
 although the explicit integration of
 Eq.~\eqref{Eq:Introduction-u} may be difficult. 
 When the equation $f\alpha+2R=0$ has two positive solutions,
 the function $f\alpha+2R$ can be formally written as
 \begin{equation}
   f\alpha+2R\, = \,
   \frac{(R-R_{\rm C}^-)(R-R_{\rm C}^+)}{2(R_{\rm C}^+-R_{\rm C}^-)}\kappa(R),
 \end{equation}
 where $\kappa(R)$ is some positive regular function of $R$ in the range
 $0\le R<\infty$. Then, the formal integration
 of Eq.~\eqref{Eq:Introduction-u} can be performed
 to give the modified equation
 of Eq.~\eqref{Eq:formula-for-u} as
 \begin{equation}
   u \, = \, \ln\tilde{v}
   -\frac{1}{\kappa_-}\ln|R-R_{\rm C}^-|+
   \frac{1}{\kappa_+}\ln|R-R_{\rm C}^+|
   +G(R),
 \end{equation}
 where $\kappa_\pm :=\kappa(R_{\rm C}^\pm)$
 and
 \begin{equation}
G(R):=\int\left[\frac{1/\kappa(R)-1/\kappa_+}{R-R_{\rm C}^+}-\frac{1/\kappa(R)-1/\kappa_-}{R-R_{\rm C}^-}\right]dR.
 \end{equation}
Note that the integrand of $G(R)$
is regular at $R=R_{\rm C}^\pm$.
The null coordinates $U_{\pm}$ that are continuous
across $R=R_{\rm C}^\pm$ are introduced locally by
Eqs.~\eqref{Continuous-coordinate-across-Rpm},
and in these coordinates the metric takes the regular form,
\begin{equation}
ds^2 \,=\, - \frac{M_0^2\alpha\tilde{v}^{1\mp\kappa_\pm}\kappa(R)\exp[\mp\kappa_\pm G(R)]\left|R-R_{\rm C}^\mp\right|^{1+\kappa_\pm/\kappa_\mp}}{2\kappa_{\pm}(R_{\rm C}^+-R_{\rm C}^-)}dU_{\pm}d\tilde{v}.
\end{equation}

Here, we discuss whether there exist spacetime singularities
in this model. The formulas of geometric quantities
in this model are presented in Appendix~\ref{Appendix:Geometrical-quantities},
and the scalar quantities constructed from the Riemann tensor
$\mathcal{R}_{\mu\nu\rho\sigma}$
behave as
\begin{subequations}
\begin{eqnarray}
\left.\mathcal{R}\right|_{R=0}  &=&-\frac{12}{(bM(v))^2},
\label{Eq:Ricci-scalar-Req0}\\
\left.\mathcal{R}_{\mu\nu}
\mathcal{R}^{\mu\nu}\right|_{R=0}  &=&\frac{36}{(bM(v))^4},
\label{Eq:Ricci-tensor-squared-Req0}\\
\left.\mathcal{R}_{\mu\nu\rho\sigma}
\mathcal{R}^{\mu\nu\rho\sigma}\right|_{R=0}  &=&\frac{24}{(bM(v))^4}.
\label{Eq:Riemann-tensor-squared-Req0}
\end{eqnarray}
\end{subequations}
Although the spacetime curvature is finite 
at $R=0$ for the region where $M(v)>0$, 
it diverges
in the limit $M(v)\to 0$ for $R=0$
(this statement holds for any fixed $R$,
see Appendix~\ref{Appendix:Geometrical-quantities}).
Hence, the regularization of the spacetime curvature 
by the formula of Eq.~\eqref{F-ingoingVaidya2}
is not very complete. For this reason, we call this model
the partly regularized ingoing charged Vaidya spacetime.

%
\begin{figure}[tb]
  \centering
  \includegraphics[width=0.49\textwidth,bb= 0 0 360 252]{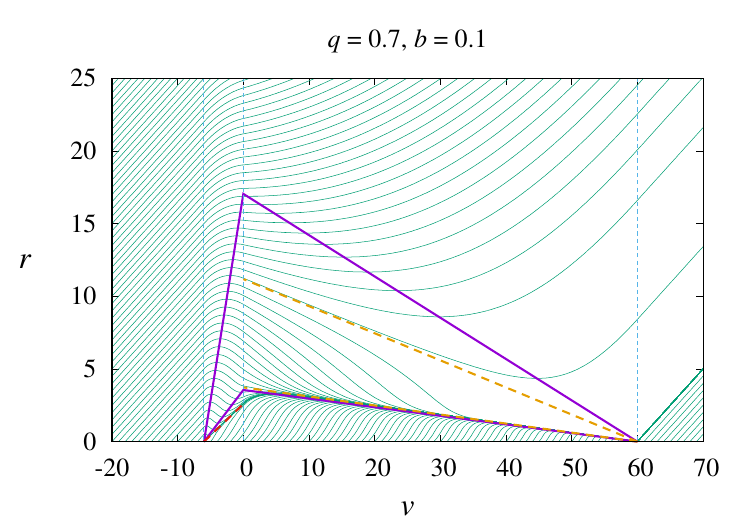}
  \includegraphics[width=0.49\textwidth,bb= 0 0 360 252]{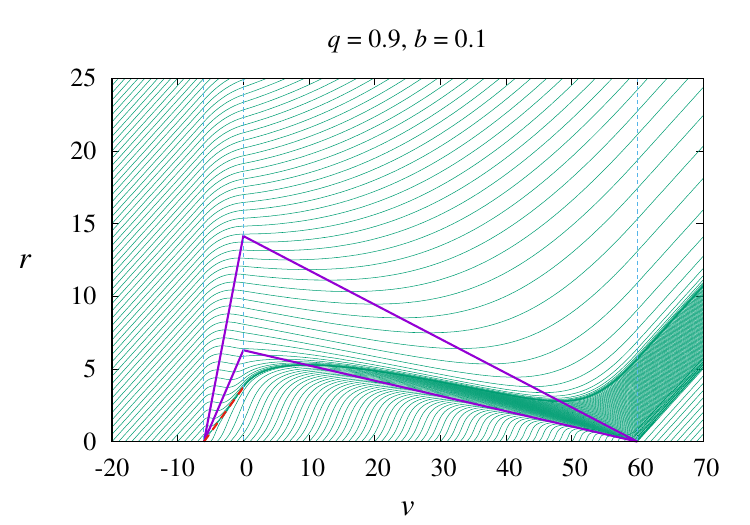}
  \caption{ The same as Fig.~\ref{geodesic_AH_ell0}
    but for the partly regularized case of $b=0.1$. 
  }
  \label{geodesic_AH_bb01}
\end{figure}
%

Figure~\ref{geodesic_AH_bb01} shows
the $(v,r)$-diagrams in the partly regularized case with $b=0.1$
for $q=0.7$ (left panel) and  $q=0.9$ (right panel). 
Similarly to Fig.~\ref{geodesic_AH_ell0} in the unregularized case with $b=0$,
the behavior 
of outgoing null geodesics (thin green curves) and the boundary of the trapped region (closed purple lines) 
are shown by selecting the same parameter values
of $M_0$, $v_{\rm i}$, and $v_{\rm f}$. 
The main difference from the $b=0$ case 
is the behavior of outgoing null geodesics
around $r=0$ in the range $-v_{\rm i}<v<v_{\rm f}$.
In the unregularized case, the value of $dr/dv$ diverges
because $r=0$ is a curvature singularity,
while in the case $b=0.1$, we have the value  $dr/dv=1/2$ 
because the geometry around $r=0$ is regular.
Except that, these $(v,r)$-diagrams are similar to those of
Fig.~\ref{geodesic_AH_ell0}.
In the case of $q=0.7$, there are two homothetic Killing horizons
in the evaporating phase as shown by the orange dashed lines.
The outer and inner homothetic Killing horizons
are the repeller and attractor of outgoing null geodesics.
All outgoing null geodesics within the outer homothetic Killing horizon 
plunge into the curvature singularity at $(v,r)=(v_{\rm f}, 0)$.
This means that the singularity is extended into $u$ direction,
and thus, it is a null singularity.
The outer homothetic Killing horizon plays the role of the event horizon
simultaneously.
In the case of $q=0.9$, no homothetic Killing horizon is present
in the evaporating phase, and thus, there is no event horizon.
The singularity at $(v,r)=(v_{\rm f}, 0)$ is not extended in the
$u$ direction because no outgoing null geodesic plunges into
this singularity from $v<v_{\rm f}$. 
In the collapsing phase, there is one homothetic Killing horizon
as indicated by the red dashed line in both cases
of $q=0.7$ and $0.9$. Since this homothetic Killing
horizon is the attractor of outgoing null geodesics in the future direction,
only one outgoing null geodesic is emitted from the singularity
at $(v,r)=(-v_{\rm i}, 0)$.
This means that this singularity is not extended in the $u$ direction
and is a pointlike singularity.

%
\begin{figure}[tb]
  \centering
  \includegraphics[width=0.8\textwidth,bb= 0 0 885 612]{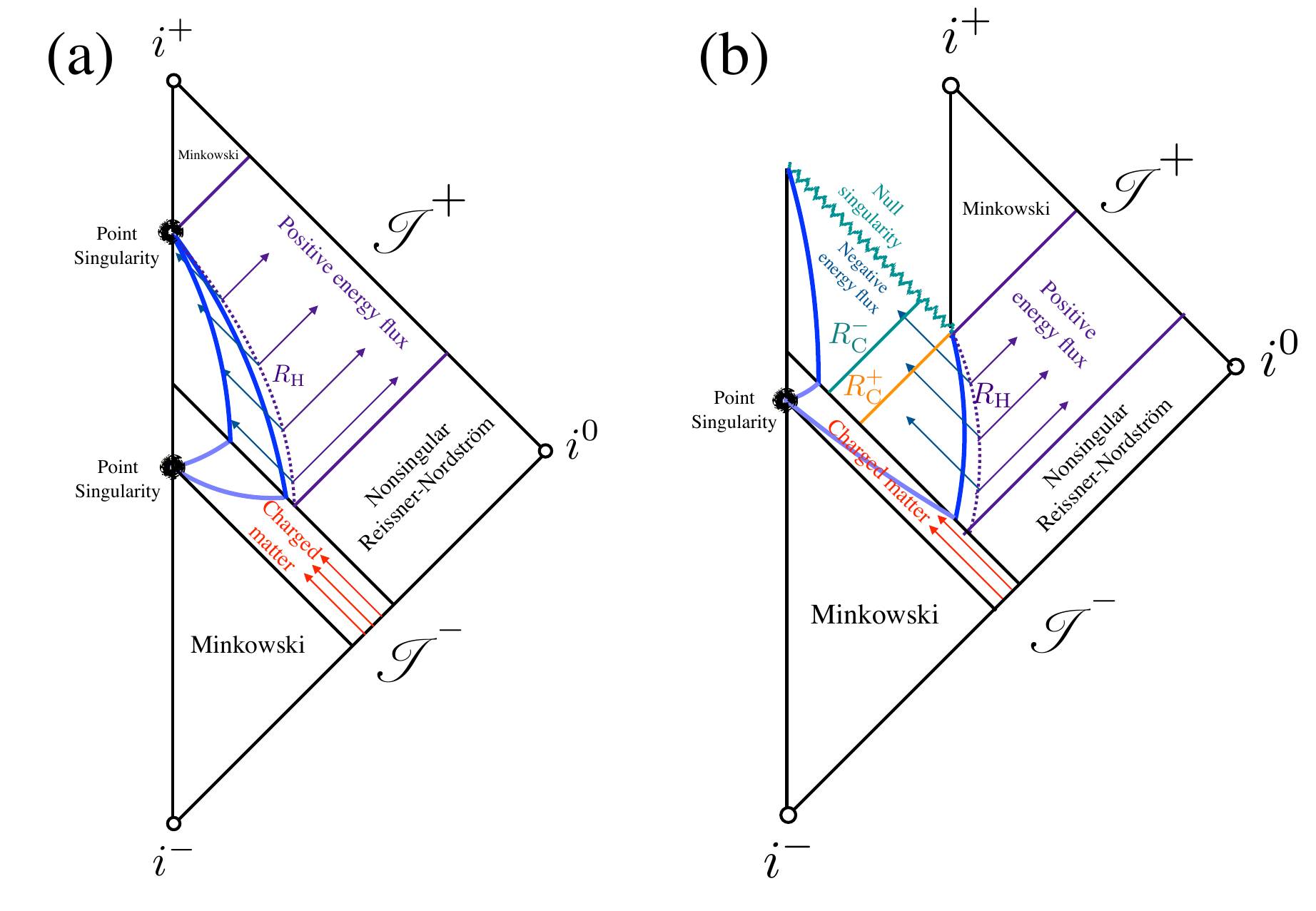}
  \caption{Penrose diagram for evaporation of
    a partly regularized Reissner-Nordstr\"om black hole.
    Left panel: The case that a homothetic Killing horizon does not exist.
    A pointlike singularity forms. Right panel: The case that two homothetic Killing horizons exist. There appears a null singularity inside the invisible region.}
  \label{Fig:Penrose-regularized-self-similar}
\end{figure}
%

We now present the Penrose diagram for the evaporation
of a partly regularized Reissner-Nordstr\"{o}m black hole
in the self-similar model, which corresponds to the right panel of
Fig.~\ref{geodesic_AH_bb01}.
The left panel of Fig.~\ref{Fig:Penrose-regularized-self-similar}
shows the Penrose diagram for the case that no homothetic Killing horizon
exists in the evaporating phase. Compared to the corresponding diagram of
Fig.~\ref{Fig:Penrose-Vaidya-wo-HKH}(b) of
Sect.~\ref{Subsec:Evaporation_model_A},
the timelike singularity at $R=0$ (hence, $r=0$) is regularized. However,
the pointlike singularity appears at $r=\tilde{v}=0$
due to Eqs.~\eqref{Eq:Ricci-scalar-Req0}--\eqref{Eq:Riemann-tensor-squared-Req0},
and a Cauchy horizon is formed on
the future light cone of $r=\tilde{v}=0$.
Except for this feature, the Penrose diagram possesses
good properties for solving the information loss problem
since the diagram 
is similar to
that of the Minkowski spacetime.

Right panel of Fig.~\ref{Fig:Penrose-regularized-self-similar}
shows the Penrose diagram for the case that two homothetic Killing horizons
exist in the evaporating phase, which
corresponds to the left panel of Fig.~\ref{geodesic_AH_bb01}.
Compared to the corresponding diagram of
Fig.~\ref{Fig:Penrose-Vaidya-w-HKH}(b) of
Sect.~\ref{Subsec:Evaporation_model_A}, the timelike singularity at $R=0$
is regularized in the domain $\tilde{v}>0$.
However, the null singularity still remains
at $\tilde{v}=0$ and $R=R_{\rm C}^-$
because of Eqs.~\eqref{Eq:Ricci-scalar-Req0}--\eqref{Eq:Riemann-tensor-squared-Req0}.
As a result, the domain $0<R<R_{\rm C}^+$ is invisible to outside observers.
This property is not good for solving the information loss problem
because the information of the invisible domain will be lost
in the process of evaporation.

To summarize, the spacetime structures
of the partly regularized evaporating  
Reissner-Nordstr\"om black holes do not
have satisfactory features for solving the information loss problem.
However, we do not consider that the Hayward-Frolov scenario fails because of
this result. Rather, the time-dependent regularization
introduced in Sect.~\ref{Time-dependent-regularization}
would not be appropriate. In fact,
the motivation for introducing time-dependent regularization
was that the 
self-similarity of the spacetime is maintained
and the analytic method of Kaminaga \cite{Kaminaga:1988pg}
can be extended straightforwardly. 
The appropriate interpretation of this result is that
the constant regularization
of Sect.~\ref{Constant-regularization} would be
a more preferred one. In the next section,
we study the global structure of an evaporating charged
black hole applying the constant regularization.

%
%
\section{Model of evaporation with regularized center: Non-self-similar case}
\label{Sec:Non-self-similar}

We now consider the constant regularization,
where the function $F_-(v,r)$ is given by Eq.~\eqref{F-ingoingVaidya}.
In Sect.~\ref{Appendix:diagram of linear mass},
we adopt the linear mass function of Eq.~\eqref{Linear-mass-function}
and compare the case $\ell=0$ (Kaminaga's model)
and the case $\ell\neq 0$ (the nonsingular model). 
In Sect.~\ref{Subsec:Non-self-similar-setup},
we study the case of the nonlinear mass function, and present the
Penrose diagram.
Section~\ref{Subsec:Frequency_shift} is devoted
to the numerical study on the frequency shift of a photon
that falls into the black hole and comes out of it
at the last stage of the evaporation.

\subsection{The case of the linear mass function and comparison with Kaminaga's model}
\label{Appendix:diagram of linear mass}

In this section, we compare the case of the constant regularization
with the unregularized case (Kaminaga's self-similar spacetime).
This helps us to understand how the constant regularization works 
in the Hayward-Frolov model.
Except for the value of $\ell$, we choose the same setups:
The linear mass
function of Eq.~\eqref{Linear-mass-function}
and the same parameters, $M_0=10$, $v_{\rm i}=6$, and $v_{\rm f}=60$,
as those of Fig.~\ref{geodesic_AH_ell0}, are adopted. 
The values of $q$ of Eq.~\eqref{Linear-charge-function}
is assumed to be constant, and 
the cases $q=0.7$ and $0.9$ are discussed.

%
\begin{figure}[tb]
  \centering
  \includegraphics[width=0.49\textwidth,bb= 0 0 360 252]{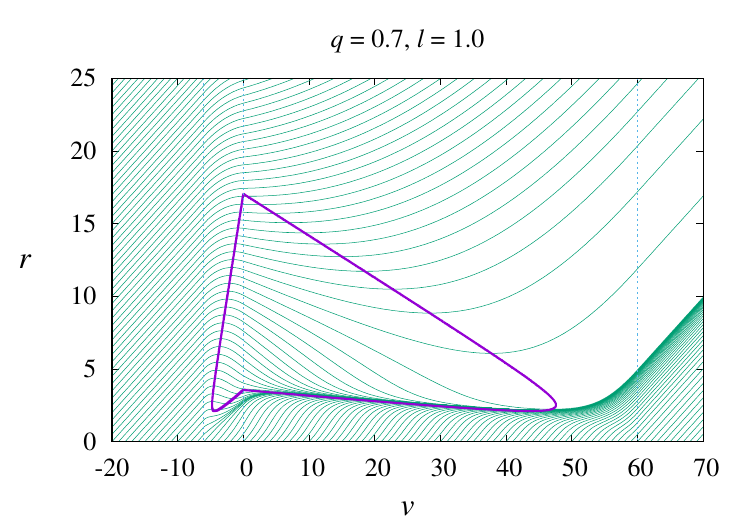}
  \includegraphics[width=0.49\textwidth,bb= 0 0 360 252]{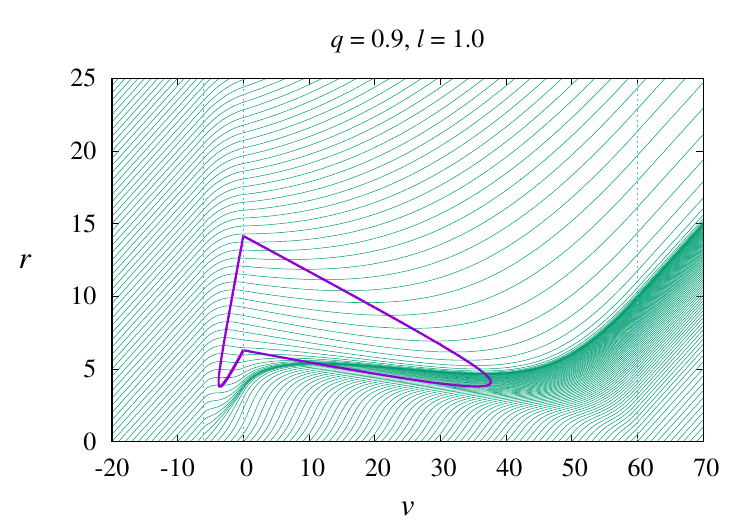}
  \caption{
The same as Fig.~\ref{geodesic_AH_ell0} and Fig.~\ref{geodesic_AH_bb01}
    but for the completely regularized case of $\ell=1.0$.
  }
  \label{geodesic_AH_linearmass}
\end{figure}
%

Figure~\ref{geodesic_AH_linearmass} shows the
$(v,r)$-diagrams for 
for $q=0.7$ (left panel) and $0.9$ (right panel) for the case $\ell=1.0$.
Let us discuss the case of $q=0.7$ first. 
In the unregularized case ($\ell=0$), the left panel of Fig.~\ref{geodesic_AH_ell0} shows that 
there exist two homothetic Killing horizons
and all outgoing null geodesics within the outer homothetic
Killing horizon are trapped and plunge into the singularity
at $(v,r)=(v_{\rm f},0)$.
By contrast, in the regularized case
(left panel of Fig.~\ref{geodesic_AH_linearmass}),
the self-similar structure is broken
by the introduction of nonzero $\ell$, and all of the null geodesics
escape to infinity. Furthermore,
the regularization of the spacetime is complete
and no singularity appears.
Therefore, the spacetime possesses the desired property
of the Hayward-Frolov scenario: the event and Cauchy horizons are not present.
Although the spacetime structure drastically changes
from the unregularized case,
the behavior of the null geodesics
at distant positions from the center is similar to the case $\ell=0$. 
The homothetic Killing horizons are absent, but
we can recognize the existence of the temporal repeller and attractor
for a certain range of $v$.
The boundary of the trapped region is topologically closed, and 
it is called the closed apparent horizon
in Ref.~\cite{Frolov:2014jva}. The temporal attractor
is located in the neighborhood of the inner boundary of the trapped region.
These repeller and attractor would be a kind of 
residues of the homothetic Killing horizons.

Next we discuss the case $q=0.9$. In the case of $\ell=0$
(the right panel of Fig.~\ref{geodesic_AH_ell0}), 
the homothetic Killing horizons do not exist and
all outgoing null geodesics escape to infinity.
The behavior of outgoing null geodesics
in the regularized spacetime 
of the right panel of Fig.~\ref{geodesic_AH_linearmass}
is similar to the unregularized case 
except that the singularity is completely resolved. 
Reflecting the absence of the homothetic Killing horizon
in the case $\ell=0$, the temporary repeller and attractor are not found.
Similarly to the left panel, 
no event horizon or no Cauchy horizon is present,
and the Hayward-Frolov scenario works well also for $q=0.9$.

To summarize, by introducing the constant regularization scale $\ell$,
the singularity completely vanishes, 
while the temporal repeller and attractor
are present for certain range of $v$
for small $q$. 
The absence of the temporary repeller/attractor
for a large $q$ in the regularized
spacetime is understood
from the absence of the homothetic Killing horizons
in the unregularized case. 
We do not present the Penrose diagrams for Fig.~\ref{geodesic_AH_linearmass}
here, because they are 
similar to the one that will be drawn
in a slightly different setup below.

\subsection{The case of the nonlinear mass function}
\label{Subsec:Non-self-similar-setup}

Since there cannot exist a self-similar configuration
in the case that $\ell\neq 0$,
it is not necessary to stick to the linear mass function
of Eq.~~\eqref{Linear-mass-function}.
Below, we study the case that the mass function is assumed as
\begin{align}
  M(v)&=
  \begin{cases}
    0 &(v\leq -v_{\rm i}),\\
    M_0\left(1+v/v_{\rm i}\right) &(-v_{\rm i}\leq v\leq 0),\\
    M_0\left(1-v/v_{\rm f}\right)^{1/3} &(0\leq v\leq v_{\rm f}),\\
     0 &(v\geq v_{\rm f}).
  \end{cases}
  \label{mass-more-realistic}
\end{align}
In this case, the time evolution of the mass in the evaporation period,
$0\leq v\leq v_{\rm f}$,  
is expected to resemble that of the realistic Hawking evaporation.
The setup here is similar to that of Frolov \cite{Frolov:2014jva},
and we can observe how the electric charge
affects the spacetime structure in the Hayward-Frolov scenario.
As for the charge function, we adopt Eq.~\eqref{Linear-charge-function}
with a constant $q$ for simplicity. 
Note that in a realistic evaporation,
the black hole does not conserve the ratio of the electric charge
to the mass, and 
the value of $q$ in Eq.~\eqref{Linear-charge-function}
depends on time in general \cite{Carter:1974,Page:1977,Hiscock:1990}. 
A more detailed study on this scenario with 
a more realistic evolution of the charge function
is left for future study.

Below, we draw the Penrose diagram of the regularized
ingoing charged Vaidya spacetime in this setup. 
For constructing the model of an evaporating black hole,
we have to cut the appropriate domain out,
and glue the regularized Reissner-Nordstr\"om spacetime
and the regularized outgoing charged Vaidya spacetime.
This method is discussed in Appendix~\ref{Appendix:Outer-domain}.

\subsubsection{Method of drawing Penrose diagram}
\label{Subsec:Method-Penrose}

The numerical calculation of the outgoing
null geodesics is performed in the same way as explained
in Sect.~\ref{subsubsec:Numerical-method-linear-mass}.
Each geodesic can be regarded as a surface of 
the constant advanced time $u$. 
The value of $u$ is naturally assigned by requiring
\begin{equation}
{u}\,=\,{v}-2{r}
\end{equation}
in the Minkowski domain $v_{\rm f}<v$ after the evaporation.
Once the ${u}$ coordinate is generated,
we specify the locations of the central point ${r}=0$
and of the boundary of the trapped region at which $F_-({v},{r})=0$
in the $({u},{v})$-plane.

In order to show the spacetime in
a compact domain, 
two compact coordinates $(\eta,\zeta)$ are introduced as
\begin{subequations}
\begin{eqnarray}
  \eta &=& \arctan{\left({v}/L\right)}+\arctan{\left({u}/L\right)},
  \label{Def:eta}
  \\
  \zeta&=&\arctan{\left({v}/L\right)}-\arctan{\left({u}/L\right)},
  \label{Def:zeta}
\end{eqnarray}
\end{subequations}
where $L$ is a positive constant with the dimension of length 
whose value can be arbitrarily chosen.
In these coordinates, whole spacetime is shown within the
diamond-shaped domain, $-\pi<\eta \pm \zeta <\pi$. 

\subsubsection{Outgoing null geodesics and the Penrose diagram}
\label{Subsec:Mass-function-one-third-power}

%
\begin{figure}[tb]
  \centering
  \includegraphics[width=0.49\textwidth,bb= 0 0 360 252]{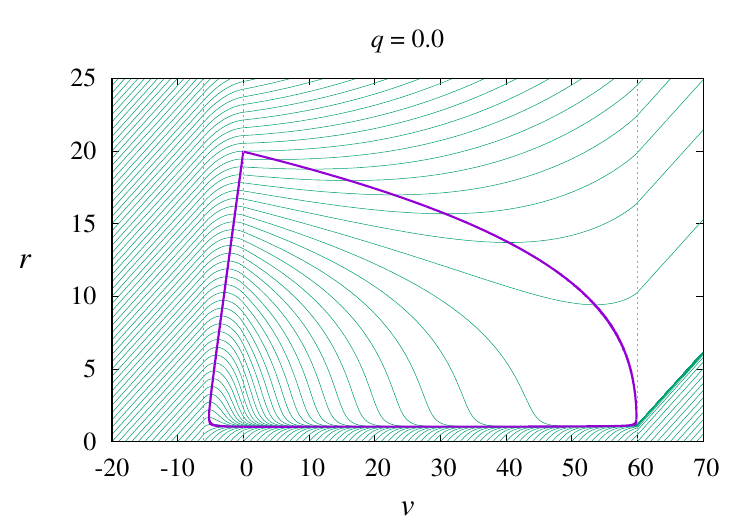}
  \includegraphics[width=0.49\textwidth,bb= 0 0 360 252]{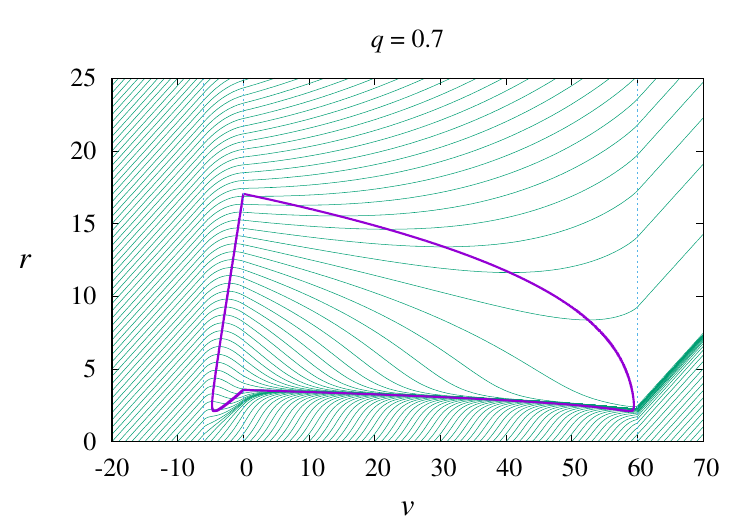}
  \includegraphics[width=0.49\textwidth,bb= 0 0 360 252]{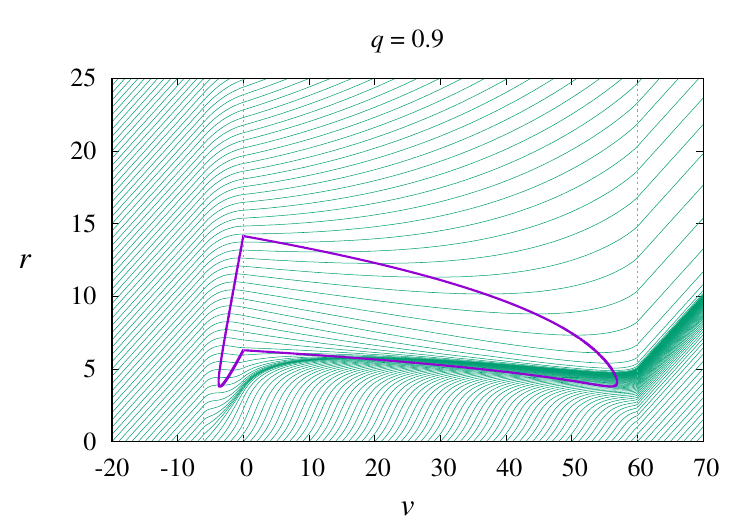}
  \caption{Outgoing null geodesics (thin green curves) and the boundary of the trapped region (closed purple curve) for $q=0.0$ (upper left panel), $0.7$ (upper right panel), and $0.9$ (lower panel) for the nonlinear mass function
    of Eq.~\eqref{mass-more-realistic}.
    The vertical dotted lines in each panel indicate
    $v=-v_{\rm i}$, $0$, and $v_{\rm f}$ from left to right.
    The unit of the length is $\ell$.
  }
  \label{NSRN AH and flow_realistic mass}
\end{figure}
%

We now show the results for the nonlinear mass function
given by Eq.~\eqref{mass-more-realistic}.
We select the same parameters
as Sect.~\ref{Appendix:diagram of linear mass}, i.e., 
$M_0/\ell =10$, $v_{\rm i}/\ell =6$, and $v_{\rm f}/\ell =60$
as an example.
Figure~\ref{NSRN AH and flow_realistic mass} shows the $(v,r)$-diagrams 
for $q=0.0$ (upper left), $0.7$ (upper right), and $0.9$ (bottom).
Each of the outgoing null geodesics gives
a $u$-constant surface. 
In all cases, all outgoing null geodesics
escape to future null infinity,
similarly to the case of the linear mass function
of Sect.~\ref{Appendix:diagram of linear mass}.

In each figure for $q=0.0$ and $q=0.7$,
we can observe the existence of an attractor
near the inner boundary of the trapped region
which attracts outgoing null geodesics for some period of time.
In the case of $q=0.0$, the radial position
of the attractor is approximately constant.
This is because the geometry near the center
is approximately equal to that of the dS spacetime, 
and the attractor is understood as the
cosmological horizon in that geometry. 
On the other hand, for $q=0.7$,
the geometry around the center is similar to that
of the AdS spacetime (see Appendix \ref{Appendix:Geometrical-quantities}).
In that case, there is no cosmological
horizon and the reason for the existence
of the attractor is different from the case $q=0.0$.
It would be understood by the similarity to the case of the linear mass function,
i.e., as the residue of the inner homothetic Killing horizons of the case $\ell=0$
(compare with the right panel of Fig.~\ref{geodesic_AH_linearmass}).
Although the gathering of outgoing null geodesics
also occurs for $q=0.9$, the attractive property
does not seem to be sufficiently strong to
form an attractor. This result is also similar
to the case of the linear mass function
(compare with the right panel of Fig.~\ref{geodesic_AH_linearmass}). 
In all cases, the gathering of
outgoing null geodesics
causes the blueshift of propagating photons.
This frequency shift will be studied
in more detail in Sect.~\ref{Subsec:Frequency_shift}.


%
\begin{figure}[tb]
  \centering
  \includegraphics[width=0.35\textwidth,bb= 0 0 505 767]{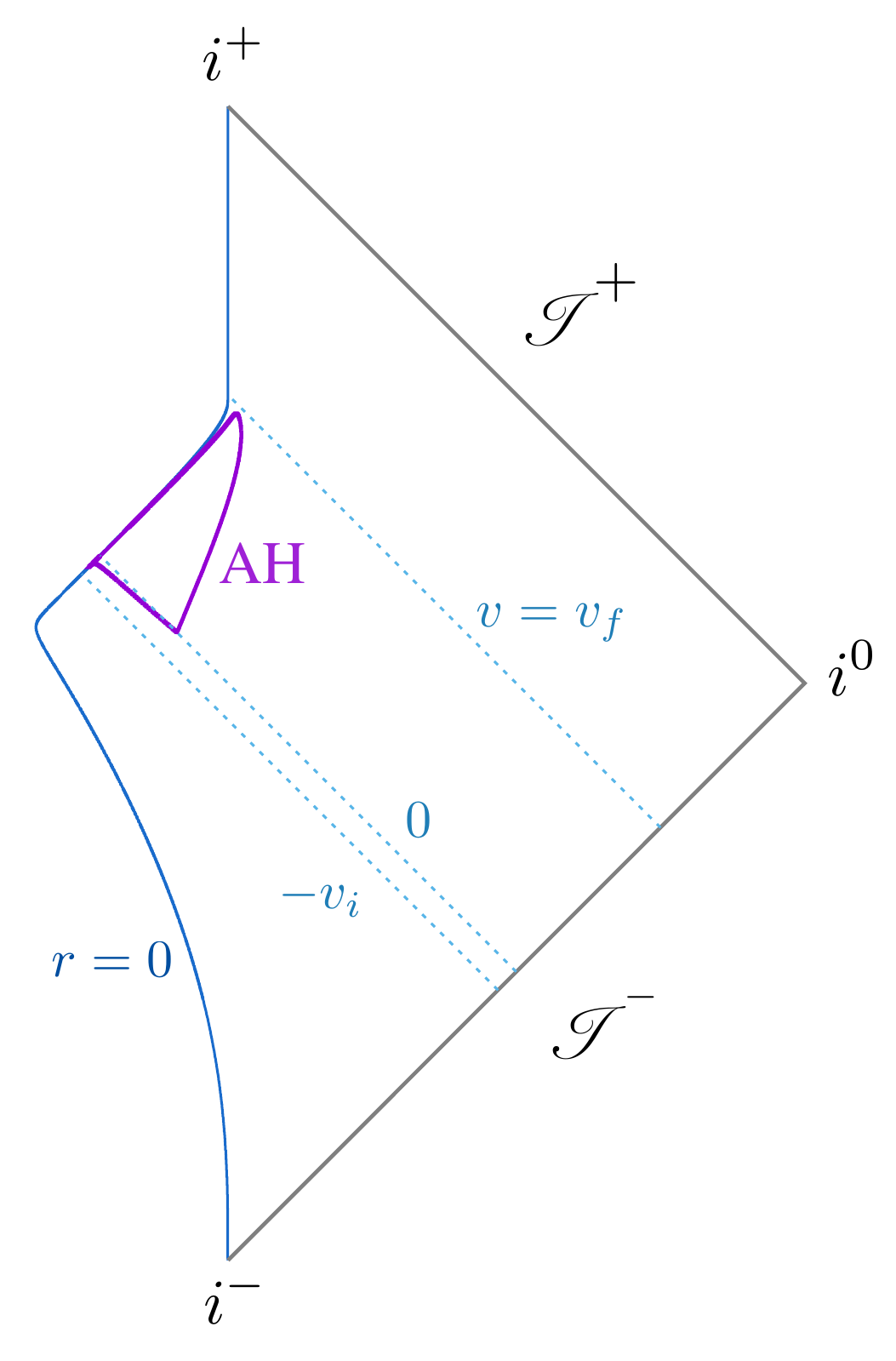}
  \caption{Penrose diagram of the collapse of a charged null fluid and the subsequent evaporation in the case that the mass function $M(v)$ is chosen as Eq.~\eqref{mass-more-realistic} for the parameters $q=0.9$, $M_0/\ell =10$, $v_{\rm i}/\ell =6$, and $v_{\rm f}/\ell =60$.}
  \label{NSRN diagram_realistic mass}
\end{figure}
%

We now show the Penrose diagram for  $q=0.9$ 
in Fig.~\ref{NSRN diagram_realistic mass}.
This diagram is drawn with  
the value $L=v_{\rm f}=60\ell$
in Eqs.~\eqref{Def:eta} and \eqref{Def:zeta}.  
The blue curve shows the worldline of the central point, $r=0$,
and the purple curve shows the closed apparent horizon. 
In this spacetime, there is no event horizon 
because the past of the future null infinity covers all spacetime regions.
Also, no singularity formation happens, and hence,
there is no Cauchy horizon either.
The Penrose diagram has the same structure as that of the Minkowski spacetime,
and therefore, no information problem occurs in this spacetime.
Thus, we have obtained the result that is consistent with the
Hayward-Frolov scenario.

In this diagram, the inner boundary of the trapped region
and the worldline of $r=0$ almost degenerate.
This is because some of the outgoing null geodesics
tend to gather around the inner boundary of the trapped region, and
such outgoing geodesics have similar $u$ values.
In Appendix~\ref{Appendix:Penrose_different},
we present a Penrose diagram for the same spacetime
drawn in a different way. There, the inner boundary of the trapped
region and the worldline of $r=0$ are clearly separated.

\subsection{Frequency shift}
\label{Subsec:Frequency_shift}

Since the spacetime possesses no event and Cauchy horizons,
a photon that falls into the black hole
will come out from it eventually.
Frolov pointed out that some of them are highly
blueshifted in the uncharged case \cite{Frolov:2014jva}.
In that paper, the blueshift experienced by
a photon propagating along the attractor
was calculated analytically.
In the charged case, it is difficult to estimate the strength
of the blueshift analytically because the attractive
property becomes vague for $q$ close to unity.
For this reason, we perform numerical calculations
in the setup of Sect.~\ref{Subsec:Non-self-similar-setup} as follows.

We consider only photons with zero angular momenta,
and suppose that a photon is incident from
past null infinity along $v=v_{\rm in}$.
The angular frequency 
for the observers staying at constant
spatial coordinates (for large $r$) 
is denoted by $\omega_{\rm in}$. 
Then, the photon comes out along $u=u_{\rm out}$,
and at the distant place, the same observers
measure its angular frequency
$\omega_{\rm out}$. We compute the ratio of the two
angular frequencies, 
$\omega_{\rm out}/\omega_{\rm in}$. 
If this quantity is smaller/greater than unity,
the outcoming photon is redshifted/blueshifted.

We derive the necessary formula for computing this quantity.
Hereafter, the affine parameter
of a null geodesic is denoted as $\lambda$
and dot indicates the derivative with respect to $\lambda$.
The tangent vector to the null geodesic is
$k^\mu \ = \ (\dot{v}, \ \dot{r}, \ 0,\ 0)$,
and the geodesic equations are
\begin{subequations}
\begin{equation}
  \ddot{r}-F_-\ddot{v}-\frac12 F_{-,v}\dot{v}^2
  -F_{-,r}\dot{r}\dot{v} \ =\ 0,
  \label{geodesic-eq1}
\end{equation}
\begin{equation}
  \ddot{v}+\frac12F_{-,r}\dot{v}^2 \ = \ 0.
  \label{geodesic-eq2}
\end{equation}
\end{subequations}
From the null condition, we have $\dot{v}=0$ for an ingoing
photon and 
$F_-(v,r)\dot{v} \ = \ 2\dot{r}$ for an outcoming photon.
The four-velocity of the observers staying at
constant $r$, $\theta$, $\phi$ coordinates is
\begin{equation}
u_{\rm (o)}^\mu \ =\ \left(\frac{1}{\sqrt{F_-}}, \ 0, \ 0, \ 0\right),
\end{equation}
and the angular frequency of a photon for these observers is
\begin{equation}
  \omega \ = \
  -k_\mu u_{\rm (o)}^\mu
  \ = \ \sqrt{F_-}\dot{v}
  -\frac{1}{\sqrt{F_-}}\dot{r}. 
\end{equation}

First, we consider the incident photon with $\dot{v}=0$.
For the observer at distant position ($r\to\infty$), the
observed angular frequency is
$\omega_{\rm in}=-\left.\dot{r}\right|_{r=\infty}$
since $F_-=1$ holds at $r\to \infty$.
We introduce an auxiliary observer staying at the center,
$r=0$. The angular frequency observed by this auxiliary observer
is $\omega_{\rm c} =  -\left.\dot{r}\right|_{r=0}$
since $F_-=1$ also holds at $r=0$.
Since Eq.~\eqref{geodesic-eq1} indicates
$\ddot{r}=0$ for $\dot{v}=0$, we have $\dot{r}=\mathrm{constant}$
Therefore,  
$\omega_{\rm in} \ = \ \omega_{\rm c}$ is obtained.

Next, we consider the outcoming photon with $F_-(v,r)\dot{v} \ = \ 2\dot{r}$.
The angular frequencies observed by the auxiliary observer
at the center $r=0$ and the distant observer at $r\to\infty$ is
$\omega_{c} = (1/2)\left.\dot{v}\right|_{r=0}$ and 
$\omega_{out} = (1/2)\left.\dot{v}\right|_{r=\infty}$,
respectively. Therefore, we have 
\begin{equation}
\frac{\omega_{\rm out}}{\omega_{\rm in}} = \frac{\omega_{\rm out}}{\omega_{c}} 
=  \frac{\left.\dot{v}\right|_{r=\infty}}{\left.\dot{v}\right|_{r=0}}.
\end{equation}
 In order to evaluate this quantity, we divide both sides of Eq.~\eqref{geodesic-eq2} with $\dot{v}$ and integrate them from $\lambda=0$ to $\infty$,
 where $\lambda=0$ is the affine parameter at the center, $r=0$.
 The result is
\begin{equation}
  \frac{\omega_{\rm out}}{\omega_{\rm in}}
   \ = \ \exp\left[-\frac12 \int_{v_{\rm in}}^{v_{\rm f}} F_{-,r}(v,r(v)) dv\right],
\end{equation}
where we used $F_-=1$ for $v\ge v_{\rm f}$.
After obtaining the numerical data of
outgoing null geodesics, $r(v)$, numerical integrations
of this formula can be computed easily.

%
\begin{figure}[tb]
  \centering
  \includegraphics[width=0.45\textwidth,bb=0 0 360 252]{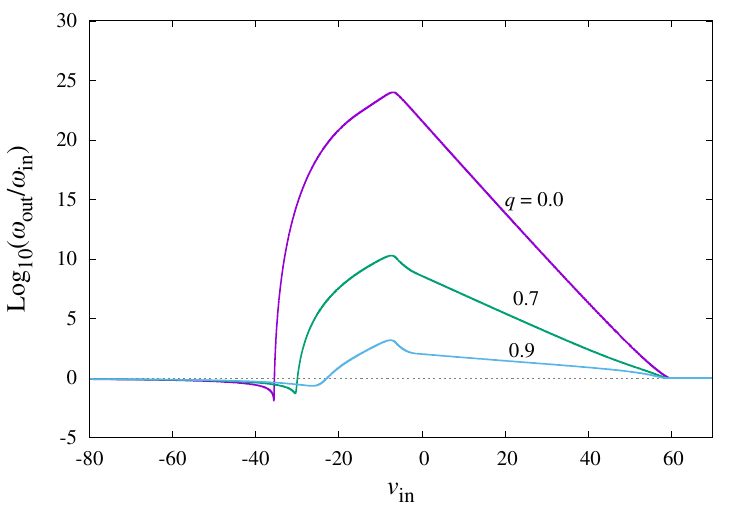}
  \includegraphics[width=0.45\textwidth,bb=0 0 360 252]{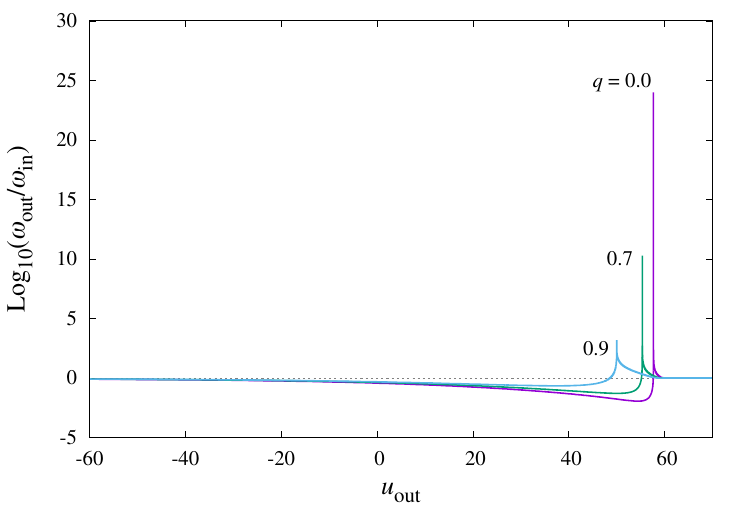}
  \caption{The factor $\omega_{\rm out}/\omega_{\rm in}$
    as a function of $v_{\rm in}$ (left panel) and of $u_{\rm out}$ (right panel)
    in the cases of $q=0.0$, $0.7$, and $0.9$. The unit of the horizontal axis is $\ell$.}
  \label{Fig:Redshift-Blueshift}
\end{figure}
%

Figure~\ref{Fig:Redshift-Blueshift} shows
the behavior of $\omega_{\rm out}/\omega_{\rm in}$
in the logarithmic scale
for the case $q=0.0$, $0.7$, and $0.9$.
Since each geodesic has the values of $v_{\rm in}$
and $u_{\rm out}$, the quantity $\omega_{\rm out}/\omega_{\rm in}$
is shown as functions of $v_{\rm in}$
and $u_{\rm out}$ in the left and right panels, respectively.
In all cases, the redshift happens initially,
and the blueshift occurs after that.
If $\omega_{\rm out}/\omega_{\rm in}$
is regarded as a function of $u_{\rm out}$, 
the curve for $q=0$ has a sharp peak with
an extremely large peak value, while the duration
of the blueshift is very short.
As the value of $q$ is increased,
the peak becomes less sharp, and the
duration becomes longer instead.
This reflects the fact that the attractive
property around the inner boundary of the trapped region
becomes weaker for larger $q$.

This blueshift at the last stage of evaporation 
may lead to interesting astrophysical phenomena.
Also, it would be interesting to study the property
of Hawking radiation caused by this blueshift,
as pointed out in \cite{Frolov:2014jva}
and analyzed in \cite{Frolov:2016,Frolov:2017} in the uncharged case.
These issues are left as
remaining problems to be discussed in future.

%
%
\section{Summary and discussion}
\label{Sec:summary}

In this paper, we have studied 
the collapse of a charged matter
and its evaporation in the situation
that the central singularity is resolved,
in order to assess whether the Hayward-Frolov scenario
possesses sufficient generality.
In Sect.~\ref{Sec:regularization},
we have constructed the simplest model of the (static)
regularized Reissner-Nordstr\"{o}m black hole spacetime 
and the regularized charged Vaidya spacetime.
The spacetime structure of an evaporating black hole
has been studied using two kinds of regularization, the
time-dependent regularization and the constant regularization,
in Sects.~\ref{Sec:Self-similar-model}
and \ref{Sec:Non-self-similar}, respectively. 
The time-dependent regularization, where the regularization scale
is proportional to mass as $\ell(v)=bM(v)$, 
has been considered since the spacetime possesses
the self-similarity and analytical construction of 
double null coordinates is possible if we choose 
a linearly decreasing mass function $M(v)$.
The result is that a point singularity or
a null singularity is formed in this spacetime,
and the resolution is not sufficient to resolve the
information loss problem.
In the study of the constant regularization, where 
the scale of the regularization $\ell$ is fixed,
we have obtained the Penrose diagrams of Fig.~\ref{NSRN diagram_realistic mass}.
The resultant spacetime possesses no event horizon,
no singularity, and no Cauchy horizon,
and therefore, the information loss problem does not arise.  
This result supports the Hayward-Frolov scenario.

We have also examined the properties of both  Kaminaga's  model and the evaporating non-singular Reissner-Nordstr\"{o}m black hole and compared the two results.
In the case of Kaminaga's model, the inner and outer homothetic Killing horizons exist for the case of a sufficiently small charge, and they coincide with the attractor and repeller of outgoing null geodesics.
For the case that the charge is sufficiently large, there is no such an attractor or a repeller because the homothetic Killing horizons are absent.
A similar phenomena could be observed in the nonsingular model. 
Although an attractor and a repeller can be temporally found in the case of a small charge, both of them vanish for the case of a sufficiently large charge.
Reflecting the absence of the attractor, the blueshift of outcoming photons
at the last stage of the evaporation 
becomes weaker as the charge is increased.
Correspondingly, the period during which
the blueshift can be observed becomes longer for
outside observers.

There remains several issues to be explored.
First, as mentioned
in Sect.~\ref{Sec:Introduction}, Levin and Ori \cite{Levin:1996qt}
suggested that there are two possible cases
for the models of collapse of charged null matter
and its evaporation,
which they called models (A) and (B).
The model (A) corresponds to Kaminaga's model
in which the accreting matter continues to
implode to the center.
On the other hand, in the model (B), 
the accreting charged null matter
experiences bounce at some radius. Although both models are
the solution to Einstein's equations,
it was suggested that the model (B)
would be more plausible. Regularizing such a spacetime
would be more difficult than the analysis of the present paper,
and we postpone this problem for the future work.

Second, whether the Hayward-Frolov model
holds for systems with angular momenta
must be investigated.
Since the Penrose diagram of a Kerr black hole
is very similar to the diagram of the Reissner-Nordstr\"{o}m black hole,
we expect that the information loss problem would be solved for the
system with angular momentum in a similar manner.
The explicit construction of the model for the gravitational
collapse of a rotating star and its evaporation
is necessary for clarifying this point.
The Kerr-Vaidya solution \cite{Senovilla:2014}
may be useful for this purpose.

Third, the mechanism of regularizing the curvature singularity
must be studied. As cited in Sect.~\ref{Sec:Introduction},
there are several existing works that realize the
spacetime of an evaporating black hole without curvature
singularity \cite{Frolov:1981,Roman:1983,Bambi:2013,Rovelli:2014,Bambi:2016,Kawai:2017,Binetruy:2018,Ho:2019,Kawai:2020}.
In addition to these works, 
a new mechanism to resolve the curvature singularity
due to running of the gravitational constant 
is proposed recently \cite{Chen:2022}. It would be interesting
to examine whether this mechanism realizes the Hayward-Frolov scenario.

Finally, we have to comment on the existing works \cite{Carballo-Rubio:2018}
(see also \cite{Bonanno:2022})
that raised doubts for the Hayward-Frolov scenario.
If the evaporation
of a black hole with the metric of Eqs.~\eqref{Static-spacetime-metric-ingoing-EF} and \eqref{Hayward-form} is considered in the quasistatic approximation,
the two horizons tend to degenerate as the mass is decreased
in the evaporation. Then, the temperature of the black hole becomes zero,
the Hawking radiation stops, 
and a remnant remains.
Inside of the remnant, the Cauchy horizon is formed
and the instability develops around it.
If this is the case, the Hayward-Frolov scenario
would not work well to resolve the information loss problem.

Our opinion on this issue is as follows.
Recall that the black hole
is supposed to evaporate completely within a finite time
in the works of Hayward and Frolov.
In such a case, both the event and Cauchy horizons do not appear
in the spacetime. The same thing has happened in the analysis
of our paper. 
In Sect.~\ref{Sec:Introduction}, we asked whether
the singularity that develops near the Cauchy horizon
must be resolved in the system with electric charge/angular momentum,
and our analysis indicates that
such a procedure is not necessary. This is because the
development of the Cauchy horizon is a phenomena associated
with infinite time (with respect to the outside domain of the black hole),
while the black hole is assumed to evaporate in a finite time
in our model. In the analyses Hayward, Frolov, and us,
the quasi-static approximation
is obviously violated 
at the last stage of the evaporation 
as indicated by the disappearance of the trapped region. In this sense,
the Hayward-Frolov scenario requires
the violation of quasistatic approximation.

From this observation, it is very important to determine whether
the regularized black hole evaporates completely in a finite time or
not. For this purpose, a more refined model of
the evaporation process must be developed
by taking account of the
backreaction effects of the Hawking radiation.
We hope to clarify this problem in near future.

\ack

The authors thank Ken-ichi Nakao and Hideki Maeda for helpful comments
and discussions. 
K.S. is in part, supported by Japan Science and Technology Agency, the establishment of university fellowships towards the creation of science technology innovation, Grant Number JPMJFS2138.
H. Y. is in part supported by JSPS KAKENHI Grant Numbers JP22H01220 and
JP21H05189,
and is partly supported by Osaka Central Advanced Mathematical Institute 
(MEXT Joint Usage/Research Center on Mathematics and Theoretical Physics JPMXP0619217849).

\appendix

%
%
\section{Homothetic Killing vector field of the self-similar model}
\label{Appendix:Self-similar}

In this Appendix, we present the detailed calculations
that show the presence of a homothetic Killing vector field
for the self-similar model of an evaporating charged black hole
in Sects.~\ref{Subsec:Evaporation_model_A} and \ref{Sec:Self-similar-model}.

For simplicity, we express the metric as $ds^2=-A(u,\tilde{v})dud\tilde{v}+r(u,\tilde{v})^2d\Omega^2$, where
\begin{equation}
  A(u,\tilde{v})=M_0^2\alpha\tilde{v}(f\alpha+2R).
  \label{Eq:definition-A}
\end{equation}
We write the covariant
components of the homothetic Killing vector as
\begin{equation}
  \xi_\mu=(\xi_u, \xi_{\tilde{v}}, 0, 0).
\end{equation}
The $(u,u)$ and $(\tilde{v},\tilde{v})$ components
of the homothetic Killing equation,
Eq.~\eqref{Eq:homothetic_Killing_equation}, 
are
\begin{subequations}
\begin{eqnarray}
  -\frac{2\xi_u\partial_u A}{A}+2\partial_u \xi_u&=&0,\\
  -\frac{2\xi_{\tilde{v}}\partial_{\tilde{v}} A}{A}+2\partial_{\tilde{v}}\xi_{\tilde{v}}&=&0,
\end{eqnarray}
\end{subequations}
respectively, and these equations are integrated as
\begin{subequations}
\begin{eqnarray}
  \xi_u(u,\tilde{v})&=&C_u(\tilde{v})A(u,\tilde{v}),\\
  \xi_{\tilde{v}}(u,\tilde{v})&=&C_{\tilde{v}}(u)A(u,\tilde{v}).
\end{eqnarray}
\end{subequations}
Substituting these formulas
into the $(u,\tilde{v})$ and $(\theta,\theta)$ components of the
homothetic Killing equation, we have
\begin{eqnarray}
      \frac{dC_u}{d\tilde{v}}+\frac{dC_{\tilde{v}}}{du}+\frac{1}{A}\frac{\partial A}{\partial {\tilde{v}}}C_u+\frac{1}{A}\frac{\partial A}{\partial u}C_v&=&-\frac{K}{2},\\
      \frac{1}{r}\partial_v rC_u+\frac{1}{r}\partial_u rC_{\tilde{v}}&=&-\frac{K}{4},
\end{eqnarray}
respectively. Note that 
the $(\phi,\phi)$ component is the same as the $(\theta,\theta)$ component
due to spherical symmetry. 
Eliminating $K$ from these equations, we obtain
\begin{align}
  \frac{dC_u}{d\tilde{v}}+\left(\frac{1}{A}\frac{\partial A}{\partial \tilde{v}}-\frac{2}{r}\frac{\partial r}{\partial \tilde{v}}\right)C_u+\frac{dC_{\tilde{v}}}{du}+\left(\frac{1}{A}\frac{\partial A}{\partial u}-\frac{2}{r}\frac{\partial r}{\partial u}\right)C_{\tilde{v}}=0.
\end{align}
Here, we rewrite the formulas in the
parentheses in this equation using
Eqs.~\eqref{Eq:definition-A} and \eqref{Eq:Introduction-u}.
The result is
\begin{equation}
  \frac{dC_u}{d\tilde{v}}+\frac{\alpha}{\tilde{v}}\left(\frac{f}{R}-\frac{1}{2}\frac{df}{dR}\right)C_u+\frac{dC_{\tilde{v}}}{du}+\left(-1+\frac{\alpha}{2}\frac{df}{dR}-\frac{f\alpha}{R}\right)C_{\tilde{v}}=0.
\end{equation}
Putting $C_u(\tilde{v})=\tilde{v}\tilde{C}_u(\tilde{v})$, this equation
becomes
\begin{align}
  \left(\tilde{v}\frac{d\tilde{C}_u}{d\tilde{v}}+\frac{dC_v}{du}\right)+\left[1+\alpha\left(\frac{f}{R}-\frac{1}{2}\frac{df}{dR}\right)\right](\tilde{C}_u-C_v)=0.
\end{align}
It follows that $C_v=\tilde{C}_u=\mathrm{constant}$ is the solution to
this equation.
With a constant parameter $\chi$, the covariant and contravariant
components
of the homothetic Killing vector field in this spacetime is
\begin{subequations}
\begin{eqnarray}
  \xi_\mu &=& \chi A(u,\tilde{v})(\tilde{v}, 1, 0, 0),\\
  \xi^\mu &=& -2\chi(1,\tilde{v}, 0, 0).
\end{eqnarray}
\end{subequations}
The norm of the homothetic Killing vector is
\begin{equation}
  g_{\mu\nu}\xi^\mu \xi^\nu
  \, =\,-4\chi^2\tilde{v}A(u,\tilde{v}).
\end{equation}
From this result, $\xi^\mu$ becomes null
at positions satisfying $A(u,\tilde{v})=0$,
and thus, 
the homothetic Killing horizons can be found by solving
the equation $f \alpha+2R=0$.
Choosing $\chi={1}/{2}$, the homothetic Killing vector field
becomes $\xi^\mu=-(1,\tilde{v}, 0, 0)$ [that is, Eq.~\eqref{homothetic_Killing_contravariant}] and satisfies
Eq.~\eqref{Eq:homothetic_Killing_equation} with $K=-2$.

%
%
\section{Apparent horizons in Vaidya spacetime}
\label{Appendix:AH}

In this Appendix, we show that the boundaries
of the trapped region (or the apparent horizon)
are located at the radial position satisfying $F_-(v,r)=0$ 
in (both regularized and nonregularized) 
Vaidya spacetimes. The following calculations hold for both charged
and uncharged cases.

Let us compute the expansion of the congruence
of the outgoing radial
null geodesics.
Since the null condition $ds^2=0$ implies
\begin{align}
  0=dv(-F_-dv+2dr),
\end{align}
we have the conditions for the ingoing null geodesics $dv=0$
and the outgoing null geodesics 
$F_-dv=2dr$. This implies that the tangent vector
$k^\mu$ of the outgoing null geodesics can be expressed as 
$k^\mu=(1,{F_-}/{2},0,0)$ in the $(v,r,\theta,\phi)$ coordinates. 
The geodesic equation expressed in terms of $k^\mu$ is
\begin{align}
  k^\nu\nabla_{\nu}k^\mu\,=\,\kappa \,k^\mu\quad
  \textrm{with}\quad\kappa=\frac{1}{2}\partial_r F_-.
\end{align}
From this formula, $k^\mu= dx^\mu/d\lambda$ 
is the tangent vector of a geodesic
parametrized by a non-affine parameter $\lambda$.
In terms of the affine parameter $\lambda^*$,
the tangent vector is expressed as $k_*^\mu = dx^\mu/d\lambda^*$
and satisfies $k_*^\nu\nabla_\nu k_*^\mu = 0$.
From a simple calculation, the relation between
$k_*^\mu$ and $k^\mu$ is shown to be given by 
\begin{align}
  k_*^\mu&=\exp{\left(-\int{\kappa d\lambda}\right)}\,k^\mu.
\end{align}
Then, the expansion $\Theta$ of the null geodesic congruence
can be computed as 
\begin{align}
  \Theta\,:=\,\nabla_{\mu}k_*^\mu
  \,=\,\exp{\left(\int{\kappa d\lambda}\right)}\frac{F_-}{r}.
\end{align}
This shows that the contour surfaces of $F_-(v,r)=0$
are the boundaries of the trapped region, and
the outermost one is the apparent horizon.

The interpretation of this result is as follows. 
The zero expansion of the outgoing null geodesic congruence
is realized at a position 
where the change in the sectional area $4\pi r^2$ along the congruence 
becomes zero, that is, $dr=0$.
This condition with the null condition $-F_-dv+2dr=0$ implies that
$F_-=0$ is the condition for zero expansion.

%
%
\section{Spacetime structure of the collapse domain}
\label{Appendix:Collapse-domain}

In this Appendix, we present the 
analysis of the spacetime structure in the collapse domain
of a charged null-matter fluid in the self-similar cases
of Sects.~\ref{Sec:review-part} and \ref{Sec:Self-similar-model}. 
We adopt the metric of the ingoing Vaidya metric
of Eq.~\eqref{Metric:IngoingVaidya}, and
consider the collapse domain,
$-v_{\rm i}\leq v\leq 0$, of Eqs.~\eqref{Linear-mass-function}
and \eqref{mass-more-realistic}.
The charge function is related to the mass function
as Eq.~\eqref{Linear-charge-function}.
In the similar manner to the analysis of the
evaporation phase, $0\leq v\leq v_{\rm f} $, we 
introduce new coordinates, $\tilde{v}=M(v)/M_0$ and $R=r/M(v)$.
Then, for the function $F_-(v,r)$
in the self-similar cases given by Eq.~\eqref{F-ingoingVaidya2},
the function can be written in the same form as
Eq.~\eqref{Definition-f(R)}, i.e. $F_-(v,r) = f(R)$.
Introducing $\alpha=v_{\rm i}/M_0$, the metric is rewritten as
\begin{align}
  ds^2=-M_0^2\alpha\tilde{v}(f\alpha-2R)d\tilde{v}\left(\frac{d\tilde{v}}{\tilde{v}}-\frac{2dR}{f\alpha-2R}\right).
\end{align}
From this equation, the null coordinate $u$ can be defined as 
\begin{align}
  \label{eq-of u}
  u=\ln{\tilde{v}}-\int{\frac{2dR}{f\alpha-2R}},
\end{align}
and then, the metric is given as $ds^2=-M_0^2\alpha\tilde{v}(f\alpha-2R)d\tilde{v}du$.
Below, we consider the case without regularization
(Kaminaga's model in Sect.~\ref{Sec:review-part}, or equivalently, the case $b=0$ of Sect.~\ref{Sec:Self-similar-model}) and
the case with time-dependent regularization ($b\neq 0$ in Sect.~\ref{Sec:Self-similar-model}), separately.

\subsection{The case without regularization}

Similarly to the
analysis in the evaporation domain of Sect.~\ref{Sec:review-part},
the homothetic Killing horizons
play an important role. As shown in Appendix~\ref{Appendix:Self-similar},
the homothetic Killing horizons 
exist at positions where $f\alpha-2R=0$ is satisfied,
and this equation is rewritten as 
\begin{align}
  \label{eq-formation}
\frac{1}{\alpha}\,=\,\frac{R^2-2R+q^2}{2R^3}=:y(R).
\end{align}
The definition of $y(R)$
is the same as Eq.~\eqref{eq-formation-evaporation}, and 
the graph of the function $y(R)$ is shown in Fig.~\ref{Fig:y}
for the case $q = 0.6$. 
As noted in the main text,
$y(R)$ takes extremal values at $R=R_\pm:=2\pm\sqrt{4-3q^2}$,
and there, $y(R_-)<0$ and $y(R_+)>0$
are satisfied.
Since $\alpha$ is a positive constant, 
Eq.~\eqref{eq-formation}
has one, two, and three solutions
in the cases that $1/\alpha>y(R_+)$,
$1/\alpha=y(R_+)$,
and $1/\alpha<y(R_+)$, respectively, where
\begin{equation}
  y(R_+)
  \,=\,
  \frac{2-q^2+\sqrt{4-3q^2}}{(2+\sqrt{4-3q^2})^3}.
\end{equation}
Therefore, the mass accretion rate $1/\alpha=M_0/v_{\rm i}$ 
and the electric charge per unit mass $q$
determine the number of the homothetic Killing horizons.

We suppose that the mass accretion rate can be controlled
and the simple case can be selected. For this reason,
the case that $f\alpha-2R=0$ has one solution (denoted as $R=R_{\rm C}$)
is considered below.
In this case, $f\alpha-2R$ can be factorized as 
\begin{eqnarray}
  f\alpha-2R &=& -\frac{2}{R^2}(R^3+\alpha R^2+2\alpha R-\alpha q^2)
  \nonumber\\
  &=&-\frac{2}{R^2}(R-R_{\rm C})(R^2+C_1R+C_2),
\end{eqnarray}
with constants $R_{\rm C}$, $C_1$, and $C_2$
satisfying $R_{\rm C}>0$ and $C_2>{C_1^2}/{4}$. 
Then, the integral of Eq.~\eqref{eq-of u}
can be carried out as
\begin{align}
  \label{u of Appendix-C}
  u\,=\,
  \ln{\tilde{v}}+\frac{1}{\kappa_A}\arctan{\left[\frac{C_1+2R}{\sqrt{-C_1^2+4C_2}}\right]}+\frac{1}{\kappa_B}\ln{(R^2+C_1R+C_2)}+\frac{1}{\kappa_C}\ln{|R-R_{\rm C}|},
\end{align}
with
\begin{subequations}
\begin{eqnarray}
    \kappa_A&=&-\frac{(R_{\rm C}^2+C_1R_{\rm C}+C_2)\sqrt{-C_1^2+4C_2}}{C_1C_2+C_1^2R_{\rm C}-2C_2R_{\rm C}},\\
    \kappa_B&=&\frac{2(R_{\rm C}^2+C_1R_{\rm C}+C_2)}{C_1R_{\rm C}+C_2},\\
    \kappa_C&=&\frac{R_{\rm C}^2+C_1R_{\rm C}+C_2}{R_{\rm C}^2}.
\end{eqnarray}
\end{subequations}
The value of $u$ diverges at $R=R_{\rm C}$, and this coordinate singularity
must be resolved. This is done by introducing the new coordinate
\begin{align}
  U=
  \begin{cases}
    -\exp({\kappa_C u})&\quad(R>R_{\rm C}),\\
    \exp({\kappa_C u})&\quad (R<R_{\rm C}),
  \end{cases}
  \label{Coordinate-U-collapse}
\end{align}
and then, the relation between $U$, $\tilde{v}$ and $R$ is given by  
\begin{align}
  \frac{U}{\tilde{v}^{\kappa_C}}=-(R-R_{\rm C})(R^2+C_1R+C_2)^{\frac{\kappa_C}{\kappa_B}}\exp{\left(\frac{\kappa_C}{\kappa_A}\arctan{\left[\frac{C_1+2R}{\sqrt{-C_1^2+4C_2}}\right]}\right)}.
  \label{relation-U-tildev-R}
\end{align}
Using $U$, the metric is expressed as 
\begin{align}
  ds^2=-\frac{2\alpha M_0^2}{\kappa_C R^2}\tilde{v}^{1-\kappa_C}(R^2+C_1R+C_2)^{1-\frac{\kappa_C}{\kappa_B}}\exp{\left(-\frac{\kappa_C}{\kappa_A}\arctan{\left[\frac{C_1+2R}{\sqrt{-C_1^2+4C_2}}\right]}\right)}dUd\tilde{v}.
\end{align}
There is no singularity other than $R=0$ (that corresponds to
the central point $r=0$)
in this metric.

Below, we consider properties
of $R$-constant surfaces.
For this purpose,
we compute the normal of the $R$-constant surfaces, $n_\mu = \nabla_\mu R$.
From Eq.~\eqref{eq-of u}, the $(u, \tilde{v}, \theta,\phi)$-components
of $n_\mu$ are
\begin{equation}
  n_\mu\,=\,\frac{f\alpha-2R}{2}\left(-1, \,\tilde{v},\, 0, \, 0\right),
\end{equation}
and the norm of this vector is computed as
\begin{align}
n_{\mu}n^{\mu} \,=\, \frac{f\alpha-2R}{M_0^2\alpha\tilde{v}^2}.
\end{align}
From this, $R$-constant surfaces with $R<R_{\rm C}$, $R=R_{\rm C}$,
and $R>R_{\rm C}$
are timelike, null, and spacelike since $n_\mu n^\mu>0$, $n_\mu n^\mu=0$,
and $n_\mu n^\mu<0$ hold, respectively. 

%
\begin{figure}[tb]
  \centering
  \includegraphics[scale=0.7]{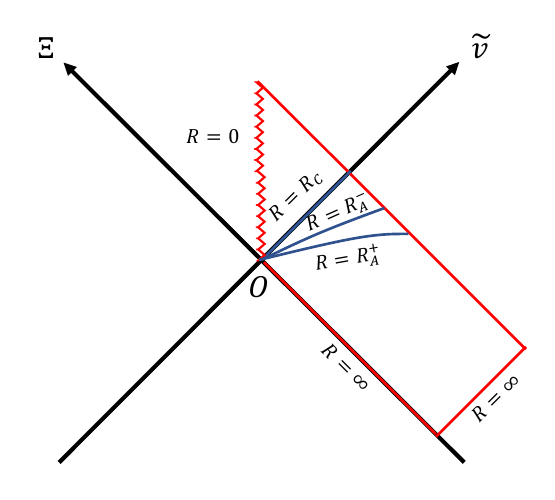}
  \caption{Penrose diagram of the collapse domain in the case of $b=0$.}
  \label{pic-diagram of collapse}
\end{figure}
%

Let us consider the location of several $R$-constant surfaces. 
First, we consider the homothetic Killing
horizon at $R=R_{\rm C}$.
Substituting $R=R_{\rm C}$ into Eq.~\eqref{relation-U-tildev-R},
we find $U=0$. 
Second, we consider $R=\infty$. Taking the limit $R\to \infty$
in Eq.~\eqref{relation-U-tildev-R}, we have 
${U}/{\tilde{v}^{\kappa_C}}=-\infty$. This means 
$U=-\infty$ or $\tilde{v}=0$.
Third, the central singularity 
$R=0$ is located at
\begin{align}
  \frac{U}{\tilde{v}^{\kappa_C}}=R_{\rm C}(C_2)^{\frac{\kappa_C}{\kappa_B}}\exp{\left(\frac{\kappa_C}{\kappa_A}\arctan{\left[\frac{C_1}{\sqrt{-C_1^2+4C_2}}\right]}\right)}. 
\end{align}
This is a timelike surface in the domain $U>0$ and $\tilde{v}>0$. 
Finally, let us consider the location of the boundary
of the trapped region, or the (inner and outer) apparent horizons.
As shown in Appendix~\ref{Appendix:AH},
apparent horizons satisfy $f(R)=0$. This is equivalent
to $y(R)=0$, and there are two solutions
$R_{\rm A}^+$ and $R_{\rm A}^-$ (here we suppose $R_{\rm A}^-<R_{\rm A}^+$).
From Fig.~\ref{Fig:y}, both $R_{\rm A}^+$ and $R_{\rm A}^-$
are larger than $R_{\rm C}$, and therefore, the apparent horizons
are spacelike hypersurfaces.
We also remark that $R$
can take arbitrary positive values at the point $(U,\tilde{v})=(0,0)$.
Therefore, all $R$-constant surfaces plunge into this point.

We now present the Penrose diagram of the collapse domain, $-v_{\rm i}\leq v\leq 0$. 
The coordinate range of $\tilde{v}$ is finite,
$0\leq\tilde{v}\leq 1$, while that of $U$ is infinite.
In order to 
present the $U$ direction in a finite domain,
we introduce the compactified coordinate by 
$U=\tan{\Xi}$ with 
$-{\pi}/{2}<\Xi<{\pi}/{2}$. 
In the coordinates $(\Xi, \tilde{v})$,
the past null infinity $\mathscr{I}^-$ is located at $\Xi=-{\pi}/{2}$,
and the homothetic Killing horizon
is located at $\Xi=0$. The central singularity $R=0$
is a timelike surface in the domain $\Xi>0$ and $\tilde{v}>0$,
and the apparent horizons at $R=R_{\rm A}^\pm$ are spacelike surfaces 
in the domain $\Xi<0$ and $\tilde{v}>0$. 
From these observations,
we obtain the diagram of Fig.~\ref{pic-diagram of collapse}.

\subsection{The case of time-dependent regularization}

We now briefly comment on the case 
that Eq.~\eqref{F-ingoingVaidya2}
with $b\neq 0$ is used for the function $F_-(v,r)$.
In this case, the function $y(R)$ of Eq.~\eqref{eq-formation}
is modified as
\begin{equation}
  y(R)\,=\, \frac{R^4-2R^3+q^2R^2+(2R+q^2)b^2}{2R\left[R^4+(2R+q^2)b^2\right]}.
  \label{eq-formation-nonzerob}
\end{equation}
The primary difference from Eq.~\eqref{eq-formation}
is that the divergence at $R\to 0$ becomes weaker.
Equation~\eqref{eq-formation} behaves as
$y(R)\approx q^2/2R^3$ while Eq.~\eqref{eq-formation-nonzerob}
behaves as $y(R)\approx 1/2R$ at $R\ll 1$.
Except that, the behavior of $y(R)$ of Eq.~\eqref{eq-formation-nonzerob}
is similar to that of $y(R)$ of Eq.~\eqref{eq-formation}
as long as $b$ is sufficiently small,
and the similar discussion holds also for the case $b\neq 0$.
Namely, $y(R)$ takes extremal values at $R=R_{\pm}$,
and Eq.~\eqref{eq-formation-nonzerob}
has one, two, and three positive solutions in the case that $1/\alpha>y(R_+)$,
$1/\alpha=y(R_+)$,
and $1/\alpha<y(R_+)$, respectively.

%
\begin{figure}[tb]
  \centering
  \includegraphics[scale=0.7]{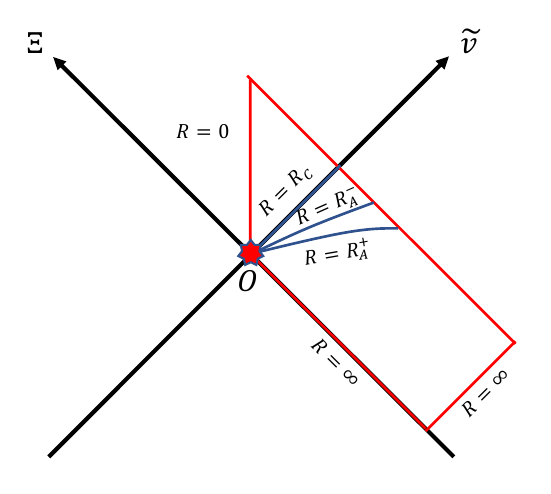}
  \caption{Penrose diagram of collapse domain in the case of $b\neq 0$.}
  \label{pic-diagram of collapse-regularized}
\end{figure}
%

We consider the case that
there is only one solution to Eq.~\eqref{eq-formation-nonzerob}
(denoted as $R=R_{\rm C}$).
Although carrying out the integral of Eq.~\eqref{eq-of u}
analytically seems to be difficult, we can
proceed a formal discussion as follows.
We can write $f\alpha-2R$ as
\begin{equation}
f\alpha-2R \,=\, -2(R-R_{\rm C})\kappa(R)
\end{equation}
with some positive definite function $\kappa(R)$
with the property $\kappa(0) = \alpha /2R_{\rm C}$ and
$\kappa(\infty)=1$. 
Equation~\eqref{eq-of u} is rewritten as
\begin{equation}
  u=\ln{\tilde{v}} + \frac{1}{\kappa_C}
  \ln|R-R_{\rm C}|
  +{G(R)}
\end{equation}
with $\kappa_C:=\kappa(R_{\rm C})$ and
\begin{equation}
G(R)=\int \frac{1/\kappa(R)-1/\kappa_C}{R-R_{\rm C}}dR.
\end{equation}
Note that the integrand of $G(R)$
is regular both at $R=0$ and $R=R_{\rm C}$.
Introducing the coordinate
$U$ in the same manner as Eq.~\eqref{Coordinate-U-collapse},
we have
\begin{equation}
\frac{U}{\tilde{v}^{\kappa_C}}=-(R-R_{\rm C})\exp\left[\kappa_C G(R)\right]. 
\end{equation}
In the $(U,\tilde{v})$-coordinates, the metric is written as
\begin{equation}
  ds^2 \,=\, -\frac{2\alpha M_0^2}{\kappa_C}
  \tilde{v}^{1-\kappa_C}
  \kappa(R)\exp\left[-\kappa_C G(R)\right]dUd\tilde{v}.
\end{equation}
In the case $b\neq 0$,
the metric is regular on the worldline of the central point, $R=0$, except
at $(U,\tilde{v})=(0,0)$. The point $(U,\tilde{v})=(0,0)$ 
is a curvature singularity by the fact that
all $R$-constant surfaces plunges into this point.
The Penrose diagram can be drawn in the similar manner to the $b=0$ case.
The only difference is that $R=0$ becomes the regular center
except at $(U,\tilde{v})=(0,0)$, and the diagram of 
Fig.~\ref{pic-diagram of collapse-regularized} is obtained.

%
%
\section{Geometrical quantities of ingoing Vaidya spacetimes and the characteristic of apparent horizons}
\label{Appendix:Geometrical-quantities}

In this appendix, we summarize the calculation
of geometric quantities of the (partly) regularized charged Vaidya spacetimes.
In Appendix~\ref{Curvature_invariants},
we present the formula for the curvature invariants
constructed from the Riemann tensor.
In Appendix~\ref{Appendix:Energy condition}, we examine the behavior
of the Einstein tensor in the ingoing Vaidya spacetimes
in detail, and whether the null and dominant energy conditions
(NEC and DEC)
are satisfied in the framework of general relativity.
We examine the validity of the characteristic of the apparent horizons
in Appendix~\ref{Characteristic_of_AH}.\footnote{Appendixes \ref{Appendix:Energy condition} and \ref{Characteristic_of_AH} have been added by a strong
suggestion by the referee, and many of the analyses were helped and inspired by referee's report.}

\subsection{Curvature invariants}
\label{Curvature_invariants}

For the metric of Eq.~\eqref{Metric:IngoingVaidya},
the representative geometric invariants
that are constructed from the Riemann tensor $\mathcal{R}_{\mu\nu\rho\sigma}$ 
are calculated as
\begin{subequations}
\begin{eqnarray}
  \mathcal{R}&=& -\frac{2\left(F_--1\right)}{r^{2}}-\frac{4F_{-,r}}{r}-F_{-,rr},
  \\
  \mathcal{R}_{\mu\nu}\mathcal{R}^{\mu\nu}&=&
  2\left(\frac{F_--1}{r^2}+\frac{F_{-,r}}{r}\right)^2
  +2\left(\frac{F_{-,r}}{r}+\frac{F_{-,rr}}{2}\right)^2,
  \\
  \mathcal{R}_{\mu\nu\rho\sigma}\mathcal{R}^{\mu\nu\rho\sigma}&=&
  4\left(\frac{F_--1}{r^2}\right)^2
    +4\left(\frac{F_{-,r}}{r}\right)^2+F_{-,rr}.
\end{eqnarray}
\end{subequations}

\subsubsection{Time-dependent regularization}

In the self-similar case studied in Sect.~\ref{Sec:Self-similar-model},
$F_-(v,r)$ is expressed as  
a function $f(R)$ with $R=r/M(v)$ as Eq.~\eqref{Definition-f(R)}.
In this case, the geometric invariants are
\begin{subequations}
\begin{eqnarray}
  \mathcal{R}&=& -\frac{1}{M(v)^{2}}
  \left[\frac{2\left(f(R)-1\right)}{R^2}
    +\frac{4f^\prime(R)}{R}+f^{\prime\prime}(R)\right],
  \\
  \mathcal{R}_{\mu\nu}\mathcal{R}^{\mu\nu}&=&
  \frac{2}{M(v)^{4}}\left[
  \left(\frac{f(R)-1}{R^2}+\frac{f^\prime(R)}{R}\right)^2
  +\left(\frac{f^\prime(R)}{R}+\frac{f^{\prime\prime}(R)}{2}\right)^2
  \right],
  \\
  \mathcal{R}_{\mu\nu\rho\sigma}\mathcal{R}^{\mu\nu\rho\sigma}&=&
  \frac{1}{M(v)^{4}}\left[
  4\left(\frac{f(R)-1}{R^2}\right)^2+4\left(\frac{f^\prime(R)}{R}\right)^2
  +\left(f^{\prime\prime}(R)\right)^2
  \right].
\end{eqnarray}
\end{subequations}
These geometric invariants diverge
in the limit $M(v)\to 0$ for a fixed $R$,
indicating the presence of the spacetime singularity.
In particular, $f(R)$ behaves as $f(R)\approx 1+R^2/b^2$ 
in the neighborhood of $R=0$, and this gives
Eqs.~\eqref{Eq:Ricci-scalar-Req0}--\eqref{Eq:Riemann-tensor-squared-Req0}
in the main text.

\subsubsection{Constant regularization}

In the constant regularization in Sect.~\ref{Sec:Non-self-similar},
the function $F_-(v,r)$ behaves as $F_-(v,r)\approx 1+r^2/\ell^2$
in the neighborhood of $r=0$. Then, the geometric invariants
take the values
\begin{subequations}
\begin{eqnarray}
\left.\mathcal{R}\right|_{r=0}  &=&-{12}/{\ell^2},
\\
\left.\mathcal{R}_{\mu\nu}
\mathcal{R}^{\mu\nu}\right|_{r=0}  &=&{36}/{\ell^4},
\\
\left.\mathcal{R}_{\mu\nu\rho\sigma}
\mathcal{R}^{\mu\nu\rho\sigma}\right|_{r=0}  &=&{24}/{\ell^4},
\end{eqnarray}
\end{subequations}
at $r=0$. 

\subsection{Einstein tensor}
\label{Appendix:Energy condition}

Although we consider that the regularization
of the Vaidya spacetime would be realized by modification
of the gravity theory, it is interesting to discuss
what kind of matter is required in order to realize the
regularized Vaidya spacetime within the framework of general relativity.
For this reason, we examine the properties of the Einstein tensor
in detail. 
For the metric given by Eq.~\eqref{Metric:IngoingVaidya}, 
the nonzero components of the Einstein tensor ${G^\mu}_{\nu}$ are
\begin{subequations}
\begin{eqnarray}
  {G^v}_v \ = \ {G^r}_r &=& \frac{F_--1}{r^2}+\frac{F_{-,r}}{r},\\
  {G^\theta}_\theta \ = \ {G^\phi}_\phi &=& \frac{F_{-,r}}{r}+\frac{F_{-,rr}}{2},\\
  {G^r}_v &=& -\frac{F_{-,v}}{r}.
\end{eqnarray}
\end{subequations}
Adopting the same tetrad $(e^a)_\mu$ as
Eqs.~\eqref{tetrad0}--\eqref{tetrad3} but $F$ being replaced by $F_-$,
the tetrad components of the Einstein tensor is written as
\begin{equation}
  G^{ab}=\left(
  \begin{array}{cccc}
    \rho & 0 & 0 & 0\\
    0 & -\rho & 0 & 0\\
    0 & 0 & p & 0\\
    0 & 0 & 0 & p
    \end{array}
  \right)+\left(
  \begin{array}{cccc}
    \nu & \nu & 0 & 0\\
    \nu & \nu & 0 & 0\\
    0 & 0 & 0 & 0\\
    0 & 0 & 0 & 0
    \end{array}
  \right),
  \label{Einstein-tensor-canonical-typeII}
\end{equation}
(which belongs to the canonical form of type II \cite{Hawking:1973},
see also \cite{Maeda:2022}),
where
\begin{subequations}
\begin{eqnarray}
\rho &=& \frac{1-F_--rF_{-,r}}{r^2},\\
p &=& \frac{F_{-,r}}{r}+\frac{F_{-,rr}}{2},\\
\nu &=& -\frac{F_{-,v}}{r}.\label{formula-nu}
\end{eqnarray}
\end{subequations}
In the case of the charged ingoing Vaidya spacetime, the first
and second
terms of the right-hand side of Eq.~\eqref{Einstein-tensor-canonical-typeII}
correspond to the electromagnetic part and the matter part
of Eq.~\eqref{charged-ingoing-Vaidya-Tmunu-spilit}, respectively
(except for the factor $8\pi$),
and the null vector $k^\mu$ of Eq.~\eqref{ingoing-radial-null-vector}
is expressed as $k^\mu = (e^0)^\mu - (e^1)^\mu$ in terms of the tetrad
basis here.
$\rho$, $p$,
and $\nu$ here are related to 
the quantities in Eqs.~\eqref{energy-momentum-tensor-EM}
and \eqref{energy-momentum-tensor-matter} as
$\rho/8\pi=\tilde{\rho}=-\tilde{p}_r$,
$p/8\pi=\tilde{p}_\theta = \tilde{p}_\phi$, and $\nu/8\pi=\tilde{\nu}$.

In terms of the quantities $\rho$, $p$, and $\nu$, the NEC is equivalent to
\begin{equation}
\nu\ge 0 \quad \textrm{and} \quad \rho+p\ge 0,
\end{equation}
while the DEC is equivalent to
\begin{equation}
\nu\ge 0 \quad \textrm{and} \quad \rho\ge |p|.
\end{equation}
Below, we examine the behavior of $\rho$, $p$, and $\nu$,
and whether the NEC and DEC are satisfied.

\subsubsection{Self-similar case}

In the self-similar case, the function $F_-$ has the form
$F_-=f(R)$ with $R=r/M(v)$. In this case, the three quantities have
the form
\begin{subequations}
\begin{eqnarray}
\rho &=& \frac{1}{M(v)^2}\frac{1-f-Rf^\prime(R)}{R^2},\\
p &=& \frac{1}{M(v)^2}\left(\frac{f^\prime(R)}{R}+\frac{f^{\prime\prime}(R)}{2}\right),\\
\nu &=& \frac{M^\prime(v)}{M(v)^2}\, f^\prime(R).
\end{eqnarray}
\end{subequations}

%
\begin{figure}[tb]
  \centering
  \includegraphics[width=0.4\textwidth,bb= 0 0 360 263]{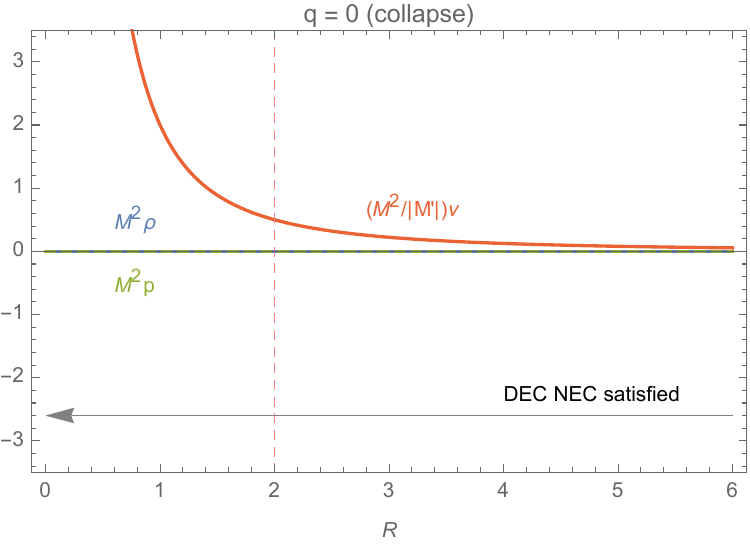}
  \includegraphics[width=0.4\textwidth,bb= 0 0 360 263]{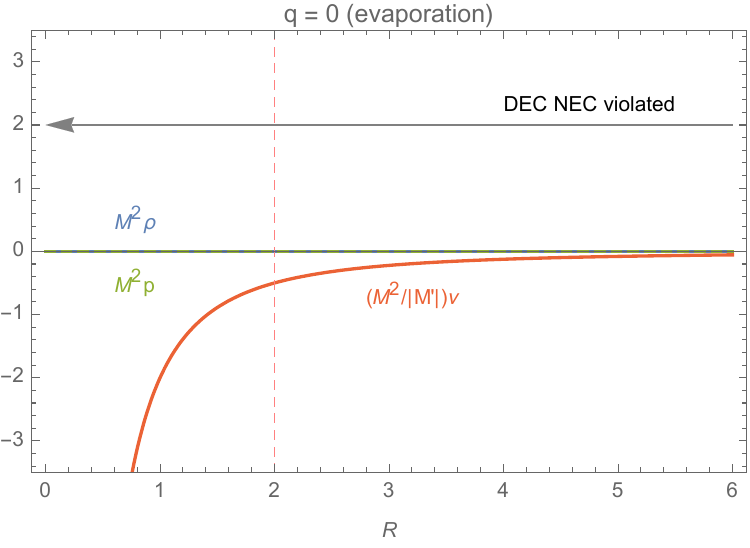}
  \caption{The behavior of $M^2\rho$, $M^2p$, and $(M^2/M^\prime)\nu$
    as functions of $R$ in the case of a uncharged ingoing Vaidya spacetime
    in the collapsing phase (left panel) and in the evaporation phase
    (right panel). The vertical red dashed line indicates the location of the apparent horizon.}
  \label{EnergyCondition-Vaidya-q000}
\end{figure}
%

In the (uncharged) ingoing Vaidya spacetime with $q=0$,
where $f(R)$ is given by Eq.~\eqref{Eq:definition-fR}, 
we have $\rho=p=0$ and
\begin{equation}
\nu \ = \ \frac{M^\prime}{M^2}\ \frac{2}{R^2}.
\end{equation}
Figure~\ref{EnergyCondition-Vaidya-q000}
shows the behavior of $M^2\rho$, $M^2p$, and $(M^2/M^\prime)\nu$
    as functions of $R$ 
    in the collapsing phase (left panel) and in the evaporation phase
    (right panel).
In the collapsing phase where $M^\prime(v)$ is positive,
$\nu$ is positive for an arbitrary $R$, and thus,
both the NEC and DEC are satisfied everywhere.
On the other hand, in the evaporating phase where $M^\prime(v)$ is negative,
$\nu$ becomes negative and the NEC and DEC are violated for arbitrary $R$.
This is because negative energy
flows into the central region.

%
\begin{figure}[tb]
  \centering
  \includegraphics[width=0.4\textwidth,bb= 0 0 360 263]{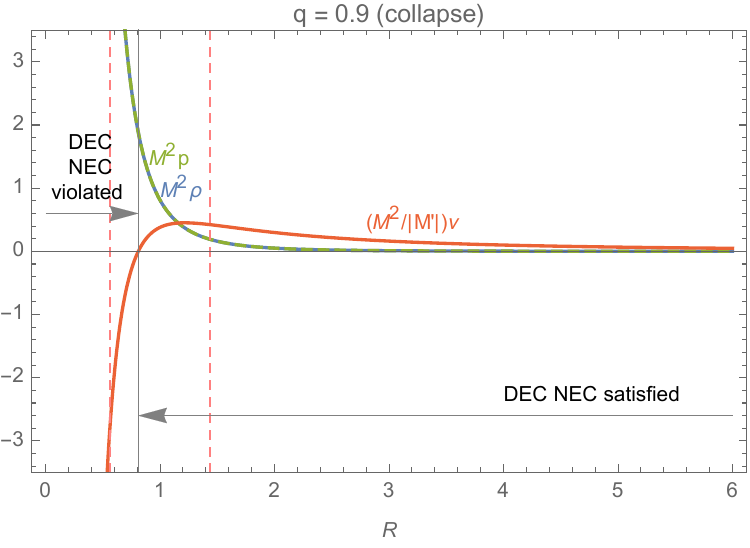}
  \includegraphics[width=0.4\textwidth,bb= 0 0 360 263]{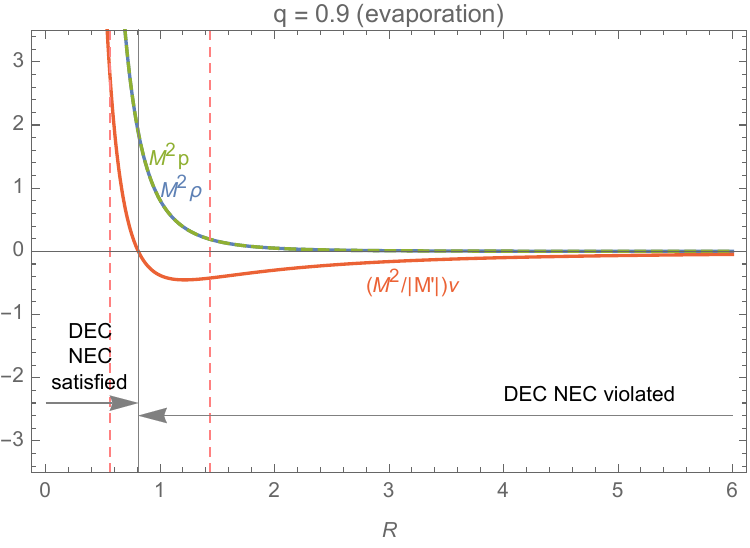}
  \caption{The same as Fig.~\ref{EnergyCondition-Vaidya-q000} but for $q=0.90$.
    The left and right vertical red dashed lines indicate the locations of the
    inner and outer apparent horizons, respectively.}
  \label{EnergyCondition-Vaidya-q090}
\end{figure}
%

In the case of charged ingoing Vaidya spacetime with
$q\neq 0$ given by Eq.~\eqref{Eq:definition-fR}, 
$p$, $\rho$, and $\nu$ are calculated as
\begin{subequations}
\begin{eqnarray}
  p \ = \ \rho & = & \frac{q^2}{M^2R^4}, \\
  \nu &=& \frac{2M^\prime}{M^2R^3}\left(R-q^2\right).
  \label{nu-charged-ingoing-Vaidya}
\end{eqnarray}
\end{subequations}
$\rho$ and $p$ have the positive values,
and this corresponds to the energy density and pressures of the
electric field. 
Figure~\ref{EnergyCondition-Vaidya-q000}
shows the case of $q=0.9$.
In the collapsing phase, $\nu$ is positive for $R>q^2$
while it becomes negative for $R<q^2$ (as Ori pointed out in \cite{Ori:1991-2}).
Correspondingly, around the center, the NEC and DEC are violated.
Hence, the NEC and DEC are violated and satisfied for $R<q^2$ and $R>q^2$,
respectively.
In the evaporating phase, the opposite happens:
negative energy flows to the center for large $R$, but
it becomes positive around the center.
Hence, the NEC and DEC are satisfied and violated for $R<q^2$ and $R>q^2$,
respectively.

%
\begin{figure}[tb]
  \centering
  \includegraphics[width=0.4\textwidth,bb= 0 0 360 263]{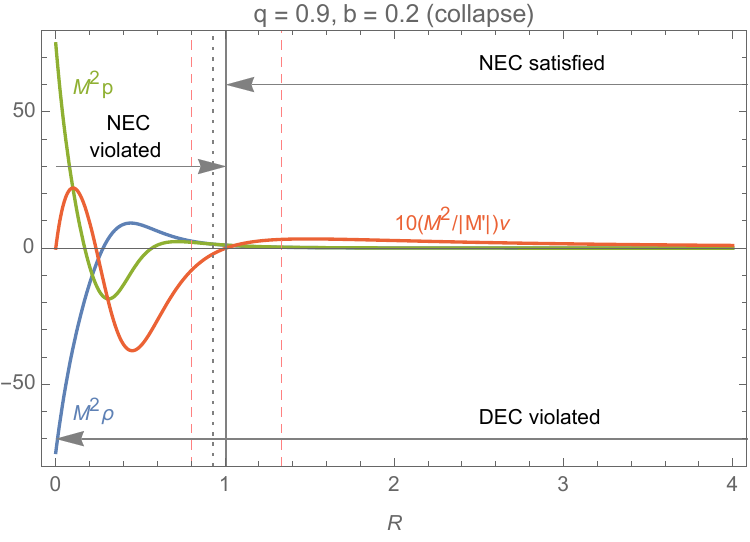}
  \includegraphics[width=0.4\textwidth,bb= 0 0 360 263]{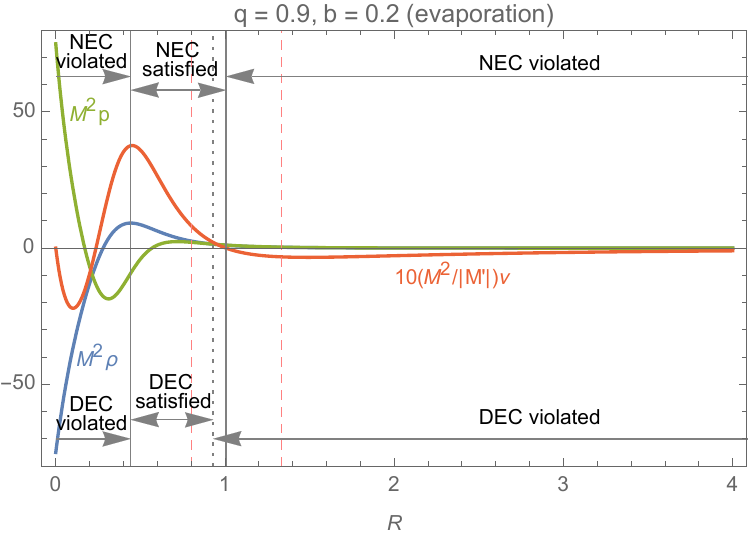}
  \caption{The same as Fig.~\ref{EnergyCondition-Vaidya-q090} but for
    a partly regularized spacetime with $q=0.90$ and $b=0.20$.
    As for the function $\nu$, we plot $10(M^2/|M^\prime|)\nu$
  for visibility.}
  \label{EnergyCondition-RegularizedVaidya-q090b020}
\end{figure}
%

We now turn to the partly regularized spacetime with $b\neq 0$
given by Eq.~\eqref{Definition-f(R)}. 
$p$, $\rho$, and $\nu$ are given as
\begin{subequations}
\begin{eqnarray}
  \rho & = & \frac{q^2R^4+b^2(12R^2+4q^2R-3q^4)}{M^2[R^4+b^2(2R+q^2)]^2}, \\
  p &=& \frac{q^2R^8+6b^2R^4(4R^2+q^2R-2q^4)+b^4(-24R^3-28q^2R^2-6q^4R+3q^6)}{M^2[R^4+b^2(2R+q^2)]^3},\\
  \nu &=& \frac{M^\prime}{M^2}
  \frac{2R\left[R^5-q^2R^4+b^2(-4R^2-2q^2R+q^4)\right]}{[R^4+b^2(2R+q^2)]^2}.
\end{eqnarray}
\end{subequations}
The behavior of $\rho$, $p$, and $\nu$ are relatively complicated
as shown in Figure~\ref{EnergyCondition-RegularizedVaidya-q090b020}.
The value of $\rho$ is positive for large $R$, and becomes negative
as $R$ is decreased. It takes the value $\rho=-3/b^2M^2$ at $r=0$.
The value of $p$ is positive for large $R$. As $R$ is decreased,
it once becomes negative,
and then, becomes positive again,
and takes the value $p=3/b^2M^2$ at $r=0$.
In the collapsing phase, 
the value of $\nu$ is positive for large $R$.
As $R$ is decreased, it once becomes negative, and then, becomes positive again,
and takes the value $\nu=0$ at $r=0$.
The DEC is violated everywhere. The NEC is satisfied
in the distant region, while it is violated around the center.
In the evaporating phase, the behavior of $\nu$ is opposite.
There are two regions where both the NEC and DEC are violated
around the center and at the distant region.
Between them, there is a region where the NEC and DEC are satisfied.
There is also a small region where the NEC is satisfied
but the DEC is violated.

\subsubsection{Completely regularized case}

In the completely regularized case, we have
\begin{subequations}
\begin{eqnarray}
  \rho & = & \frac{M^2\left[q^2r^4+\ell^2(12r^2+4q^2Mr-3q^4M^2)\right]}
       {[r^4+\ell^2(2Mr+q^2M^2)]^2}, \\
       p &=& \frac{M^2\left[q^2r^8+6\ell^2r^4(4r^2+q^2Mr-2q^4M^2)\right]}
       {[r^4+\ell^2(2Mr+q^2M^2)]^3},\nonumber\\&&
       \qquad+\frac{\ell^4M^3[-24r^3-28q^2Mr^2+3q^4M^2(-2r+q^2M)]}
       {[r^4+\ell^2(2Mr+q^2M^2)]^3},\\
  \nu &=& 
  \frac{2r^2M^\prime\left(r^4-q^2Mr^3-2\ell^2q^2M^2\right)}
       {[r^4+\ell^2(2Mr+q^2M^2)]^2},
\end{eqnarray}
\end{subequations}
from Eq.~\eqref{F-ingoingVaidya}. 
Here, we consider the case of $q=0.9$ with the linear mass function
considered in the right panel of Fig.~\ref{geodesic_AH_linearmass}.
The collapsing phase is $-v_{\rm i}\le v\le 0$ and the evaporation phase
is $0\le v\le v_{\rm f}$, where $v_{\rm i}=6\ell$ and $v_{\rm f}=60\ell$.

%
\begin{figure}[tb]
  \centering
  \includegraphics[width=0.4\textwidth,bb= 0 0 360 256]{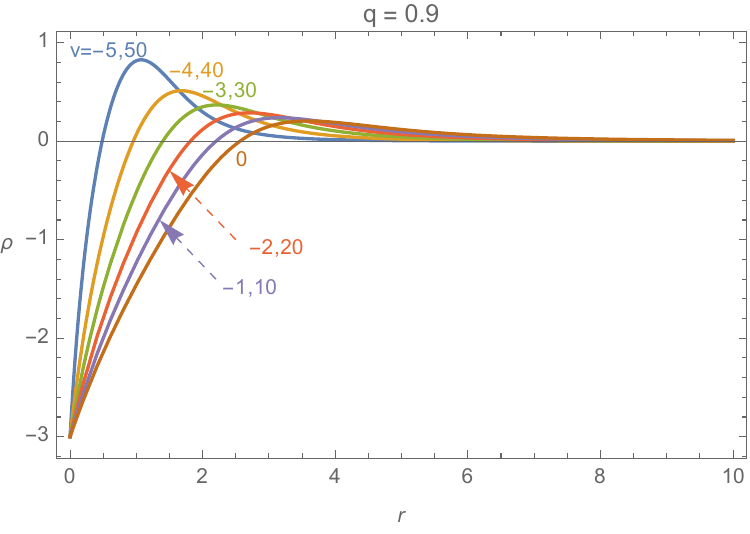}
  \includegraphics[width=0.4\textwidth,bb=  0 0 360 256]{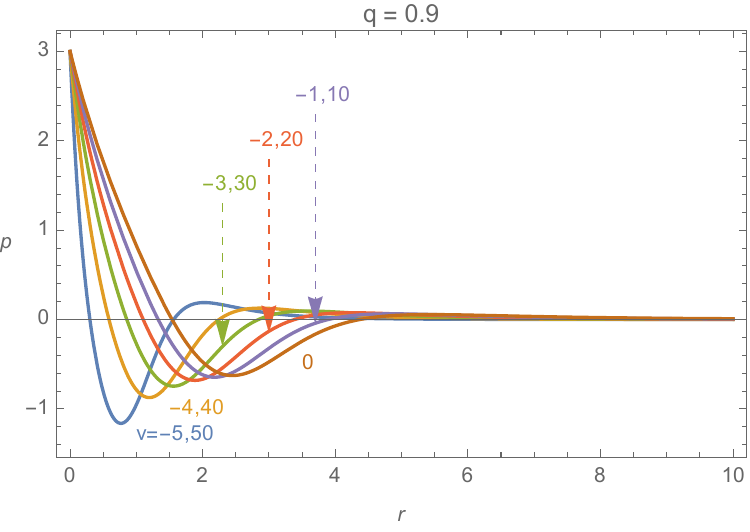}
  \includegraphics[width=0.4\textwidth,bb=  0 0 360 256]{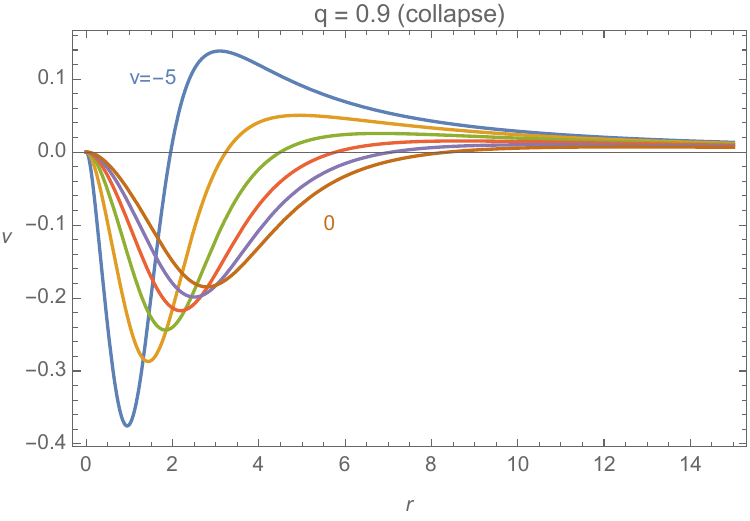}
  \includegraphics[width=0.4\textwidth,bb=  0 0 360 256]{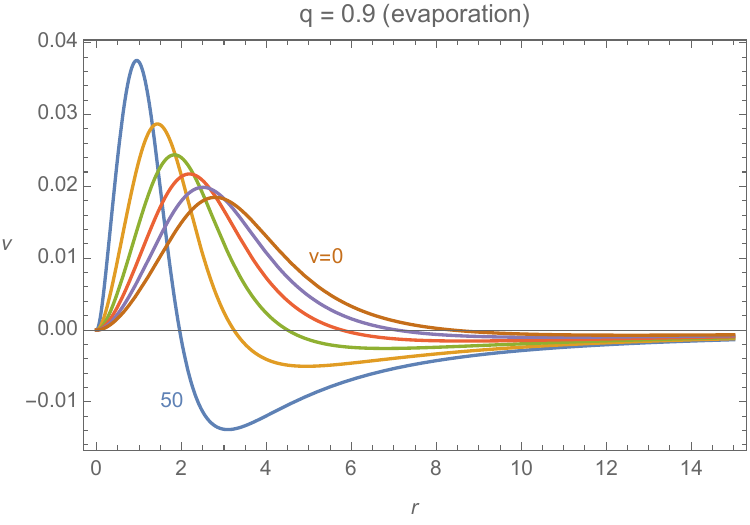}
  \caption{The snapshots of $\rho$ (top left), $p$ (top right), and $\nu$ (bottom left and bottom right) as functions of $r$ for $v=-5$, $-4$, $-3$, $-2$, $-1$, $0$, $10$, $20$, $30$, $40$, and $50$ in the unit $\ell=1$.}
  \label{EnergyCondition-CompletelyRegularizedVaidya-q090}
\end{figure}
%

Figure~\ref{EnergyCondition-CompletelyRegularizedVaidya-q090} shows the snapshots of $\rho$ (top left), $p$ (top right), and $\nu$ (bottom)
as functions of $r$ for $v=-5$, $-4$, $-3$, $-2$, $-1$, $0$,
$10$, $20$, $30$, $40$, $50$
in the unit $\ell=1$.
The value of $\rho$ is positive for large $r$, becomes negative
as $r$ is decreased, and has the value $\rho=-3/\ell^2$ at $r=0$.
The value of $p$ is positive for large $r$,
once becomes negative as $r$ is decreased, and then, 
becomes positive again, and has the value 
$p=3/\ell^2$ at $r=0$.
The functions $\rho(r)$ and $p(r)$ for some $v>0$
in the evaporation phase are the same as
those at time $-(v_{\rm i}/v_{\rm f})v$ in the collapsing phase,
because these functions are determined by the value of $M$.
In the collapsing phase, 
the value of $\nu$ is positive for large $r$ and negative for small $r$.
The function $\nu(r)$ for some $v>0$ in the evaporation phase is the same
as that at $-(v_{\rm i}/v_{\rm f})v$ except that it is flipped
about the horizontal axis (i.e., the sign is changed) and the absolute value
is decreased by the factor of $v_{\rm i}/v_{\rm f}$.

%
\begin{figure}[tb]
  \centering
  \includegraphics[width=0.9\textwidth,bb= 0 0 1920 938]{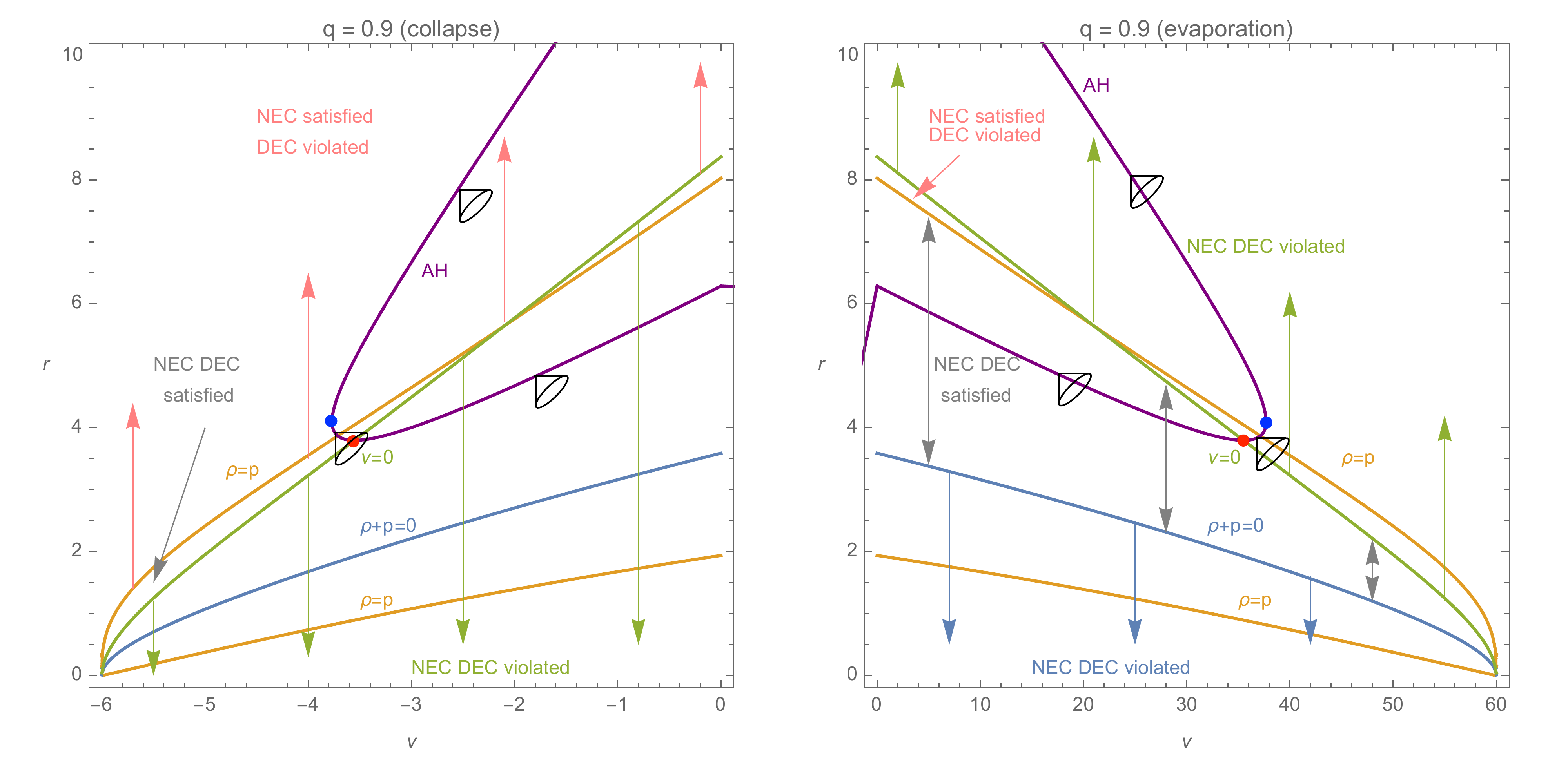}
  \caption{The domains where the NEC and DEC are satisfied/violated in the
    $(v,r)$-plane in the collapsing phase (left panel) and in the evaporation phase (right panel). The purple curve is the location of the apparent horizons.
    At blue and red points, the apparent horizon changes its characteristic.
  The light cones at some points on the apparent horizon are also indicated.}
  \label{EnergyCondition-CompletelyRegularizedVaidya-v-r-plane}
\end{figure}
%

Figure~\ref{EnergyCondition-CompletelyRegularizedVaidya-v-r-plane} 
shows the domains where the NEC and DEC are satisfied/violated in the
$(v,r)$-plane in the collapsing phase (left panel) and in the evaporation phase (right panel).
In the collapsing phase, the NEC is violated
around the center (as indicated by green arrows)
and is satisfied at the distant region (as indicated by red arrows).
The DEC is violated in almost all regions, except the narrow region
indicated by the grey arrow.
In the evaporation phase, the NEC and DEC are violated
around the center (as indicated by blue arrows) and at the distant
region (as indicated by green arrows).
Between them, there is a region where the NEC and DEC
are satisfied as indicated by grey arrows.
There is also a narrow region (indicated by the red arrow) where the NEC is 
satisfied but the DEC is violated.

\subsection{Characteristic of apparent horizons}
\label{Characteristic_of_AH}

We discuss the condition that determines
the characteristic 
of the sequence of the apparent horizon
(i.e., whether it is spacelike, null, timelike),
and provide a support for the validity of our Penrose diagrams in the main text.

\subsubsection{General criteria}
\label{Caracteristic:Genreal_criterion}

The criteria for determining the characteristics of the apparent horizons
can be derived as follows.
Since $F_-=0$ holds on the apparent horizon,
we have 
\begin{equation}
F_{-,v}dv+F_{-,r}dr \, = \, 0
\end{equation}
on the sequence of the apparent horizons.
Substituting this relation into the metric together with
$F_-=0$, we have the induced metric on the apparent horizon,
\begin{equation}
  ds^2 \,=\, -\frac{2F_{-,v}}{F_{-,r}}dv^2+r^2d\Omega^2.
  \label{induced-metric-AH}
\end{equation}

In the case of self-similar spacetimes,
the function $F_-$ is given by $F_-(v,r)=f(R)$ with $R=r/M(v)$.
Substituting $F_{-,v} = -(M^\prime/M)Rf^\prime(R)$ and $F_{-,r}=f^\prime(R)/M$ 
into Eq.~\eqref{induced-metric-AH}, we have
\begin{equation}
ds^2 \, = \, 2M^\prime Rdv^2+r^2d\Omega^2,
\end{equation}
that is, the apparent horizon is spacelike if $M^\prime>0$
and is timelike if $M^\prime < 0$.
This means that the apparent horizon is spacelike
in the collapsing phase, and is timelike in the evaporation phase.
This result is consistent with our Penrose diagrams
for the self-similar spacetimes.

In the case that the spacetime is not self-similar,
we have to use Eq.~\eqref{induced-metric-AH}.
Let us recall the fact that $\nu$ is given by Eq.~\eqref{formula-nu},
and hence, the metric on the apparent horizons is rewritten as
\begin{equation}
ds^2 \, = \, \frac{2r\nu}{F_{-,r}}dv^2+r^2d\Omega^2.
\end{equation}
We now consider the case where $F_-=1$ holds at infinity $r=\infty$ and
at the center $r=0$, and the inner and outer apparent horizons exist.
In such a situation, $F_{-,r}<0$ holds on the inner apparent horizon,
while $F_{-,r}>0$ holds on the outer apparent horizon.
This discussion gives the following criteria:
\begin{itemize}
\item The outer apparent horizon is spacelike if $\nu>0$ and is timelike if $\nu<0$;
\item The inner apparent horizon is timelike if $\nu>0$ and is spacelike if $\nu<0$.
\end{itemize}
There are two possibilities where the sequence of the apparent horizons
becomes null: The case  $F_{-,r}=0$ and the case $\nu=0$.
In the former case, the sequence of the apparent horizons becomes tangential to
a $v=\mathrm{constant}$ surface, and 
it is the point where the inner and outer
apparent horizons are connected. In the latter case,
the sequence of the apparent horizons becomes tangential to a $u=\mathrm{constant}$ surface,
and it occurs
at the point where the apparent horizon changes its characteristic in each
of the outer and inner sequences.
Let us examine whether Fig.~\ref{EnergyCondition-CompletelyRegularizedVaidya-v-r-plane}
is consistent with these criteria.
The outer apparent horizon is spacelike (timelike) for $v<0$ ($v>0$) and
it is located in the region $\nu>0$ ($\nu<0$).
In the region around $v=0$ (i.e., between the blue and red points in each panel),
the inner apparent horizon
is spacelike (timelike) for $v<0$ ($v>0$) and
it is located in the region $\nu<0$ ($\nu>0$).
At some values of $v$ (i.e., at red points),
the inner apparent horizon crosses the
contour surface of $\nu=0$,
and change of the characteristic occurs there
(i.e., $dr/dv$ of the apparent horizon becomes zero).
Therefore, our results satisfy the above criteria, indicating
the validity of our results.

\subsubsection{Consistency with Hayward's theorem}

Here, we briefly comment on
the consistency of our 
results on the apparent horizons with Hayward's theorem \cite{Hayward:1994}.
In the framework of general relativity, 
Hayward has proven the fact that if the NEC holds,
an outer apparent horizon is spacelike,
and an inner apparent horizon is timelike.

In the unregularized charged ingoing Vaidya spacetime where $f(R)$
is given by Eq.~\eqref{Eq:definition-fR},
$\rho$ and $p$ are both positive,
and hence, $\rho+p$ is also always positive.
Therefore, whether the NEC is satisfied or not is solely determined by
the sign of $\nu$. From Eq.~\eqref{Eq:Location-AH-charged-Vaidya},
the location of the apparent horizons, $R_{\rm A}^\pm$, are written as
\begin{equation}
R_{\rm A}^\pm = \frac{q^2}{1\mp\sqrt{1-q^2}},
\end{equation}
which indicates $R_{\rm A}^+ > q^2$ and $R_{\rm A}^- < q^2$.
Since Eq.~\eqref{nu-charged-ingoing-Vaidya} tells that
$\nu$ is positive and negative in the regions $R>q^2$ and $R<q^2$,
respectively, 
$\nu$ is positive (negative) at $R=R_{\rm A}^+$ in the collapsing (evaporating) phase,
and 
$\nu$ is negative (positive) at $R=R_{\rm A}^-$ in the collapsing (evaporating) phase.
Therefore, Hayward's theorem states that
the outer (inner) apparent horizon must be spacelike
(timelike) in the collapsing (evaporating)
phase. 
On the characteristic of the outer (inner) apparent horizon
in the evaporating (collapsing) phase,
Hayward's theorem tells nothing. These are consistent with 
Fig.~\ref{Fig:Penrose-Vaidya-w-HKH}.

In the partly regularized model where $f(R)$ is given by 
Eq.~\eqref{Definition-f(R)},
it is difficult to discuss whether the null energy condition
is satisfied based on the exact formulas for the location
of the apparent horizon, $R=R_{\rm A}^\pm$,
because $f(R)$ is complicated.
However, as long as the value of $b^2$ is small,
the values of $R_{\rm A}^\pm$ are shifted
by $O(b^2)$ from the $b=0$ case. Thus, the values of $\rho+p$ and $\nu$
at $R=R_{\rm A}^\pm$ are also different from the $b=0$ case
by $O(b^2)$. Therefore, the same conclusion as the $b=0$ case must hold,
and Fig.~\ref{Fig:Penrose-regularized-self-similar} is consistent
with Hayward's theorem.

Finally, we discuss the completely regularized case
using Fig.~\ref{EnergyCondition-CompletelyRegularizedVaidya-v-r-plane}
as an example.
Although one of the null energy conditions,
$\rho+p\ge 0$, is violated around the center,
this condition is satisfied on the apparent horizons.
Therefore, whether the NEC is satisfied or not is solely determined by
the sign of $\nu$. Then, from the criteria for the
characteristic of the apparent horizons in terms of $\nu$ obtained in
Appendix~\ref{Caracteristic:Genreal_criterion}, we conclude
that the characteristics of the apparent horizons
are consistent with Hayward's theorem.

\section{Connecting the ingoing and outgoing charged Vaidya spacetimes}
\label{Appendix:Outer-domain}

In Sects.~\ref{Sec:review-part}, \ref{Sec:Self-similar-model}
and \ref{Sec:Non-self-similar},
we studied the structure of the ingoing charged Vaidya spacetimes.
In order to make a model for the evaporation of a black hole,
we must cut and glue the ingoing and outgoing Vaidya spacetimes
on some timelike hypersurface 
(e.g., the procedure to make the right panel
from the left panel in Fig.~\ref{Fig:Penrose-Vaidya-w-HKH}).
We discuss the method of this procedure here. 
The metrics of ingoing and outgoing Vaidya spacetimes
are given by Eqs.~\eqref{Metric:IngoingVaidya}
and \eqref{Metric:OutgoingVaidya}, respectively. 
The two spacetimes are connected on a timelike hypersurface,
$\Gamma$. The surface $\Gamma$ is specified by
$v=v_-(r)$ and $u=u_+(r)$
in the ingoing and outgoing Vaidya spacetimes, respectively.
For the connection to be continuous, 
the two induced metrics on the hypersurface $\Gamma$ must coincide,
\begin{equation}
  F_-(v_-(r),r)(v^\prime_-)^2-2v^\prime_-
  \ = \  F_+(u_+(r),r)(u^\prime_+)^2+2u^\prime_+.
  \label{induced-metric_continuous2}
\end{equation}

\subsection{Self-similar case}

Here we consider the self-similar spacetimes
studied in Sects.~\ref{Sec:review-part} and \ref{Sec:Self-similar-model}.
As for the ingoing Vaidya spacetime, 
the function $F_-(v,r)$ is chosen to be the same as
Eq.~\eqref{F-ingoingVaidya}, but here, we add the subscript
``$-$'' to $M(v)$ and $Q(v)$ as $M_-(v)$ and $Q_-(v)$, respectively.
The function $F_-(v,r)$ is
\begin{equation}
F_-(v,r) = 1-\frac{(2M_-(v)r-Q_-(v)^2)r^2}{r^4+(2M_-(v)r+Q_-(v)^2)M_-(v)^2b^2}, 
\end{equation}
with $Q_-(v) = q M_-(v)$.
The mass function is the same as 
Eq.~\eqref{Linear-mass-function}, and we focus on the range $0\leq v\le v_{\rm f}$,
\begin{equation}
  \qquad M_-(v) = M_0\left(1-\frac{v}{v_{\rm f}}\right).
\end{equation}
In the same manner to Secs~\ref{Sec:review-part} and \ref{Sec:Self-similar-model}, the coordinate $R$ is introduced by $R=r/M_-(v)$,
and the surface $\Gamma$ on which the outgoing Vaidya spacetime is glued
is supposed to be $R=R_{\rm H}$. Then, the function $v_-(r)$ becomes
\begin{equation}
  v_-(r) = v_{\rm f} \left(1-\frac{r}{M_0R_{\rm H}}\right).
  \quad 
\end{equation}
The derivative of $v_-(r)$ is constant,
$v_-^\prime = -{v_{\rm f}}/{M_0R_{\rm H}}$.
The function $F_-(v,r)$ is expressed only in terms of $R$, and it is
denoted as $f(R)$ like Eq.~\eqref{Definition-f(R)}. Note that
on the surface $\Gamma$, the relations $F_{-}(v_-(r),r) = f(R_{\rm H})$
and $M_-(v_-(r)) = {r}/{R_{\rm H}}$ hold.

\subsubsection{Procedure for connecting two spacetimes}

For the function $F_+(u,r)$ in the outer domain,
we assume the same formula as $F_-(v,r)$ but
$M_-(v)$ and $Q_-(v)$ being replaced by $M_+(u)$ and $Q_+(u)$,
respectively:
\begin{equation}
  F_{+}(u,r) =
  1-\frac{(2M_+(u)r-Q_+(u)^2)r^2}{r^4+(2M_+(u)r+Q_+(u)^2)M_+(u)^2b^2},
\end{equation}
with $Q_+(u) =q M_+(u)$.
We would like to determine the two functions
$u_+(r)$ and $M_+(u)$. Since there is only one condition
given by Eq.~\eqref{induced-metric_continuous2}
for these functions, we have to impose one more
condition. Here, we assume the continuity of the two mass
functions on $\Gamma$, 
$M_-(v_-(r))=M_+(u_+(r))$, as a natural condition.\footnote{In Ref.~\cite{Frolov:2014jva}, the condition $v_-(r)=u_+(r)$ is used and this condition leads to a different model. The degree of freedom for choosing such conditions
  corresponds to what kind of discontinuity in the extrinsic curvature
  appears on $\Gamma$. }
This means that 
$F_-(v_-(r),r) = F_+(u_+(r),r) = f(R_{\rm H})$ holds on $\Gamma$.
Using this condition, Eq.~\eqref{induced-metric_continuous2} is solved as
\begin{equation}
u_+^\prime =
  v_-^\prime -\frac{2}{f(R_{\rm H})},
\end{equation}
where we chose the solution that satisfies $u_+^\prime<0$.
Integrating this equation, $u_+(r) = u_+^\prime (r-M_0R_{\rm H})$
is obtained.
The advanced time $u_{\rm f}:=u_+(0)$ at which the charged star completely
evaporates is given as
\begin{equation}
    u_{\rm f} 
    \ =\ v_{\rm f} +\frac{2M_0R_{\rm H}}{f(R_{\rm H})}.
    \label{Formula-for-uf}
\end{equation}
Using $u_{\rm f}$, the function $u_+(r)$ is expressed as
\begin{equation}
  u_+(r) = u_{\rm f}\left(1-\frac{r}{M_0R_{\rm H}}\right),
\end{equation}
and the mass function $M_+(u)$ that realizes the condition
$M_-(v_-(r))=M_+(u_+(r))=r/R_{\rm H}$
is 
\begin{equation}
M_+(u) = M_0\left(1-\frac{u}{u_{\rm f}}\right).
\end{equation}

\subsubsection{The surface stress tensor}
The surface $\Gamma$ is a singular hypersurface
in the sense of Israel's definition \cite{Israel:1966}.
To see this, we calculate the extrinsic curvature
of $\Gamma$ for both ingoing and outgoing Vaidya spacetime.
Recall that the surface $\Gamma$ is given by $R=R_{\rm H}$
in the ingoing Vaidya spacetime. In the outgoing Vaidya spacetime,
$\Gamma$ is also given by $R=R_{\rm H}$
introducing the coordinate $R$ by $R=r/M_+(u)$,
since $M_-(v)=M_+(u)$ holds on $\Gamma$. 
To simplify the presentation,
we calculate the extrinsic curvature
in a unified manner 
by introducing $z_-=v$ and $z_+=u$. The two metrics are
\begin{equation}
  ds^2 \ = \ -F_{\pm}(z_{\pm},r)dz_{\pm}^2\mp 2dz_{\pm} dr+r^2d\Omega^2.
  \label{metric_both}
\end{equation}
Introducing $R=r/M_{\pm}(z_{\pm})$, the metric components in the $(z_{\pm}, R, \theta, \phi)$ coordinates are expressed as
\begin{equation}
  g_{\mu\nu} =
  \left(
  \begin{array}{cccc}
    -\left(f(R)\pm 2M_{\pm}^\prime R\right) & \mp M_{\pm}(z_{\pm}) & 0 & 0\\
    \mp M_{\pm}(z_{\pm}) & 0 & 0 & 0\\
    0 & 0 & R^2M_{\pm}(z_{\pm})^2 & 0\\
    0 & 0 & 0 & R^2M_{\pm}(z_{\pm})^2\sin^2\theta 
    \end{array}
  \right).
\end{equation}
The inverse metrics are
\begin{equation}
  g^{\mu\nu} =\frac{1}{M_{\pm}(z_{\pm})^2}
  \left(
  \begin{array}{cccc}
    0 & \mp M_{\pm}(z_{\pm}) & 0 & 0\\
    \mp M_{\pm}(z_{\pm}) & f(R)\pm 2M_{\pm}^\prime R  & 0 & 0\\
    0 & 0 & R^{-2} & 0\\
    0 & 0 & 0 & R^{-2}\sin^{-2}\theta 
    \end{array}
  \right),
\end{equation}
and the unit normal to the hypersurface $R=\mathrm{constant}$
is given by
\begin{equation}
  n^\pm_\mu \ =\ \frac{M_-}{\sqrt{f(R)\pm 2M_\pm^\prime R}}\ (0,\ 1,\ 0,\ 0).
\end{equation}
The unit timelike tangent vector to $\Gamma$ is calculated as
\begin{equation}
  \hat{t}^\mu \ =\ \frac{1}{\sqrt{f(R)\pm 2M_\pm^\prime R}}\ (1,\ 0,\ 0,\ 0).
\end{equation}
The two spacelike unit tangent vectors are given as
\begin{subequations}
\begin{eqnarray}
  \hat{\theta}^\mu &=& \frac{1}{RM_\pm(z_\pm)}\left(0,\ 0,\ 1,\ 0\right),\\
  \hat{\phi}^\mu &=& \frac{1}{RM_\pm(z_\pm)\sin\theta}\left(0,\ 0,\ 0, \ 1\right).
\end{eqnarray}
\end{subequations}
The vectors $\hat{t}^\mu$, $\hat{\theta}^\mu$, and $\hat{\phi}^\mu$
consist of the triad basis of the hypersurface $\Gamma$.

The extrinsic curvatures
$K^{\pm}_{ij}$ are calculated below, where $i,j$ denotes
the components with respect to the triad basis, $\hat{t}^\mu$,
$\hat{\theta}^\mu$, and $\hat{\phi}^\mu$. 
Due to the spherical symmetry, the only nonzero components
are $K^{\pm}_{\hat{t}\hat{t}}$, $K^{\pm}_{\hat{\theta}\hat{\theta}}$, and
$K^{\pm}_{\hat{\phi}\hat{\phi}}$, where $K^{\pm}_{\hat{t}\hat{t}}=\hat{t}^\mu\hat{t}^\nu\nabla_\mu n^\pm_\nu$ and so on. 
A calculation gives
\begin{subequations}
\begin{eqnarray}
  K^\pm_{\hat{t}\hat{t}} &=& 
  -\frac{f^\prime(R)}{2M_{\pm}(z_\pm)\sqrt{f(R)\pm 2M_{\pm}^\prime R}},
  \\
  K^\pm_{\hat{\theta}\hat{\theta}} \ =\ K^\pm_{\hat{\phi}\hat{\phi}} &=&
  \frac{f(R)-M_{\pm}^\prime R}{RM_{\pm}(z_\pm)\sqrt{f(R)\pm 2M_{\pm}^\prime R}}.
\end{eqnarray}
\end{subequations}
Evaluating these values on $R=R_{\rm H}$ and substituting
$M_-^\prime =-M_0/v_{\rm f}$, 
$M_+^\prime =-M_0/u_{\rm f}$,
and Eq.~\eqref{Formula-for-uf}, we have
\begin{subequations}
  \begin{eqnarray}
K^+_{\hat{t}\hat{t}} & =&
  -\frac{f^\prime(R_{\rm H})}{2f(R_{\rm H})M_+(u)}
  \sqrt{\frac{v_{\rm f}f(R_{\rm H})+2M_0R_{\rm H}}{v_{\rm f}}},\\
K^-_{\hat{t}\hat{t}} &=& -\frac{f^\prime(R_{\rm H})}{2M_-(v)}
  \sqrt{\frac{v_{\rm f}}{v_{\rm f}f(R_{\rm H})+2M_0 R_{\rm H}}},
  \end{eqnarray}
  \begin{equation}
K^\pm_{\hat{\theta}\hat{\theta}} = K^\pm_{\hat{\phi}\hat{\phi}} \ =\
    \frac{v_{\rm f}f(R_{\rm H})+M_0 R_{\rm H}}{R_{\rm H}M_-(v)\sqrt{v_{\rm f}\left(v_{\rm f}f(R_{\rm H})+2M_0 R_{\rm H}\right)}}.
  \end{equation}
  \end{subequations}
  Introducing the discontinuity in the extrinsic curvature as
  $\left[K_{ij}\right] :=  K^+_{ij}-K^-_{ij}$, we have
  \begin{equation}
  \left[K_{\hat{t}\hat{t}}\right] \ = \ -\frac{M_0R_{\rm H}f^\prime(R_{\rm H})}{M_\Gamma f(R_{\rm H})
    \sqrt{v_{\rm f}\left[v_{\rm f}f(R_{\rm H})+2M_0R_{\rm H}\right]}},
  \end{equation}
  and $[K_{\hat{\theta}\hat{\theta}}]=  [K_{\hat{\phi}\hat{\phi}}]  = 0$,
  where $M_{\Gamma}$ denotes the mass function on $\Gamma$,
  i.e., $M_-(v)=M_+(u)=M_{\Gamma}$.
  If we choose $R_{\rm H}$ that satisfies $f(R_{\rm H})>0$ and
  $f^\prime(R_{\rm H})>0$, the value of $[K_{\hat{t}\hat{t}}]$
  is negative.

  On the singular hypersurface $\Gamma$, there exists distributional energy-momentum tensor (i.e. the surface stress tensor), and its specific form is determined once a theory of gravity is given. To realize the regularized spacetimes, we consider that the theory of gravity must be modified from the theory of general relativity , but an explicit calculation for the surface stress tensor in such a theory is postponed as a future issue.Instead, here we present the formula for the surface stress tensor in the framework of general relativity. This formula is expected to hold at the distant region $r\gg \ell$, because the theory of general relativity gives a good approximation in the domain where the curvature is small.

  The surface stress tensor $S_{ij}$ can be calculated 
  using Israel's junction condition \cite{Israel:1966},

\begin{equation}
  -8\pi GS_{ij}
  \ = \ \left[K_{ij}\right]
  -[K]\eta_{ij}.
\end{equation}
A calculation gives $S_{\hat{t}\hat{t}}=0$ and
$S_{\hat{\theta}\hat{\theta}}=S_{\hat{\phi}\hat{\phi}}=-[K_{\hat{t}\hat{t}}]/8\pi$.
Therefore, the surface $\Gamma$ has no surface energy density
and has negative tension. 
This situation is similar to
Hayward's model of Ref.~\cite{Hayward:2005}.

\subsection{Non-self-similar case}

We now turn our attention 
to the non-self-similar case.
The explicit construction of the evaporation model would
require numerical calculation, and here, we just
discuss the basic method for this procedure.
The function $F_-(v,r)$ is the same as
Eq.~\eqref{F-ingoingVaidya} but $M(v)$ and $Q(v)$ are
replaced by $M_-(v)$ and $Q_{-}(v)$ with $Q_-(v)=q M_-(v)$.
The mass function $M_-(v)$ is given by Eq.~\eqref{mass-more-realistic},
and the duration of the Hawking radiation measured in the
advanced time $v$ is denoted by $v_{\rm f}$.
We assume $F_+(u,r)$ to have the same form
as $F_-(v,r)$ but $M_-(v)$ and $Q_{-}(v)$
being replaced by $M_+(u)$ and $Q_{+}(u)$, respectively:
\begin{equation}
  F_+(u,r)=1-\frac{(2M_+(u)r-Q_+(u)^2)r^2}{r^4+(2M_+(u)r+Q_+(u)^2)\ell^2},
\end{equation}
with $Q_-(u)=q M_-(u)$.

One must first specify the timelike
hypersurface $\Gamma$ on which two spacetimes
are glued by choosing the function $v=v_-(r)$
in the ingoing Vaidya region.
A natural choice of $\Gamma$ would be a timelike hypersurface
outside of the apparent horizon
that satisfies $v_-(r)\to v_{\rm f}$ at $r\to 0$. 
Once such a hypersurface $\Gamma$ is specified,
we require the continuity of the mass function on it,
i.e. $M_-(v_-(r))=M_+(u_+(r))$.
This leads to the continuity
of the functions
$F_-(v,r)$ and $F_+(u,r)$ on $\Gamma$, $F_-(v_-(r),r) = F_+(u_+(r),r)$.
Then, Eq.~\eqref{induced-metric_continuous2}
is formally solved as
\begin{equation}
u_+(r) \ = \ v_-(r) - 2\int_{r_0}^{r}\frac{dr}{F_-(v_-(r),r)},
\end{equation}
where $r_0$ denotes the radius of $\Gamma$ at $v=0$ and 
the solution that satisfies $u_+^\prime(r)<0$ is chosen. 
The duration $u_{\rm f}$ of the Hawking radiation measured in the
retarded time is
\begin{equation}
  u_{\rm f}:=u_+(0) =
  v_{\rm f} + 2\int_{0}^{r_0}\frac{dr}{F_-(v_-(r),r)}.
\end{equation}  
Once the function $u=u_+(r)$ is obtained, one can
consider its inverse function $r=r_+(u)$. Then,
the mass function $M_+(u)$ is determined as
\begin{equation}
M_+(u) = M_-(v_-(r_+(u))).
\end{equation}

We calculate the surface stress tensor of the hypersurface $\Gamma$.
In the same way as the self-similar case, we introduce
the coordinates $z_+=u$ and $z_-=v$, and then,
the metrics are given by Eq.~\eqref{metric_both}.
In addition, we introduce
the function $\tilde{z}_+(r)=u_+(r)$ and $\tilde{z}_-(r)=v_-(r)$.
In the $(z_\pm, r, \theta, \phi)$ coordinates,
the unit normal to $\Gamma$ is calculated as
\begin{equation}
  n^\pm_\mu \ =\ \frac{1}{\sqrt{\tilde{z}_\pm^\prime(F_\pm \tilde{z}_{\pm}^\prime\pm 2)}}\ (1,\ -\tilde{z}_\pm^\prime,\ 0,\ 0),
\end{equation}
and the triad basis on $\Gamma$ is introduced as
\begin{eqnarray}
  \hat{t}^\mu &=& -\frac{1}{\sqrt{\tilde{z}_\pm^\prime(F_\pm \tilde{z}_{\pm}^\prime\pm 2)}}\ (\tilde{z}_\pm^\prime,\ 1,\ 0,\ 0),
\end{eqnarray}
\begin{subequations}
\begin{eqnarray}
  \hat{\theta}^\mu &=& \frac{1}{r}\left(0,\ 0,\ 1,\ 0\right),\\
  \hat{\phi}^\mu &=& \frac{1}{r\sin\theta}\left(0,\ 0,\ 0, \ 1\right).
\end{eqnarray}
\end{subequations}
The extrinsic curvature on $\Gamma$ calculated with the outgoing and ingoing
Vaidya metrics, $K^+_{ij}$ and $K^-_{ij}$ respectively, are
\begin{subequations}
\begin{equation}
  K^\pm_{\hat{t}\hat{t}} \ =\ -\frac{2\tilde{z}_{\pm}^{\prime\prime} \mp
  \left(\tilde{z}_{\pm}^{\prime}\right)^2\left[3F_{\pm,r} +\left(
    F_{\pm,z_{\pm}}\pm F_{\pm}F_{\pm,r}
    \right)\tilde{z}_{\pm}^\prime\right]}{2\left[
  \tilde{z}_\pm^\prime(F_{\pm}\tilde{z}_\pm^\prime\pm 2)
  \right]^{3/2}},
\end{equation}
\begin{equation}
  K^\pm_{\hat{\theta}\hat{\theta}} \ =\ K^\pm_{\hat{\phi}\hat{\phi}} \ =\
  -\frac{F_{\pm}\tilde{z}_{\pm}^\prime \pm 1}{r\sqrt{\tilde{z}_\pm^\prime(F_\pm \tilde{z}_{\pm}^\prime\pm 2)}}.
\end{equation}
\end{subequations}
Using the relation ${u}_+^\prime = {v}_-^\prime -2/F_-$,
$F_-=F_+$, $F_{-,r}=F_{+,r}$, and
$F_{-,v}v_-^\prime=F_{+,u}u_+^\prime$ that hold on $\Gamma$,
the discontinuity in the
extrinsic curvature is calculated as
  \begin{equation}
  \left[K_{\hat{t}\hat{t}}\right] \ = \ \frac{F_{-,v}v_{-}^\prime}{F_-\sqrt{v_-^\prime(F_-v_-^\prime-2)}},
  \end{equation}
  and $[K_{\hat{\theta}\hat{\theta}}]=  [K_{\hat{\phi}\hat{\phi}}]  = 0$.
 Then, similarly to the self-similar case, Israel's junction condition
  indicates that 
the surface $\Gamma$ has no surface energy density
and has nonzero tension in the framework of general relativity.

%
%
\section{Penrose diagram drawn in a different way}
\label{Appendix:Penrose_different}

%
\begin{figure}[tb]
  \centering
  \includegraphics[width=0.3\textwidth,bb= 0 0 432 767]{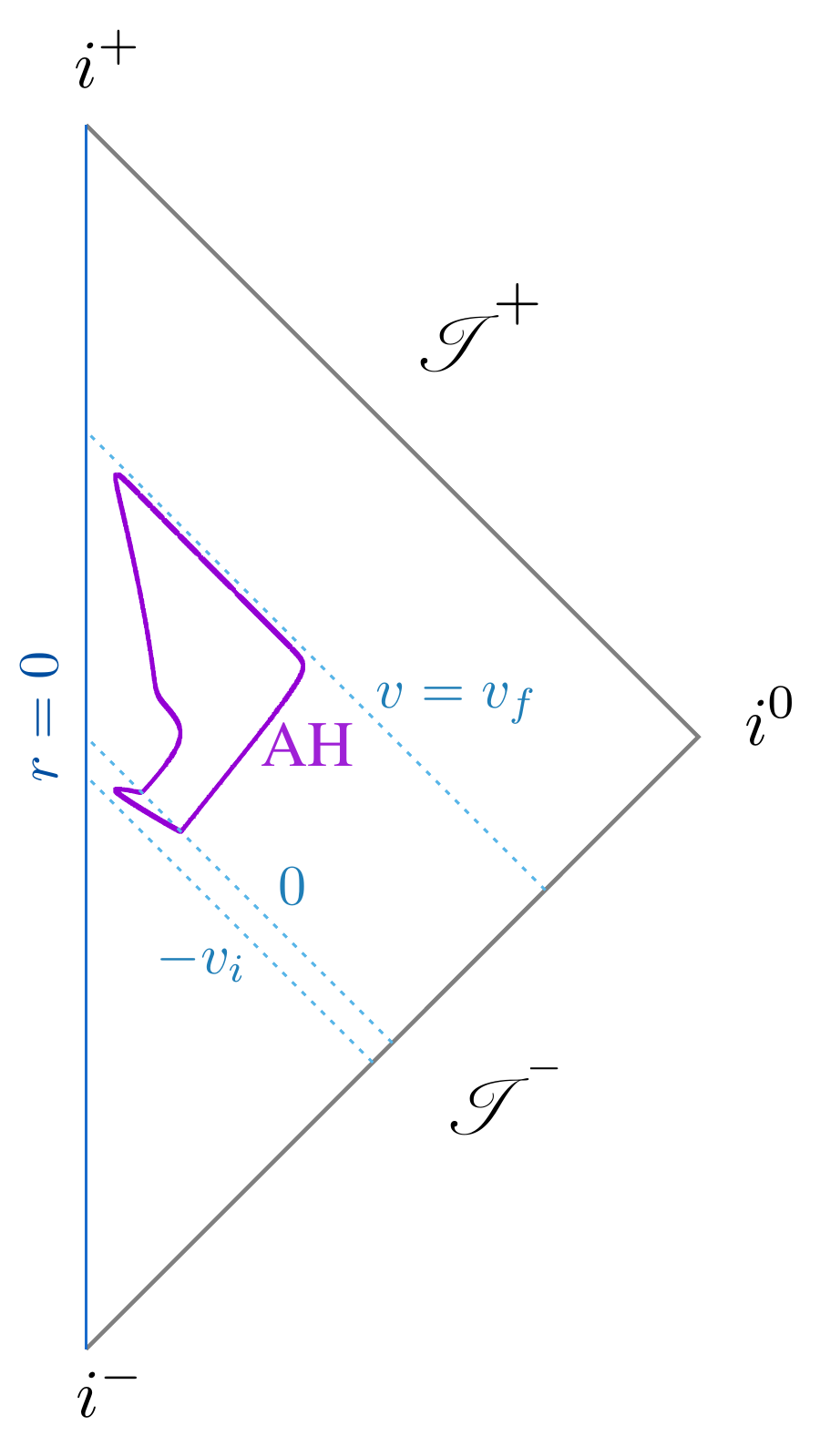}
  \caption{Penrose diagram for the same setup as that of Fig.~\ref{NSRN diagram_realistic mass} but drawn in a different way.}
  \label{Penrose_tilu_lam09_edited}
\end{figure}
%

In the Penrose diagram of Fig.~\ref{NSRN diagram_realistic mass},
the location of the inner boundary of the trapped region 
and that of $r=0$ seem to approximately coincide.
This is because of the effect of the adopted
advanced time $u$, and these two are, in fact,
located at different positions.
In this Appendix, we present the Penrose diagram
drawn in a different way to show this point.

In this Appendix, we denote
the advanced time as $\tilde{u}$.
Each outgoing null geodesic gives the 
$\tilde{u}$-constant surface,
and the value of $\tilde{u}$ is determined
by requiring $\tilde{u}=v$ at the center $r=0$.
Then, the compactified coordinates $\eta$ and $\zeta$
are introduced by the same formulas
as Eqs.~\eqref{Def:eta} and \eqref{Def:zeta}
except that $u$ is replaced by $\tilde{u}$. 
The worldline of the center $r=0$
becomes a straight vertical line in this method.
The Penrose diagram drawn in this way is presented in
Fig.~\ref{Penrose_tilu_lam09_edited}.
In this diagram, the inner boundary of the trapped region
and the center $r=0$ are clearly separated.
A similarity to Fig.~\ref{Fig:Penrose-Vaidya-wo-HKH}(a)
can be recognized.


\end{document}